\documentclass[vecphys]{svmult}

\usepackage{makeidx}         
\usepackage{graphicx}        
\usepackage{multicol}        
\usepackage[bottom]{footmisc}

\makeindex             

\begin{document}

\title*{Chemical evolution of the Milky Way and its Satellites
}
\author{Francesca Matteucci\inst{1,2}}
\institute{Astronomy Department, Trieste University
\texttt{matteucci@ts.astro.it}
\and Osservatorio Astronomico  (INAF), Trieste}
%
%
\maketitle

\section{How to model galactic chemical evolution}
\label{sec:1}
Before going into the detailed chemical evolution history of the Milky Way and its satellites, it is necessary to understand how to model, in general, galactic chemical evolution.
The basic ingredients to build a model of galactic chemical evolution 
can be summarized as :
\begin{itemize}
\item Initial conditions;
\item Stellar birthrate function (the rate at which stars are formed from the gas and their mass spectrum);
\item Stellar yields (how elements are produced in stars and restored into the interstellar medium);
\item Gas flows (infall, outflow, radial flow).
\end{itemize}

When all these ingredients are ready, we need to write a set of equations describing the evolution of the gas and its chemical abundances which include all of them.
These equations will describe the temporal variation of the gas content and its abundances by mass (see next sections). The chemical abundance of a generic chemical species $i$ is defined as:
\begin{equation}
X_i= {M_i \over M_{gas}}. 
\end{equation}
According to this definition it holds:
\begin{equation}
\sum_{i=1,n}{X_i}=1,
\end{equation}
where n represents the total number of chemical species.
Generally, in theoretical studies of stellar evolution it is common to adopt X, Y and Z as indicative of the abundances by mass of hydrogen (H), helium (He) and metals (Z), respectively.
The baryonic universe is madeup mainly of H and some He while only a very small fraction resides in metals (all the elements heavier than He), roughly 2\%. 
However, the history of the growth of this small fraction of metals is crucial 
for understanding how stars and galaxies were formed and subsequently evolved; and last but not least, because human beings exist only because of this small 
amount of metals!
We will focus then our attention is studying how the metals were formed and 
evolved in galaxies, with particular attention to our own Galaxy.


\subsection{The initial conditions}
\label{sec:2}
The initial conditions for a model of galactic chemical evolution consist in establishing whether :a) the chemical composition of the initial gas is
primordial or pre-enriched by a pre-galactic stellar generation; b)
the studied system is a  closed box or an open system (infall and/or outflow).

\subsection{Birthrate function}
The birthrate function, can be defined as:
\begin{equation}
B(M,t)=\psi(t) \varphi(m)
\end{equation}
where the quantity:
\begin{equation}
\psi(t)=SFR
\end{equation}
is called the star formation rate (SFR), namely the rate at which the gas is turned into stars, and the quantity:
\begin{equation}
\varphi(m)=IMF
\end{equation}
is the initial mass function (IMF), namely the mass distribution of the stars at birth.

\subsubsection{The star formation rate}
The most common parametrization of the SFR is the Schimdt (1959) law:
\begin{equation}
\psi(t) = \nu \sigma_{gas}^{k},
\end{equation}
where k=1-2 with a preference for $k=1.4 \pm 0.15$, as suggested by Kennicutt (1998a)
for spiral disks (see Figure 1), and $\nu$ is a parameter describing the star formation 
efficiency, in other words, the SFR per unit mass of gas, and it has the dimensions of the inverse of a time. Other physical quantities such as gas 
temperature, viscosity and magnetic field are usually ignored.

\begin{figure}
\centering
\includegraphics[width=8cm,height=8cm]{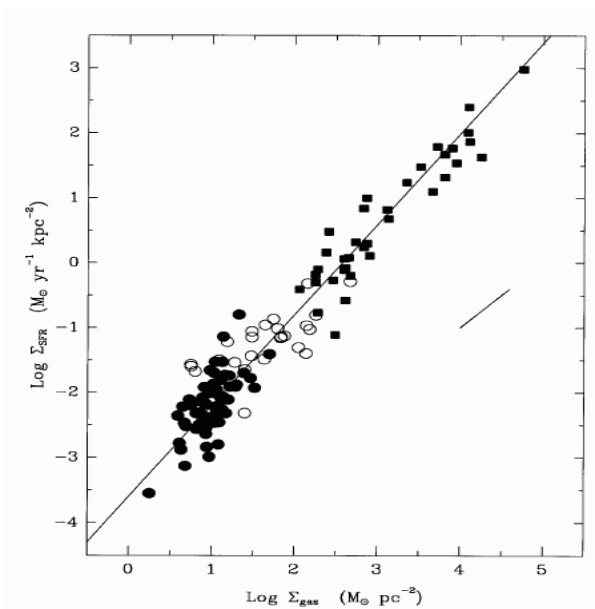}
\caption{The SFR as measured by Kennicutt (1998a) in star forming galaxies. The continuous line represents the best fit to the data and it can be achieved either with the SF law in eq. (6) with $k=1.4$   or with the SF law in eq. (9). The short, diagonal line shows the effect of changing the scaling radius by a factor of 2. Figure from Kennicutt (1998a).}
\label{fig:1}
\end{figure}

Other common parametrizations of the SFR include a dependence on the total 
surface mass density besides the surface gas density: 
\begin{equation}
\psi(t)= \nu \sigma_{tot}^{k_1} \sigma_{gas}^{k_2},
\end{equation}
as suggested by observational results of Dopita \& Ryder (1994) and taking into account the influence of the potential well in the star formation process (i.e. feedback between SN energy input and star formation, see also Talbot \& Arnett 1975).
Other suggestions concern the star formation  induced by spiral density waves
(Wyse \& Silk 1998) with expressions like:
\begin{equation}
\psi(t)=\nu V(R)R^{-1} \sigma_{gas}^{1.5},
\end{equation}
or
\begin{equation}
\psi(t)= 0.017 \Omega_{gas} \sigma_{gas}\propto R^{-1} \sigma_{gas}
\end{equation}
with $\Omega_{gas}$ being the angular rotation speed of gas (Kennicutt 1998a).
Also this law provides a good fit to the data of Figure 1.

\subsubsection{The initial mass function}

The most common parametrization of the IMF is a one-slope (Salpeter 1955) 
or multi-slope (Scalo 1986,1998; Kroupa et al. 1993; Chabrier 2003) power law.
The most simple example of a one-slope power law is:
\begin{equation}
\varphi(m)=am^{-(1+x)},
\end{equation}
generally defined in a mass range of 0.1-100 $M_{\odot}$, where $a$ is the normalization constant derived by imposing that $\int^{100}_{0.1}{m \varphi(m) dm} =1$.

The Scalo and Kroupa IMFs were derived from stellar counts in the solar vicinity and suggest a three-slope function. Unfortunately, the same analysis cannot be done in other galaxies and we cannot test if the IMF is the same everywhere. 
Kroupa (2001) suggested that the IMF in stellar clusters is a universal one, very similar to the Salpeter IMF for stars with masses  larger 
than $0.5M_{\odot}$. In particular, this universal IMF is:
\begin{eqnarray}
x_1= 0.3 \,\,\,\, for \,\,\, 0.08 \le M/M_{\odot} \le 0.50 \nonumber
\end{eqnarray}

\begin{eqnarray}
x_2= 1.3 \,\,\,\, for \,\,\, M/M_{\odot} > 0.5  
\end{eqnarray}

However, Weidner \& Kroupa (2005) suggested that the IMF integrated over galaxies, which controls the distribution of stellar remnants, the number of SNe and the chemical enrichment of a galaxy is generally different from the IMF in stellar clusters. This galaxial IMF is given by the integral of the stellar IMF over the embedded star cluster mass function which varies from galaxy to 
galaxy. Therefore, we should expect that the chemical enrichment histories of different galaxies cannot be reproduced by an unique invariant Salpeter-like IMF.
In any case, this galaxial IMF is always steeper than the universal IMF in the range of massive stars.

\subsubsection{How to derive the IMF}

We define the current mass distribution of local Main Sequence (MS) stars as 
the present day mass function (PDMF), $n(m)$. Let us suppose that we know 
$n(m)$ from observations.
Then, the quantity $n(m)$ can be expressed  as follows: for stars with initial 
masses in the range 0.1-1.0 $M_{\odot}$ which have lifetimes larger than a 
Hubble time we can write:
\begin{equation}
n(m)= \int^{t_{G}}_{0}{\varphi(m) \psi(t) dt}
\end{equation}
where $t_{G} \sim$ 14 Gyr (the age of the Universe).
The IMF, $\varphi(m)$, can be taken out of the integral if assumed to be 
constant in time, and the PDMF becomes:
\begin{equation}
n(m)=\varphi(m) <\psi> t_{G}
\end{equation}
where $<\psi>$ is the average SFR in the past.

For stars with lifetimes negligible relative to the age of the Universe, namely
for all the stars with $m > 2 M_{\odot}$, we can write:

\begin{equation}
n(m)= \int^{t_{G}}_{t_G -\tau_m}{\varphi(m) \psi(t) dt},
\end{equation}
where $\tau_m$ is the lifetime of a star of mass m.
Again, if we assume that the IMF is constant in time we can write:
\begin{equation}
n(m)= \varphi(m) \psi(t_G) \tau_m
\end{equation}
having assumed that the SFR did not change during the time interval between 
$(t_G -\tau_m)$ and $t_G$. The quantity $\psi(t_G)$ is the SFR  at the present time.

We cannot derive the IMF betwen 1 and 2 $M_{\odot}$ because none of the previous semplifying hypotheses can be applied.
Therefore, the IMF in this mass range will depend on a quantity, $b(t_G)$:
\begin{equation}
b(t_G)= {\psi(t_G) \over <\psi>}
\end{equation}
Scalo (1986) assumed:
\begin{equation}
0.5 \le b(t_G) \le 1.5
\end{equation}
in order to fit the two branches of the IMF in the solar vicinity.
In Figure 2 we show the differences between a single-slope IMF and multi-slope IMFs, which are preferred according to the last studies.
\begin{figure}
\centering
\includegraphics[width=8cm,height=8cm]{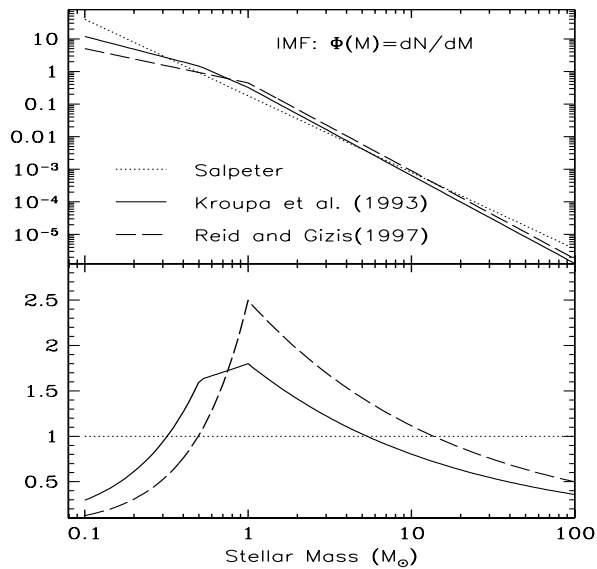}
\caption{Upper panel: different IMFs. Lower panel: normalization of the 
multi-slope IMFs to the Salpeter IMF. Figure from Boissier \& Prantzos (1999).}
\label{fig:3}
\end{figure}

\subsection{Stellar yields}
The stellar yields, namely the 
amount of newly formed  and pre-existing elements ejected by stars of 
all masses at their death, represent a fundamental ingredient to compute galactic chemical evolution. They can be calculated by knowing stellar evolution and nucleosynthesis.

I recall here the various stellar mass ranges and their nucleosynthesis 
products. In particular:

\begin{itemize}

\item Brown dwarfs: are stars with masses $M< 0.1 M_{\odot}$ which never 
ignite H. They do not enrich the interstellar medium (ISM) in chemical elements but only lock up gas.

\item Low and Intermediate mass stars ($0.8 \le M/M_{\odot} \le 8.0$).
Calculations are available from 
Marigo et al. (1996), van den Hoeck \& Groenewegen (1997), Forestini \& Charbonnel (1997), Marigo (2001), 
Meynet \& Maeder (2002),  Ventura et al. (2002),
Siess et al. (2002), Karakas \& Lattanzio (2007). These stars
produce
mainly $^{4}He$,
$^{12}C$, $^{14}N$ plus some CNO isotopes and  s-process ($A>90$)
elements. In Figure 3 we show an example of integrated yields from stars in this mass range.

\item Massive stars ($8 < M/M_{\odot} \le 40$).
In the mass range 10-40 $M_{\odot}$,
available calculations are from Woosley \& Weaver (1995, hereafter WW95), 
Langer \& Henkel (1995),  Thielemann et al. (1996), Nomoto et al. (1997),
Limongi \& Chieffi (2003), Rauscher et al. (2002), Meynet \& Maeder (2002), Nomoto et al. (2006), among others. 
These stars end their life as Type II SNe and explode by  core-collapse; 
they  produce  mainly $\alpha$-elements 
(O, Ne, Mg, Si, S, Ca), some Fe-peak elements,  
s-process elements ($A< 90$)
and r-process elements. Stars more massive than $40 M_{\odot}$ can end up
as Type Ib/c SNe: they are also core-collapse SNe and are linked to 
$\gamma$-ray bursts (GRB).

\item Type Ia SNe (white dwarfs in binary systems, see later).
Calculations are available from Nomoto et al. (1997),
Iwamoto et al. (1999). They produce mainly Fe-peak elements.

\item Very massive objects ($M> 100 M_{\odot}$). Calculations are available
from e.g.
Portinari et al. (1998), Umeda \& Nomoto (2001). 
They should produce mainly oxygen although many uncertainties are still 
present.

\end{itemize}

All the elements with mass number $A$ from 12 to 60 have 
been formed in stars during
the quiescent burnings.
Stars transform H into He and then He into 
heaviers until the 
Fe-peak elements, where the binding energy per nucleon reaches a maximum 
and the nuclear fusion reactions stop.
H is transformed into He through the proton-proton 
chain or the 
CNO-cycle, then $^{4}He$ is transformed into $^{12}C$ through the 
triple- $\alpha$
reaction.

Elements heavier than $^{12}C$ are then produced by synthesis 
of $\alpha$-particles: they are called $\alpha$-elements 
(O, Ne, Mg, Si and others).

The last main burning in stars is the $^{28}Si$ -burning which produces
$^{56}Ni$, which then decays into $^{56}Co$ and $^{56}Fe$.
Si-burning can be quiescent or explosive (depending on the temperature).

Explosive nucleosynthesis 
occurring during SN explosions 
mainly produces Fe-peak elements. Elements
originating from s- and r-processes (with A$> 60$ up to Th and U)
are formed by means of slow or rapid (relative to the $\beta$- decay)
neutron capture by Fe seed nuclei;
s-processing occurs during quiescent He-burning, 
whereas r-processing occurs during SN explosions.

In Figures 4, 5, 6, 7 and 8 we show a comparison between stellar yields for massive stars
computed for different initial stellar metallicities and with different assumptions concerning the mass loss. In particular, some yields are obtained by assuming mass loss by stellar winds with a strong dependence on metallicity (e.g. Maeder, 1992), whereas others (e.g. WW95) are computed by means of conservative models without mass loss. 
One important difference arises for oxygen in massive stars for solar metallicity and mass loss: in this case, the O yield is strongly depressed as a consequence of mass loss. In fact, the stars with masses $>25 M_{\odot}$ and solar metallicity lose a large amount of matter rich of He and C, thus subctracting these elements to further processing which would lead to O and heavier elements. So the net effect of mass loss is to increase the production of He and C and to depress that of oxygen (see Figure 9). More recently, Meynet \& Mader (2002, 2003, 2005) have computed a grid of models for stars with $M > 20 M_{\odot}$ including rotation and metallicity dependent mass loss. The effect of metallicity dependent mass loss in decreasing the O production in massive stars was confirmed, although they employed significantly lower mass loss rates compared with Maeder (1992). With these models they were able to reproduce the frequency of WR stars and the observed WN/WC ratio, as was the case for the previous Maeder results.
Therefore, it appears that the earlier mass loss rates made-up for the omission of rotation in the stellar models.
On the other hand, the dependence upon metallicities of the yields computed with conservative stellar models, such as those of WW95, is not very strong except perhaps for the yields computed with zero intial stellar metallicity (Pop III stars).

\begin{figure}
\centering
\includegraphics[width=10cm,height=8cm]{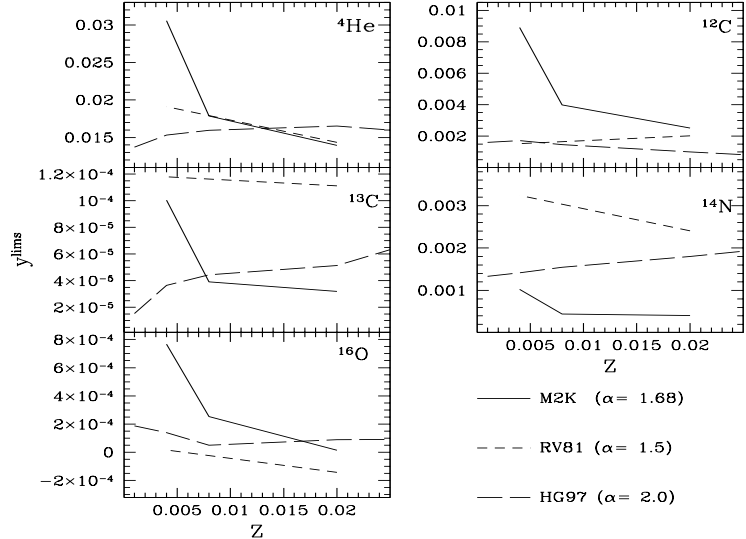}
\caption{The yields integrated over the Salpeter (1955) IMF of He, C and N produced by low and intermediate mass stars
as functions of the initial stellar metallicity. Different results are compared here: those of RV81 (Renzini \& Voli 1981), those of HG97 (van den Hoeck \& Groenewegen 1997) and those of M2K (Marigo 2001). The mixing length parameters ($\alpha$) adopted by the authors are indicated. Figure from Marigo (2001).}
\label{fig:3}
\end{figure}

\begin{figure}
\centering
\includegraphics[width=7cm,height=7cm]{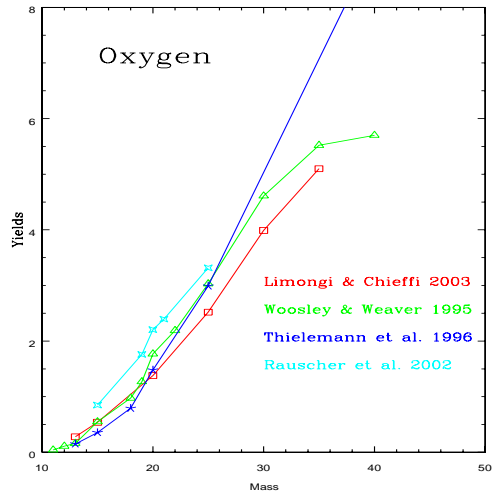}
\caption{The yields of oxygen for massive stars as computed by several authors, as indicated in the Figure. None of these calculations takes into account mass loss by stellar wind. }
\label{fig:3}
\end{figure}

\begin{figure}
\centering
\includegraphics[width=7cm,height=7cm]{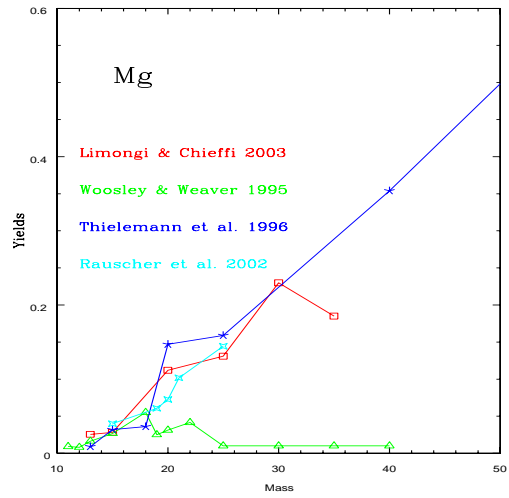}
\caption{The same as Fig. 4 for magnesium.}
\label{fig:4}
\end{figure}

\begin{figure}
\centering
\includegraphics[width=7cm,height=7cm]{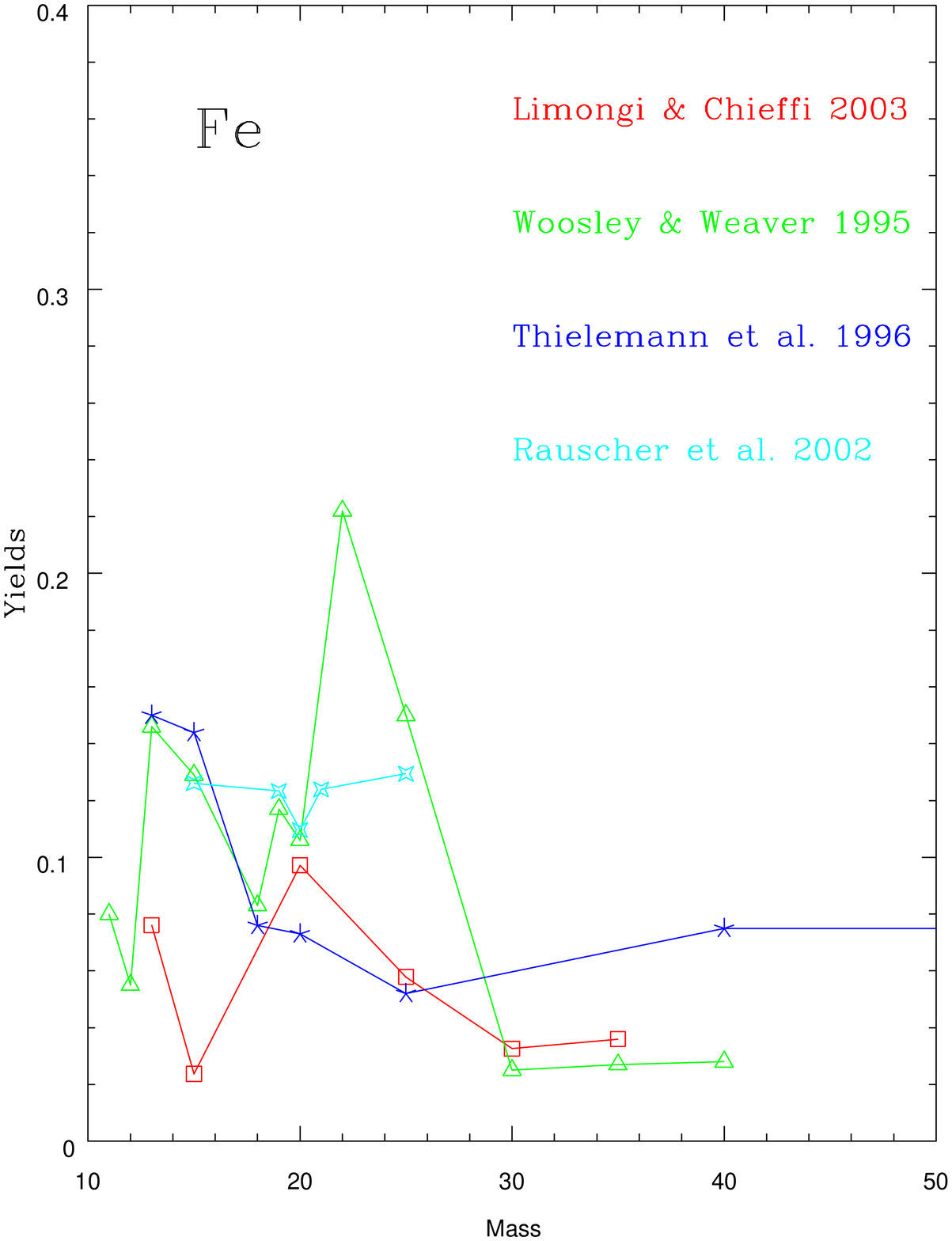}
\caption{The same as Fig. 4 for Fe.}
\label{fig:5}
\end{figure}

In Figures 7 and 8 we show the most recent results of Nomoto et al. (2006) for conservative stellar models of massive stars at different metallicities. While the O yields are not much dependent upon the initial stellar metallicity, as in WW95 , the Fe yields seem to change dramatically with the stellar metallicity.

\begin{figure}
\centering
\includegraphics[width=7cm,height=7cm]{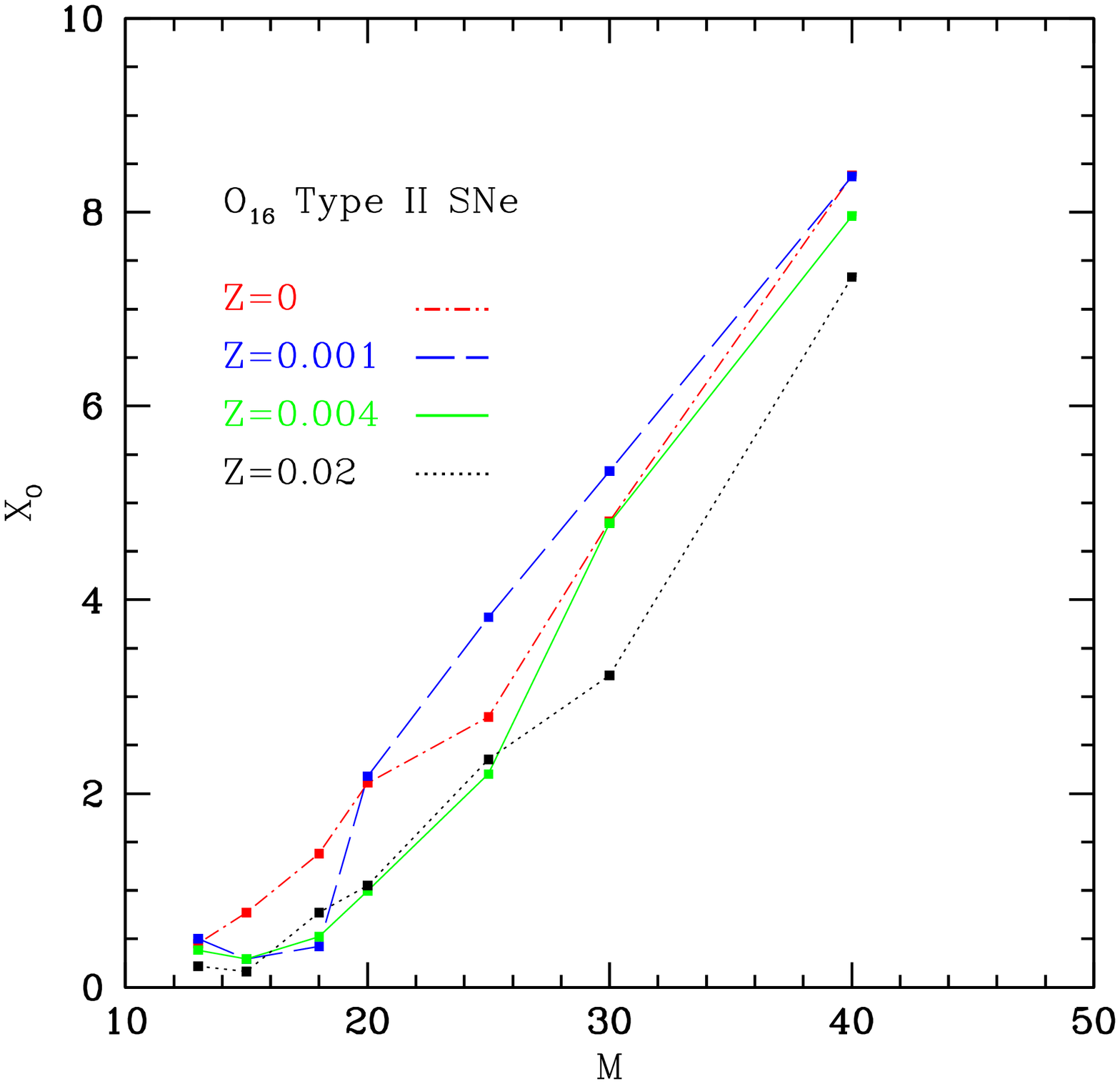}
\caption{The O yields as calculated by Nomoto et al. (2006) for different metallicities. These calculations do not take into account mass loss by stellar wind.}
\label{fig:6}
\end{figure}

\begin{figure}
\centering
\includegraphics[width=7cm,height=7cm]{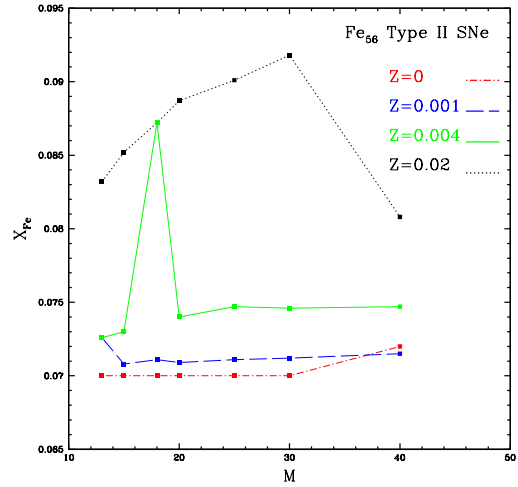}
\caption{ The same as Figure 7 for Fe.}
\label{fig:7}
\end{figure}

\begin{figure}
\centering
\includegraphics[width=7cm,height=7cm]{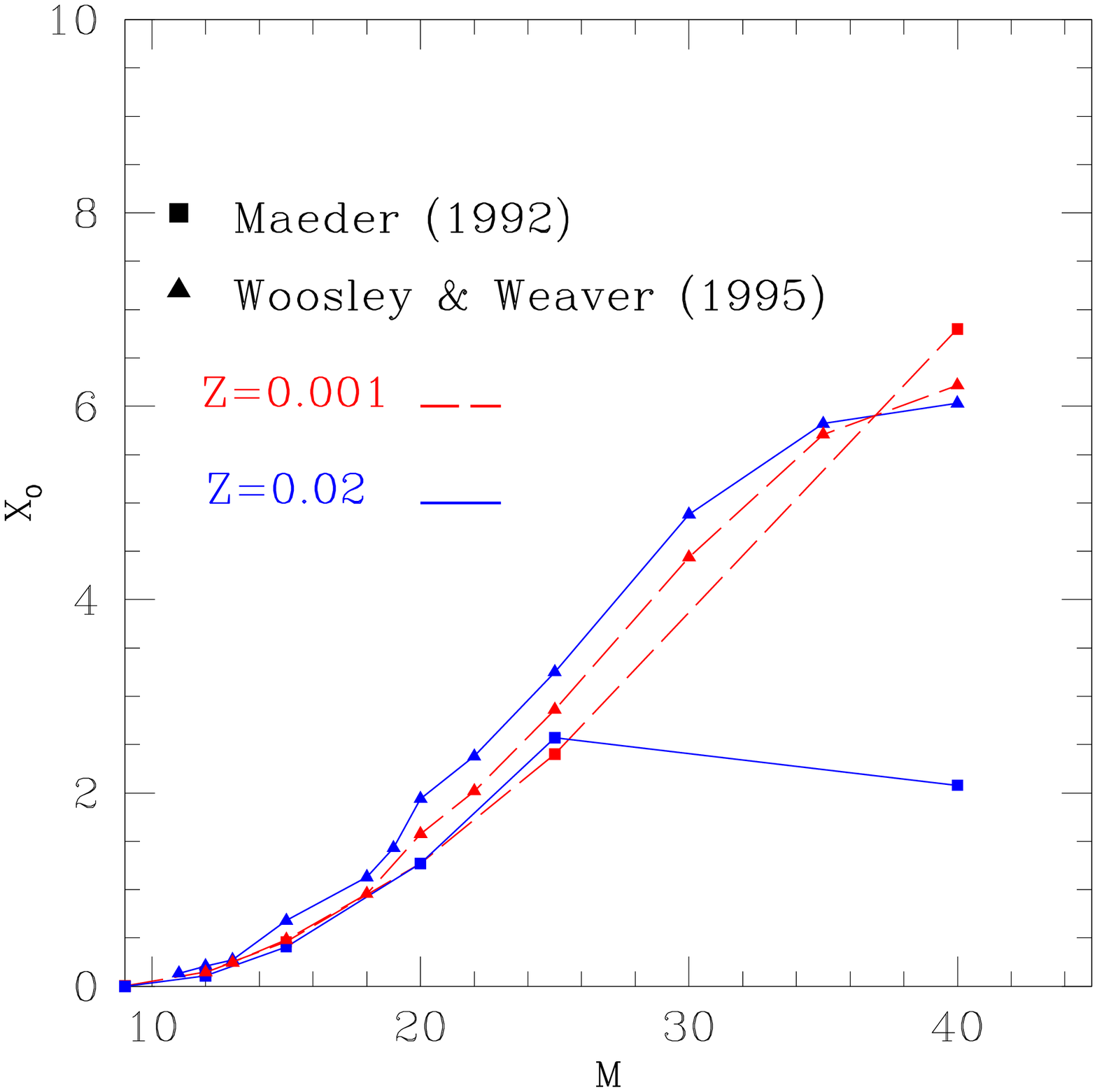}
\caption{The effect of metallicity dependent mass loss on the oxygen yield. The comparison is between the conservative yields of WW95 for Z=0.001 and Z=0.02 and the yields with mass loss of Maeder (1992) for the same metallicity. As one can see the effect of mass loss for a solar metallicity is a quite important one.}
\label{fig:8}
\end{figure}

\subsubsection{Type Ia SN progenitors}
There is a general consensus about the fact that SNeIa originate from C-deflagration in C-O white dwarfs (WD) in binary systems, but several evolutionary paths can lead to such an event.
The C-deflagration produces $\sim 0.6-0.7 M_{\odot}$ of Fe plus traces of other elements from C to Si, as observed in the spectra of Type Ia SNe.

Two main evolutionary scenarios for the progenitors of Type Ia SNe have been proposed:

\begin{itemize}
\item
Single Degenerate (SD) scenario (see Figure 10): the
classical scenario of Whelan and Iben (1973), recently revised by Han \& Podsiadlowsky (2004),  namely C-deflagration in
a C-O WD reaching the Chandrasekhar mass $M_{Ch}\sim 1.44 M_{\odot}$  after accreting material from a red giant
companion. One of the limitations of this scenario is that the accretion rate should be defined in a quite narrow range of values.
 To avoid this problem, Kobayashi et al. (1998) 
had proposed a similar scenario, based on the model of Hachisu et al. (1996), 
where the companion can be either a red giant or a main sequence star, 
including 
a metallicity effect which suggests that no Type Ia systems can form 
for [Fe/H]$< -1.0$ dex. This is due to the development of a strong radiative wind from the C-O WD which stabilizes the accretion from the companion, allowing for larger mass accretion rates than the previous scenario. 
The clock to the explosion is given by the lifetime of the secondary star in the binary system, where the WD is the primary (the originally more massive one). Therefore, the largest mass for a secondary is $8 M_{\odot}$, which is the maximum mass for the formation of a C-O WD. As a consequence, the minimum timescale for the occurrence of Type Ia SNe is $\sim 30$ Myr (i.e. the lifetime of a $8M_{\odot}$) after the beginning of star formation. Recent observations in radio-galaxies by Mannucci et al. (2005;2006) seem to confirm the existence of such prompt Type Ia SNe.

The minimum mass for the secondary is $0.8 M_{\odot}$, which is the star with lifetime equal to the age of the universe. Stars with masses below this limit are obviously not considered.
In summary, the mass range for both primary and secondary stars is, in 
principle,  between 0.8 and 8$M_{\odot}$, although two stars of $0.8 M_{\odot}$ are too small to give rise to a WD with a Chandrasekhar mass, and therefore the mass of the primary star should be assumed to be high enough to ensure that, even after accretion from a $0.8M_{\odot}$ star secondary, it will reach the Chandrasekhar mass. 

\item Double Degenerate (DD) scenario:
the merging of two C-O white dwarfs, due to loss of angular momentum caused
by gravitational wave radiation,
which explode by C-deflagration when $M_{Ch}$ is reached (Iben
and Tutukov 1984). In this scenario, the two C-O WDs should be of $\sim 0.7 M_{\odot}$ in order to give rise to a Chandrasekhar mass after they merge, therefore their progenitors should be in the range (5-8)$M_{\odot}$. The clock to the explosion here is given by the lifetime of the secondary star plus the gravitational time delay which depends on the original separation of the two WDs. The minimum timescale for the appearance of the first Type Ia SNe in this scenario is a few million years more than in the SD scenario (e.g. $\sim 40$ Myr in  Tornamb\'e \& Matteucci 1986). At the same time, the maximum gravitational time delay can be as long as more than a Hubble time. For more recent results on the DD scenario see Greggio (2005). 

\end{itemize}
Within any scenario the explosion can occur either when the C-O WD reaches the Chandrasekhar mass and carbon deflagrates at the center or when a massive enough helium layer is accumulated on top of the C-O WD. In this last case there is He-detonation which induces an off-center carbon deflagration before the Chandrasekhar mass is reached (sub-chandra exploders, e.g. Woosley \& Weaver 1994).

While the chandra-exploders are supposed to produce the same nucleosynthesis (C-deflagration of a Chandrasekhar mass), they
predict a different evolution of the Type Ia SN rate and different typical timescales for the
SNe Ia enrichment. A way of defining the typical Type Ia SN timescale is to assume it as the
time when the maximum in the Type Ia SN rate is reached (Matteucci \& Recchi, 2001). This timescale varies according to the chosen progenitor model and to the assumed star formation history, which varies from galaxy to galaxy.
For the solar vicinity, this timescale  is at least 1 Gyr,
if the SD scenario is assumed, whereas for elliptical galaxies, where the stars formed much more quickly, this timescale is only 0.5 Gyr (Matteucci \& Greggio, 1986; Matteucci \& Recchi 2001).

\begin{figure}
\centering
\includegraphics[width=10cm,height=14cm]{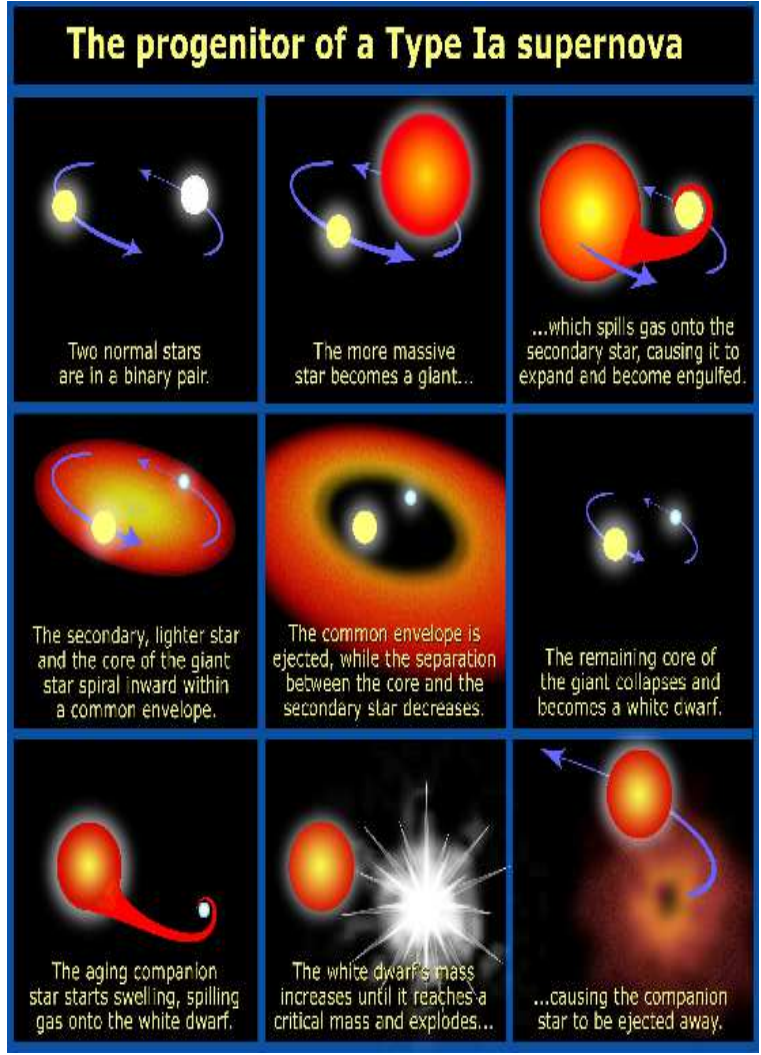}
\caption{The progenitor of a Type Ia SN in the context of the single-degenerate model (Illustration credit: NASA, ESA, and A. Field (STSci)).}
\label{fig:7}
\end{figure}

\subsection{Gas flows}

Various parametrizations have been suggested for gas flows and the most common is an exponential law for the gas infall rate:
\begin{equation}
IR \propto e^{-t/\tau}
\end{equation}
with the timescale $\tau$ being a free parameter, whereas for the galactic outflows the wind rate is generally assumed to be proportional to the SFR:
\begin{equation}
WR=- \lambda SFR
\end{equation}
where $\lambda$ is again a free parameter. Both $\tau$ and $\lambda$ should 
be fixed by reproducing the majority of observational constraints.

\section{Basic Equations for chemical evolution}

\subsection{Yields per Stellar Generation}

Under the assumption of Instantaneous Recycling Approximation  (IRA)
which states that all stars more massive than $1M_{\odot}$ die immediately, whereas all stars with masses lower than $1M_{\odot}$ live forever, one can define the yield per stellar generation (Tinsley, 1980):
\begin{equation}
y_{i}={1 \over 1-R} \int^{\infty}_{1}{m p_{im} \varphi(m) dm} 
\end{equation}

where $p_{im}$ is the stellar {\it new} yield of the element $i$, namely the newly 
formed and ejected element $i$ by a star of mass $m$, and $\varphi(m)$ is the IMF.

The quantity $R$ is the so-called Returned Fraction:
\begin{equation}
R=\int^{\infty}_{1}{(m-M_{rem}) \varphi(m) dm} 
\end{equation}
and is the mass fraction of gas restored into the ISM by an entire stellar 
generation. The term fraction derives from the fact that in its definition R is divided by $\int^{\infty}_{0}{m \varphi(m) dm }=1$, which is the normalization condition for the IMF. Therefore, (1-R) is the fraction of mass locked up in very low mass stars and remnants.

\subsection{Analytical models}

\subsubsection {\it Simple Model} 
The Simple Model for
the chemical evolution of the solar neighbourhood is the simplest approach to 
model chemical evolution. The solar neighbourhood is assumed to be a cylinder of 1 Kpc radius centered around the Sun.

The basic assumptions of the Simple Model are:\par

- the system is one-zone and closed, no inflows 
or outflows are considered,

- the initial gas is primordial (no metals),

- IRA holds, 
 
- the IMF, $\varphi(m)$, is assumed to be constant in time,

- the gas is well mixed at any time (instantaneous mixing approximation, IMA).
\par

The Simple Model fails in describing the evolution of the Milky Way 
(G-dwarf metallicity distribution, elements produced on long timescales and 
abundance ratios) and the reason is that at least
two of the above assumptions are manifestly wrong, epecially if one intends to model the evolution of the abundance of elements produced on long timescales, such as Fe. In particular, the 
assumptions of the closed box and the IRA. 

However, it is interesting 
to know the solution of the Simple Model and its implications.
Let $Z$ be the abundance by mass of metals,
then we obtain the analytical solution
of the Simple Model by ignoring the stellar lifetimes:

\begin{equation}
Z= y_{Z} ln({ 1 \over G}) 
\end{equation}

where $G=M_{gas}/M_{tot}$ is the gas mass fraction of the system
and  $y_Z$ is the yield per stellar generation, as defined above, otherwise 
called {\it effective yield}.

In particular, the effective yield is defined as:
\begin{equation}
y_{Z_{eff}}={Z \over ln(1/G)} 
\end{equation}
namely the yield that the system would have if behaving as the simple 
closed-box model.
This means that if $y_{Z_{eff}} > y_{Z}$, then the actual system has 
attained a higher metallicity at a given gas fraction G.
Generally, given two chemical elements $i$ and $j$, the solution of the Simple Model for primary elements (eq.22) implies:

\begin{equation}
{X_i \over X_j}= {y_i \over y_j}
\end{equation}

which means that the ratio of two element abundances is always equal to the 
ratio of their yields.
This is no more  true when IRA is relaxed. In fact, relaxing IRA is 
necessary  to study in detail the evolution of the abundances of single 
elements produced on long timescales (e.g. Fe, N).

\subsubsection{Analytical models in the presence of outflows}

One can obtain analytical solutions also in the presence of infall and/or outflow but the necessary condition is to assume IRA, as well as precise forms for the infall and outflow rates.

Matteucci \& Chiosi (1983) found solutions for models with outflow and infall 
and Matteucci (2001) found it for a model with infall and outflow acting at 
the same time.
The main assumption in the model with outflow but no infall
is that the outflow rate is:

\begin{equation}
W(t)=-\lambda (1-R) \psi(t)
\end{equation}

where $\lambda > 0$ is the wind parameter.

The solution of this model is:

\begin{equation}
Z = {y_{Z} \over (1+ \lambda)} ln[(1+ \lambda) G^{-1} - \lambda]
\end{equation}

for $\lambda = 0$ the equation becomes the one of the Simple Model (eq. 22).

As one can see from eq. (26), the presence of an outflow decreases the effective yield, in the sense that the true yield of a system is lower than the effective yield. Models with galactic winds or outflows in general are suitable for ellipticals, irregulars and for the Galactic halo.
A popular analytical model with outflow is that suggested by Hartwick (1976) for the evolution of the Galactic halo, under the assumption that during the halo collapse stars were forming while the gas was dissipating energy and  falling into the bulge and disk, thus producing a net gas loss from the halo. This hypothesis was suggested by the fact that the stellar metallicity distribution of the halo can be reproduced only with an effective yield lower than that of the disk.
In Hartwick's model the ouflow rate is assumed to be simply proportional to the SFR:

\begin{equation}
W(t)= -\lambda \psi(t)
\end{equation}

which is similar to eq. (25). Hartwick  used this model to reproduce the metallicity distribution of halo stars and also to alleviate the G-dwarf problem in the disk, namely the fact that the Simple Model of chemical evolution  predicts too many disk stars than observed (see Tinsley 1980 for a review on the subject). However, the gas lost from the halo cannot have contributed to form the whole disk since the distribution of the specific angular momentum of halo and disk stars are quite different, thus indicating that only a negligible amount of halo gas can have formed the disk. On the other hand, the similarity of the distributions for the halo and bulge indicates that the bulge must have formed out of gas lost from the halo (see Wyse \& Gilmore 1992).
The G-dwarf problem is instead easily solved if one assumes that the Galactic disk has formed by means of slow infall of extragalactic material, as we will see in the next sections.
Recently, Hartwick's model has been revisited by Prantzos (2003) to interpret the most recent metallicity distribution of halo stars, which is quite different 
with respect to the G-dwarf metallicity distribution in the local disk. In particular, the halo metallicity distribution is peaked at around [Fe/H]=-1.6 dex, whereas the G-dwarf distribution is peaked at around $\sim -0.2$ dex. 
Prantzos (2003) suggested that an outflow with $\lambda=8$ as well as a 
formation of the halo by early infall
are necessary to reproduce the observed halo metallicity distribution.
In Figure 11 we show the results of Prantzos (2003) compared with observations.

\begin{figure}
\centering
\includegraphics[width=10cm,height=10cm]{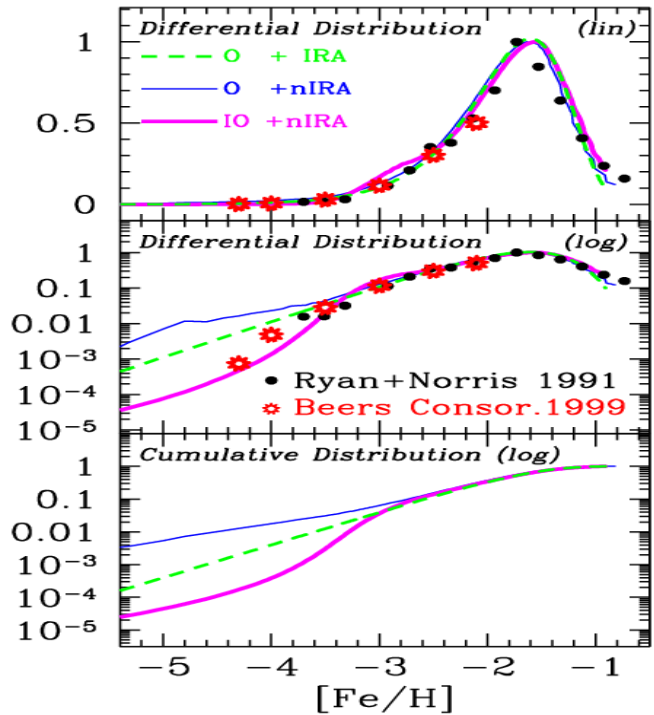}
\caption{Metallicity distribution for the halo stars. Upper panel: observed and predicted metallicity distributions. The models are: pure outflow with IRA (dashed curve),
pure outflow without IRA (thin solid curve) and early infall +outflow without 
IRA (thick solid curve). The distribution is on a linear scale. Middle panel: the same as above but the distribution is on a logarithmic scale. Lower panel: predicted cumulative distributions. Figure from Prantzos (2003).}
\label{fig:7}
\end{figure}

\subsubsection{Analytical models in presence of infall}

The solution of the equation of metals for a model without a wind but with
a primordial infalling material ($Z_{A}=0$) at a rate: 

\begin{equation}
A(t)=\Lambda (1-R) \psi(t)
\end{equation}

and $\Lambda \ne 1$  is :
\begin{equation}
Z= {y_Z \over \Lambda}[1-(\Lambda-(\Lambda-1)G^{-1})^
{-\Lambda/(1-\Lambda)}]
\end{equation}
For $\Lambda=1$ one obtains the well known case of {\it extreme infall} studied by Larson (1972) whose solution is:
\begin{equation}
Z=y_Z[1-e^{-(G^{-1} -1)}]
\end{equation}
This extreme infall solution shows that when $G \rightarrow 0$ then $Z \rightarrow y_Z$.
The infall can solve the G-dwarf problem
for disk stars except for the extreme infall solution which predicts too few low metallicity stars below [Fe/H]=-1.0 (see Tinsley 1980). Finally, we can conclude that the infall is very important for explaining both the halo and the disk formation.

\subsubsection{Analytical models in presence of infall and outflow}

Matteucci (2001) presented an analytical solution for infall and outflow present at the same time.
The solution refers to the outflow and infall rates of eq. (25) and eq.
(28), respectively.

In particular:

\begin{equation}
Z={y_Z  \over \Lambda}\{1-[(\Lambda - \lambda)-(\Lambda - \lambda - 1) 
G^{-1}]
^{{\Lambda \over \Lambda - \lambda -1}}\},
\end{equation}
for a primordial infalling gas ($Z_{A}=0$). This solution is velid for $\lambda >0$ and $\Lambda >0$ and $\Lambda \ne 1$.

\subsection{Detailed numerical models}

Detailed models of galactic chemical evolution require consideration of the stellar lifetimes, namely they should relax IRA.  However, the majority of them still retain the instantaneous mixing approximation (IMA), which assumes that the material ejected by stars at their death  is instantaneously mixed with the surrounding interstellar medium (ISM).
This approximation seems to be good in the majority of the cases with perhaps the exception of the very early phases of galactic evolution.

The basic equations of chemical evolution follow the evolution of the abundances of single chemical species and the gas as a whole.

If $\sigma_i$ is the surface mass density of an element $i$, with
$\sigma_{gas}= \sum_{i=1,n} {\sigma_i}$ being the total surface gas density and n the total number of chemical elements,
we can write:

\begin{eqnarray}
 & & \dot \sigma_i(t)  =  -\psi(t)X_i(t)\nonumber \\
& & + \int_{M_{L}}^{M_{Bm}}\psi(t-\tau_m)
Q_{mi}(t-\tau_m)\varphi(m)dm\nonumber \\ 
& & + A\int_{M_{Bm}}^{M_{BM}}
\phi(m)\nonumber \\
& & \cdot[\int_{\gamma_{min}}
^{0.5}f(\gamma)\psi(t-\tau_{m2}) 
Q_{mi}(t-\tau_{m2})d\gamma]dm\nonumber \\ 
& & + B\int_{M_{Bm}}^
{M_{BM}}\psi(t-\tau_{m})Q_{mi}(t-\tau_m)\varphi(m)dm\nonumber \\
& & + \int_{M_{BM}}^{M_U}\psi(t-\tau_m)Q_{mi}(t-\tau_m) 
\varphi(m)dm\nonumber \\ 
& & + X_{A_{i}} A(t) - X_{i}(t) W(t)
\end{eqnarray}

for any ggiven chemical element.
These equations can be solved only numerically.
The quantities $X_i(t)$ are the abundances as defined in  eq. (1). The quantity $Q_{mi}$ contains all the information about stellar evolution and nucleosynthesis: in practice it gives the 
mass of gas produced and ejected in the form of an element $i$ by a
star of initial mass $m$, together with the mass of that element which was already present in the star at birth. The various integrals represent the rates at which the mass of a given element is restored into the ISM by stars of different masses which can evolve into WDs or supernovae (II, Ia, Ib).
The integral representing the rate of matter restoration by Type Ia SNe is the second one on the right hand side. The quantity A is a constant: it is the fraction, in the IMF, of binary systems with those specific features required to give rise to Type Ia SNe, whereas B=1-A is the fraction of all the single stars and binary systems in the same mass range of definition of the progenitors of Type Ia SNe (third integral). The parameter A is obtained by imposing that the predicted Type Ia SN rate reproduces the observed rate at the present time (14 Gyr).
Values of A =0.05-0.09 are found for the evolution of the solar vicinity when an IMF of Scalo (1986, 1989) or Kroupa et al. (1993) is adopted.
If one adopts a flatter IMF such as the  Salpeter (1955) one,  then A is 
different. The integral of the Type Ia SN contribution is made over a range of mass going from $M_{Bm}=3 M_{\odot}$ to $M_{BM}=16 M_{\odot}$, which represents the total masses of binary systems able to produce Type Ia SNe in the framework of the single degenerate scenario. There is also an integration over the mass distribution of binary systems; in particular, one considers the function $f(\gamma)$ where $\gamma={M_2 \over M_1 +M_2}$,
with $M_1$ and $M_2$ being the primary and secondary mass of the binary system, respectively (for more details see Matteucci \& Greggio 1986 and 
Matteucci 2001).
The third and fourth integrals represent the rates of Type II and Type Ib/c SNe, respectively. The occurrence of Type Ib SNe seems to be partly related to Wolf-Rayet stars which have original masses larger tham 25$M_{\odot}$ and depends on the mass loss rate which is more active at high metallicities. However, it has been proposed  that Type Ib SNe can also originate from massive stars in binary systems.
Finally, the functions A(t) and W(t) are the infall and wind rate, 
respectively.

\section{The Milky Way}
We will first analyze the chemical evolution of our Galaxy, the Milky Way.

\begin{figure}
\centering
\includegraphics[width=7cm,height=7cm]{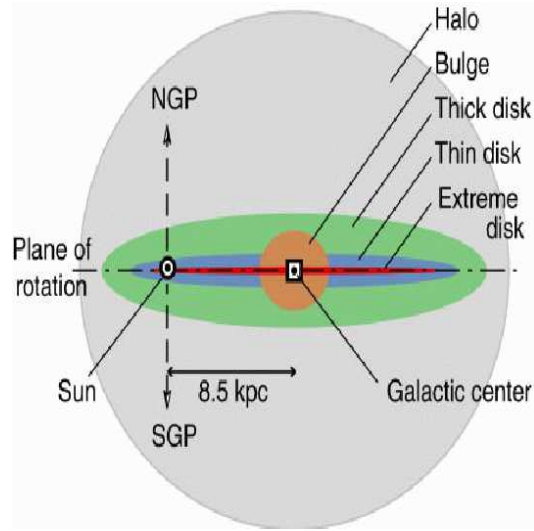}
\caption{Schematic edge-on view of the major components of the Milky Way. Illustration credit from R. Buser, www.astro.unibas.ch/forschung/rb/structure.shtml.}

\end{figure}

\subsection{The formation of the Milky Way}
\subsubsection{Observational evidence}
The Milky Way galaxy has four main stellar populations: 1) the halo stars with low metallicities \footnote{the most common stellar metallicity indicator in stars is [Fe/H]= $log(Fe/H)_{*}- log(Fe/H)_{\odot}$} and eccentric orbits, 2) the bulge population with a large range of metallicities and dominated by random motions, 3) the thin disk stars with an average metallicity $<[Fe/H]>$=-0.5 dex and circular orbits, and finally 4) the thick disk stars which possess chemical and kinematical properties intermediate between those of the halo and those of the thin disk. The halo stars have average metallicities of $<[Fe/H]>$=-1.5 dex and a maximum metallicity of $\sim -1.0$ dex,  although stars with [Fe/H] as high as -0.6 dex and halo kinematics are observed. 
The average metallicity of thin disk stars is $\sim -0.6$ dex, whereas the one of bulge stars is $\sim -0.2$ dex.

The kinematical and chemical properties of the different Galactic stellar populations can be interpreted in terms of 
the Galaxy formation mechanism.
Eggen, Lynden-Bell \& Sandage (1962),
 in a cornerstone paper suggested a rapid monolithic collapse for the formation of the Galaxy lasting $\sim 2 \cdot 10^{8}$ years. This suggestion was based on a kinematical and chemical study of solar neighbourhood stars and the value of the suggested timescale was  chosen to allow for the orbital eccentricities to vary in a potential not yet in equilibrium but sufficiently long so that massive stars forming in the collapsing gas could have time to die and enrich the gas with heavy elements.

Later on, Searle \& Zinn (1978)  measured Fe abundances and horizontal branch morphologies of 50 globular clusters and studied their properties as a function of the galactocentric distance.
As a result of this, they proposed a central collapse like the one envisaged by Eggen et al.,  but also that the outer halo formed by merging of large fragments taking place over a considerable timescale $> 1$ Gyr.
The Searle \& Zinn scenario is close to what is predicted by modern cosmological theories of galaxy formation. In particular, in the framework of the hierarchical galaxy formation scenario, galaxies form by accretion of smaller building blocks (e.g. White \& Rees, 1978, Navarro \& al. 1997). Obvious candidates for these building blocks are either dwarf spheroidal (dSph) or dwarf 
irregular (dIrr) galaxies. However, as we will see in detail later, the chemical composition and in particular the chemical abundance patterns in dSphs or dIrrs are not compatible with the same abundance patterns in the Milky Way 
(see Geisler et al. 2007), thus arguing against the identification of the building blocks with these galaxies.
On the other hand, very recently, Carollo \& al. (2007) have obtained medium resolution spectroscopy of 20,336 stars from the Sloan Digital Sky Survey (SDSS). They showed that the Galactic halo is divisible into two broadly overlapping structural components.
In particular, they find that the inner halo is dominated by stars with very eccentric orbits, exhibits a peak at [Fe/H]=-1.6 dex and has a flattened density distribution with a modest net prograde rotation. The outer halo includes stars with a wide range of eccentricities, exhibits a peak at [Fe/H]=-2.2 dex and a spherical density distribution with highly statistically significant net retrograde rotation. They conclude that most of the Galactic halo should have formed  by accrection of multiple distinct sub-systems. However, an analysis of the abundance ratios of these stars is still missing.

\subsubsection{Theoretical models}
From an historical point of view, the modelization of the Galactic chemical 
evolution has passed through different phases that I summarize in the 
following:
\begin{itemize} 
\item SERIAL FORMATION

The Galaxy is modeled by means of one accretion episode lasting for the entire Galactic lifetime, where halo, thick and thin disk form in sequence as a continuous process. The obvious limit of this approach 
is that it does not allow us to predict the observed overlapping in metallicity between halo and thick disk stars and between thick and thin disk stars, but it gives a fair representation of our Galaxy  
(e.g. Matteucci  \& Fran\c cois 1989).
\item PARALLEL FORMATION

In this formulation, the various Galactic components start at the same time 
and 
from the same gas but evolve at different rates (e.g. Pardi et al. 1995). It
predicts overlapping of stars belonging to the different components
but implies that the thick disk formed out of gas shed by the halo and that the thin disk formed out of gas shed by the thick disk, and this is at variance with the distribution of the stellar angular momentum per unit mass (Wyse \& Gilmore 1992), which indicates that the disk did not form out of gas shed by the halo.

\item TWO-INFALL FORMATION

In this scenario, halo and disk 
formed  out of two separate infall episodes (overlapping in metallicity 
is also predicted)
(e.g. Chiappini et al. 1997; Chang et al. 1999; Alib\'es et al. 2001). 
The first infall 
episode lasted no more than 1-2 Gyr whereas the second, where the thin 
disk formed, 
lasted much longer with a timescale for the formation of the solar 
vicinity of 6-8 Gyr 
(Chiappini et al. 1997; Boissier\& Prantzos 1999).

\item STOCHASTIC APPROACH

Here the hypothesis is that in the early halo phases ([Fe/H] $< -3.0$ dex), 
mixing was not 
efficient and, as a consequence, one should observe, in low metallicity halo 
stars, the 
effects of pollution from single SNe (e.g. Tsujimoto et al. 1999; Argast et 
al. 2000; Oey 2000). 
These models predict a large spread for  [Fe/H] $< -3.0$ dex for all the 
$\alpha$-elements,  
which is not observed, as shown 
by stellar data with metallicities down to -4.0 dex 
(Cayrel et al. 2004). However, inhomogeneities could explain 
the observed spread of s- and r-elements at low metallicities (see later).

\end{itemize}
\subsection{The two-infall model}
The two-infall model of Chiappini, Matteucci \& Gratton (1997) predicts two 
main episodes of gas accretion: during the first one, the halo the bulge and most of the thick disk formed, while the second gave rise to the thin disk. 
In Figure 13 we show an artistic representation of the formation of the Milky Way in the two-infall scenario. In the upper panel we see the sequence of the formation of the stellar halo, in particular the inner halo, following a monolithic-like collapse of gas (first infall episode) but with a longer timescale than originally suggested by Eggen et al. (1962): here the time scale is 0.8-1.0 Gyr. During the halo formation also the bulge is formed on a very short timescale in the range 0.1-0.5 Gyr.  During this phase also the thick disk assembles or at least part of it, since part of the thick disk, like the outer halo, could have been accreted. The second panel from left to right shows the beginning of the thin disk formation, namely the assembly of the innermost disk regions just around the bulge; this is due to the second infall episode. The thin-disk assembles inside-out, in the sense that the outermost regions take a much longer time to form. This is shown in the third panel. In Fig. 13 each panel is connected to temporal phases where the Type II and then the Type Ia SN rates are present. So, it is clear that the early phases of the halo and bulge formation are dominated by Type II SNe (and also  by Type Ib/c SNe) producing mostly $\alpha$-elements such as O and Mg and part of Fe. On the other hand, Type Ia SNe start to be non negligible only after 1Gyr and they pollute the gas during the thick and thin disk phases. The minimum shown in the Type II SN rate is due to a gap in the star formation rate occurring as a consequence of the adoption of a threshold density in the star formation process, as we will see next (Figure 14).

\begin{figure}
\centering
\includegraphics[width=12cm,height=12cm]{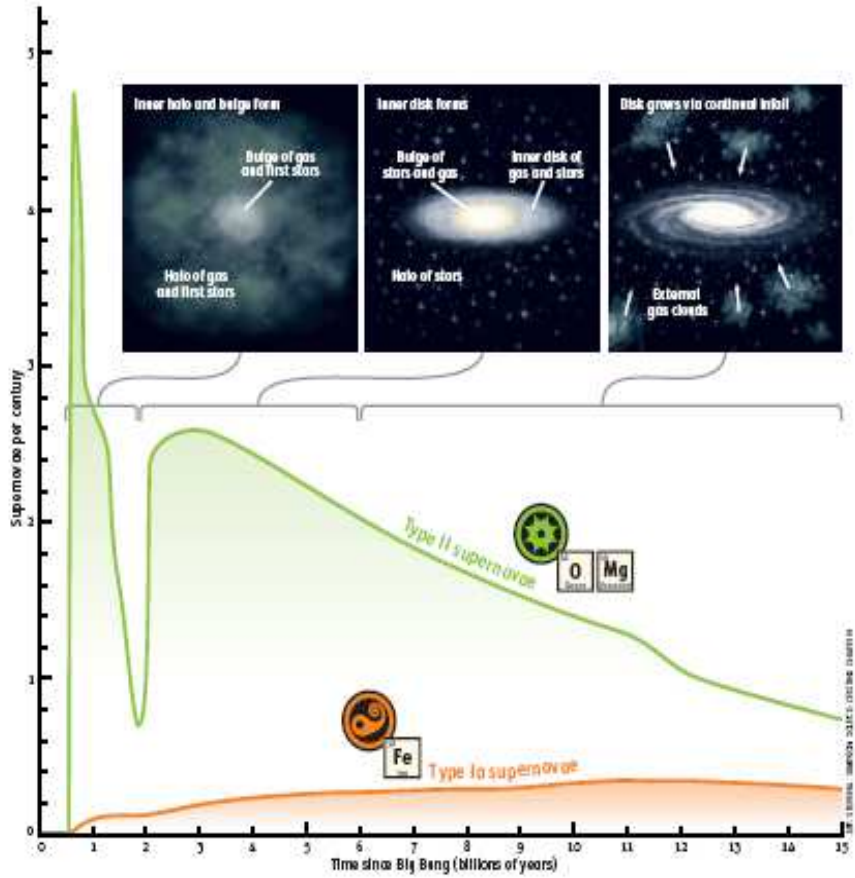}
\caption{Artistic view of the two-infall model by Chiappini et al. (1997). The pèredicted SN II and Ia rates per century are also sketched, together with the fact that Type II SNe produce mostly $\alpha$-elements (e.g. O, Mg), whereas Type Ia SNe produce mostly Fe. (Illustration credit: C. Chiappini, Sky \& Telescope,
2004, Vol. 108, number 4, p.32 ).}
\label{fig:3}
\end{figure}

\subsection{Detailed recipes for the two-infall model}

The main assumption of 
this model are:
\begin{itemize}

\item The IMF is that of Scalo (1986) normalized over a mass range of 
0.1-100$M_{\odot}$.

\item 
The infall law is:
\begin{equation}
A(r,t)= a(r) e^{-t/ \tau_{H}(r)}+ b(r) e^{-(t-t_{max})/ \tau_{D}(r)}
\end{equation}
where $A(r,t)= ({d \sigma(r,t) \over dt})_{infall}$ is the rate at which the total surface mass density changes because of the infalling gas. The quantities
$a(r)$ and $b(r)$ are two parameters fixed by reproducing the total present 
time surface mass density in the solar 
vicinity ($\sigma_{tot}$= 51 $\pm$ 6 $M_{\odot} \; pc^{-2}$ , see Boissier \& 
Prantzos 1999), $t_{max}=1$Gyr
is the time for the maximum infall on the thin  disk, $\tau_{H}= 0.8$ Gyr
is the time scale for the formation of the halo thick-disk (which means a total duration of 2 Gyr for the complete halo-thick disk formation) and $\tau_{D}(r)$
is the timescale for the formation of the thin disk and it is a function of 
the galactocentric distance (formation inside-out, Matteucci \& 
Fran\c cois 1989; Chiappini et al. 2001).

In particular, it is assumed  that:
\begin{equation}
\tau_{D}=1.033 r (Kpc) - 1.267 \,\, (Gyr)
\end{equation}

where $r$ is the galocentric distance.

\item The SFR is the Kennicutt law with a dependence on the surface gas 
density and also on the total surface mass density
(see Dopita \& Ryder 1994).
In particular, the SFR is based on the law originally suggested by Talbot \& Arnett (1975) and then adopted by Chiosi (1980):

\begin{equation}
\psi(r,t)=\nu\left(\frac{\sigma(r,t) \sigma_{gas}(r,t)}{\sigma(r_{\odot},t)^{2}}\right)^{(k-1)}\sigma_{gas}(r,t)^{k}.
\end{equation}

where the constant $\nu$ is the efficiency of the SF process, as defined in eq. (6),  and is expressed in $Gyr^{-1}$: in particular,  $\nu= 2Gyr^{-1}$ for the the halo and $1 Gyr^{-1}$ for the disk ($t\ge1Gyr$).
The total surface mass density is represented by $\sigma(r,t)$, whereas $\sigma(r_{\odot},t)$ is the total surface mass density at the 
solar position, assumed to be $r_{\odot}=8$ Kpc  
(Reid 1993). The quantity $\sigma_{gas}(r,t)$ represents the 
surface gas density. 
The exponent of the surface gas density, $k$, is set equal to 1.5, similar to what suggested by Kennicutt (1998a).
These choices for the parameters allow the model to fit very well the observational constraints, in particular in the solar vicinity.
We recall that below a critical threshold for the surface gas density 
($7M_{\odot}pc^{-2}$ for the thin disk and $4M_{\odot}pc^{-2}$ for the halo phase)
we assume that the star formation is halted. 
The existence of a threshold for the star formation has been suggested by Kennicutt (1998a,b) and Martin \& Kennicutt (2001).

The predicted behaviour of the SFR, obtained by adopting eq.(35) with the threshold is shown in Figure 14.

\begin{figure}
\includegraphics[width=4.5in,height=3.0in]{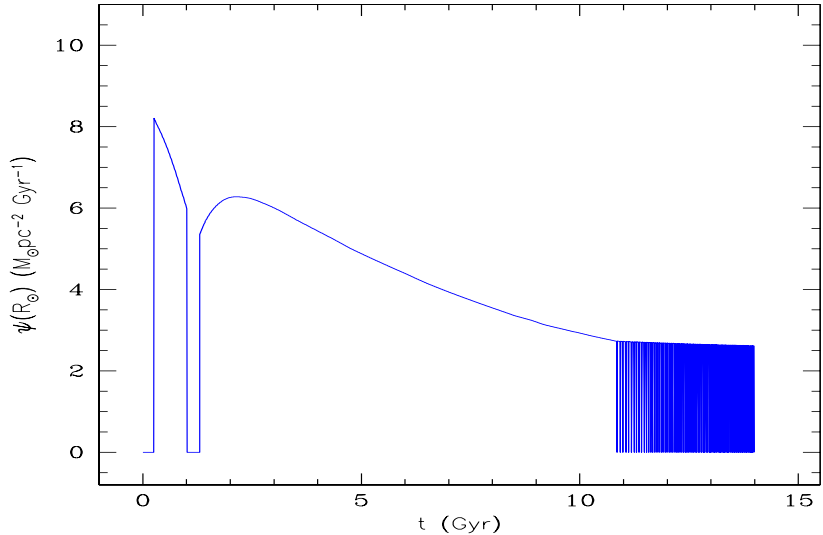}
\hfill
\caption{The SFR in the solar vicinity as predicted by the two-infall model. 
Figure from Chiappini et al. (1997). The oscillating behaviour in the 
SFR at late times 
is due to the assumed threshold density for SF. The threshold gas density is 
also responsible for the gap in the SFR seen at around 1 Gyr.}\label{fig} 
\end{figure}

\item The assumed Type Ia SN model is the single-degenerate one with the recipe first adopted in Greggio \& Renzini (1983a) and Matteucci \& Greggio (1986) and more recently in Matteucci \& Recchi (2001). The minimum time for the explosion is 30 Myr, whereas the the timescale for restoring the bulk of Fe is 1 Gyr, for the SFR adopted in  the solar vicinity. It is worth recalling that this timescale is not universal since it depends on the assumed SNIa progenitor model but also on the assumed star formation history. The SN rates in the solar vicinity
are shown in Figure 15.

\begin{figure}
\includegraphics[width=4.5in,height=3.0in]{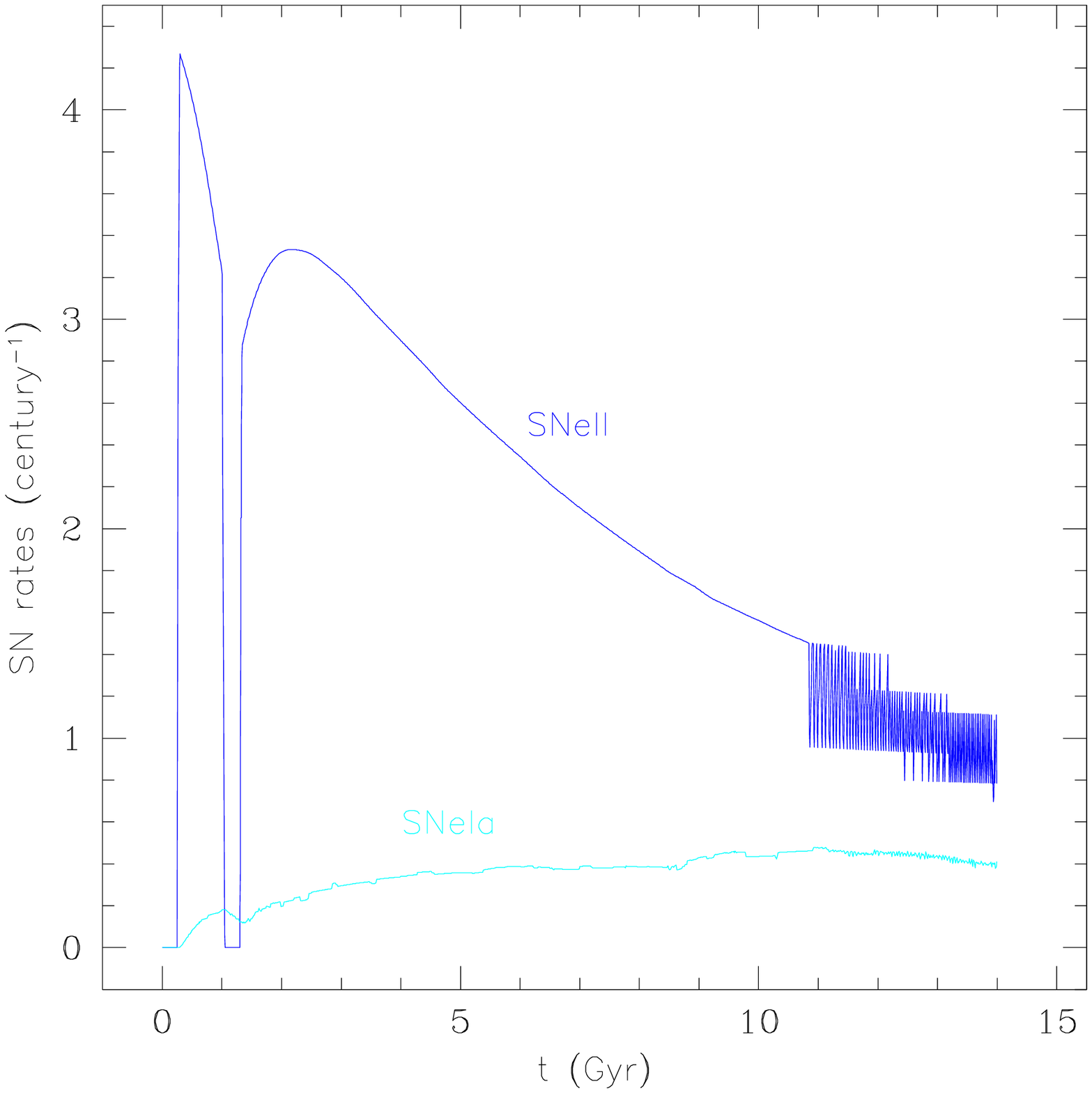}
\hfill
\caption{The Type II and Ia rate in the solar vicinity as predicted by the 
two-infall model. 
Figure from Chiappini et al. (1997). The oscillating behaviour of the Type II SN rate at late times 
is due to the assumed threshold density for SF. The threshold gas density is 
also responsible for the gap in the SFR seen at around 1 Gyr.}\label{fig} 
\end{figure}

\end{itemize}

\subsection{The chemical enrichment history of the solar vicinity}

We study first the solar vicinity, namely the local ring at 8 Kpc from the galactic center.
By integrating eq.(32) without the wind term we obtain the evolution of the abundances of several 
chemical species (H, D, He, Li, C, N, O, $\alpha$-elements, 
Fe, Fe-peak elements, s-and r- process elements). 
In Figure 16 we show the smallest mass dying at any cosmic time corresponding to a given predicted 
abundance of [Fe/H] in the ISM. This is because there is an age-metallicity relation and the 
[Fe/H] abundance increases with time. We recall that, for a generic chemical element $i$, with 
abundance $X_i$, one defines:
\begin{equation}
[X_i/H]= log(X_i/H)_{*} - log(X_i/H)_{\odot},
\end{equation}
where  $log(X_i/H)_{\odot}$ refers to the solar abundance of the element $i$.

\subsubsection{The observational constraints}
A good model of chemical evolution should be able to reproduce a minimum number of observational constraints and the number of observational constraints should be larger than the number of free parameters which are: $\tau_H$, $\tau_D$, 
$k_1$, $k_2$, $\nu$ and $A$ (the fraction of binary systems which can give rise to Type Ia SNe).

The main observational constraints in the solar vicinity that a good model should reproduce (see Chiappini et al. 2001, Boissier \& Prantzos, 1999 and references therein) are: \par

\begin{itemize}
\item The present time surface gas density:
$\sigma_{gas}= 13 \pm 3 M_{\odot} pc^{-2}$

\item The present time surface star density $\sigma_{*}= 43 \pm 5 M_{\odot} pc^{-2}$

\item The present time total surface mass density: $\sigma_{tot} = 51 \pm 6 M_{\odot} pc^{-2}$

\item The present time SFR:  $\psi_o=2-5 M_{\odot} pc^{-2} Gyr^{-1}$

\item The present time infall rate: $0.3-1.5 M_{\odot} pc^{-2} Gyr^{-1}$

\item The present day mass function (PDMF)

\item The solar abundances, namely the chemical abundances of the ISM at the time of birth of the solar system  4.5 Gyr ago as well as at the present time abundances

\item The observed  [$X_i$/Fe] vs. [Fe/H] relations

\item The  G-dwarf metallicity distribution

\item The  age-metallicity relation
\end{itemize}
And finally, a good model of chemical evolution of the Milky Way should reproduce the
distributions of
abundances, gas and star formation rate along the disk as well as 
the average observed SNII and Ia rates (SNII=$1.2 \pm 0.8 \,\, 100yr^{-1}$ and SNIa=$0.3 \pm 0.2 \,\, 100yr^{-1}$).

\begin{figure}
\includegraphics[width=4.5in,height=3.0in]{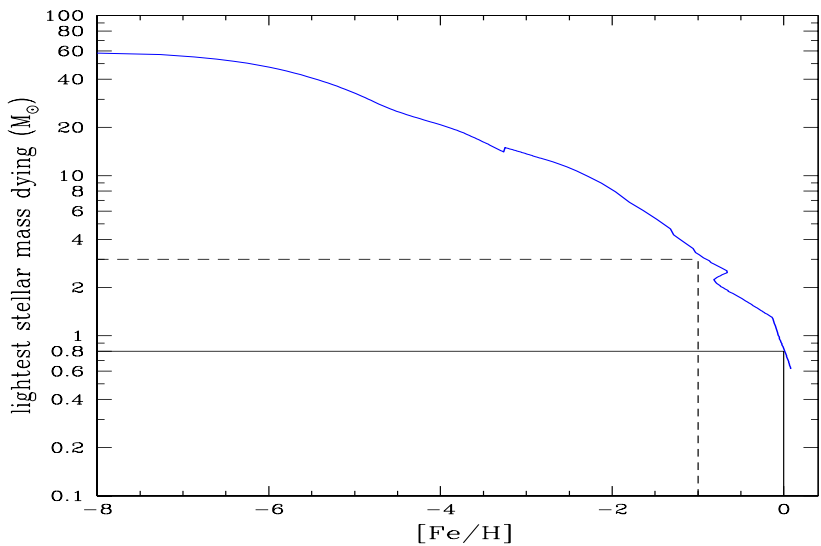}
\hfill
\caption{In this figure we show the smallest stellar mass which dies at any given [Fe/H] achieved by the ISM as a consequence of chemical evolution. Thus, it is clear that in the early phases of the halo only massive stars are dying and contributing to the chemical enrichment process. Clearly this graph depends upon the assumed stellar lifetimes and upon the age-[Fe/H] relation.
It is worth noting that the Fe production from Type Ia SNe appears before the gas has reached [Fe/H] =-1.0, therefore during the halo and thick disk phase. This clearly depends upon the assumed Type Ia SN progenitors (in this case the single degenerate model).}
\end{figure}

\begin{figure}
\hfill
\includegraphics[width=12cm,height=10cm]{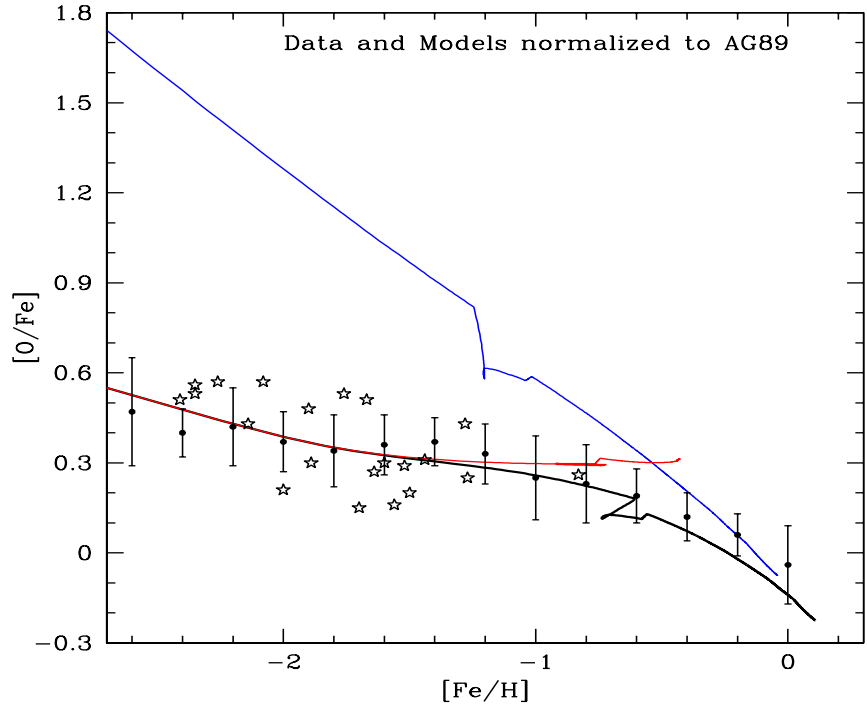}
\caption{The relation between [O/Fe] vs. [Fe/H] for Galactic stars in the solar vicinity. The models and the data are normalized to the 
solar meteoritic abundances of Anders \& Grevesse (1989). 
The thick curve represents the predictions of the two-infall 
model where Type Ia SNe produce $\sim 70\%$
of Fe and Type II SNe the remaining $\sim 30\%$. The upper thin curve represents the case where all 
the Fe is assumed to be produced by Type Ia SNe, whereas the thin lower line refers to the 
case where all the Fe is assumed to be produced in Type II SNe. The data are from 
Melendez \& Barbuy (2002).}
\end{figure}

\subsubsection{The time-delay model}

What we call time-delay model is the interpretation of the behaviour of abundance ratios such [$\alpha$/Fe] (where $\alpha$-elements are O, Mg, Ne, Si, S, Ca and Ti) versus [Fe/H], a typical way of plotting the abundances measured in the stars. The time-delay refers to the delay with which Fe is ejected into the ISM by SNe Ia relative to the fast production of $\alpha$-elements by core-collapse SNe. Tinsley (1979) first suggested that this time delay would have produced a typical signature in the [$\alpha$/Fe] vs. [Fe/H] diagram. In the following years, Greggio \& Renzini (1983b), by means of simple models (star formation burst or constant star formation) studied the effects of the delayed Fe production by Type Ia SNe on the [O/Fe] vs. [Fe/H] diagram. Matteucci \& Greggio (1986) included for the first time the Type Ia SN rate formulated by Greggio \& Renzini (1983a) in a detailed numerical model for the chemical evolution of the Milky Way. The effect of the delayed Fe production is to create an overabundance 
of O relative to Fe ([O/Fe]$>$ 0) at low [Fe/H] values, and a continuous decline  of the [O/Fe] ratio until the solar value ([O/Fe]$_{\odot}=0.0$) is reached for [Fe/H]$>-1.0$ dex. This is what is observed and indicates that during the halo phase the [O/Fe] ratio is due only to the production of O and Fe by SNe II. However, since the bulk of Fe is produced by Type Ia SNe, when these latter start to be important then the [O/Fe] ratio begins to decline. This effect was predicted by Matteucci \& Greggio (1986) to occur also for other $\alpha$-elements (e.g. Mg, Si). At the present time, a great amount of stellar abundances is available and the trend of the $\alpha$-elements has been confirmed. Before showing some of the most recent data, it is worth
showing better the time-delay model. 
In Figure 17 it is shown that a good fit of the [O/Fe] ratio as a function of [Fe/H] is obtained only if the $\alpha$-elements are mainly produced by Type II SNe and the Fe by Type Ia SNe. If one assumes that only SNe Ia produce Fe as well as if one assumes that only Type II SNe produce Fe, the agreement with observations is lost. Therefore, the conclusion is that both Types of SNe should produce Fe in the proportions of 1/3 for Type II SNe and 2/3 for Type Ia SNe. The 
IMF also plays a role in this game and these proportions are obtained for ``normal'' Salpeter-like IMFs, which includes both Salpeter (1955) and Scalo (1986) or Kroupa et al. (1993) IMFs.

As a further illustration of the time-delay model we show in Figures 18, 19 and 20 the [$X_i$/Fe] vs. [Fe/H] relations both observed and predicted for stars in the solar vicinity belonging to halo, thick- and thin-disk. The adopted yields for massive stars are those suggested by Fran\c cois et al. (2004) in order to best fit these relations and the solar abundances (namely the abundances in the ISM 4.5 Gyr ago).
These yields are obtained  by applying some corrections to the yields of WW95,
as shown in Figure 21, where the ratios between the suggested and WW95 yields are reported. 

In Figure 22 we show the predictions of a chemical evolution model for the solar vicinity where the recent yields from massive stars of Nomoto et al. (2006) have been adopted. As one can see, although  some of the problems present in the previous yields have been alleviated, for other elements the disagreement still persists.

\begin{figure}
\includegraphics[width=12cm,height=12cm]{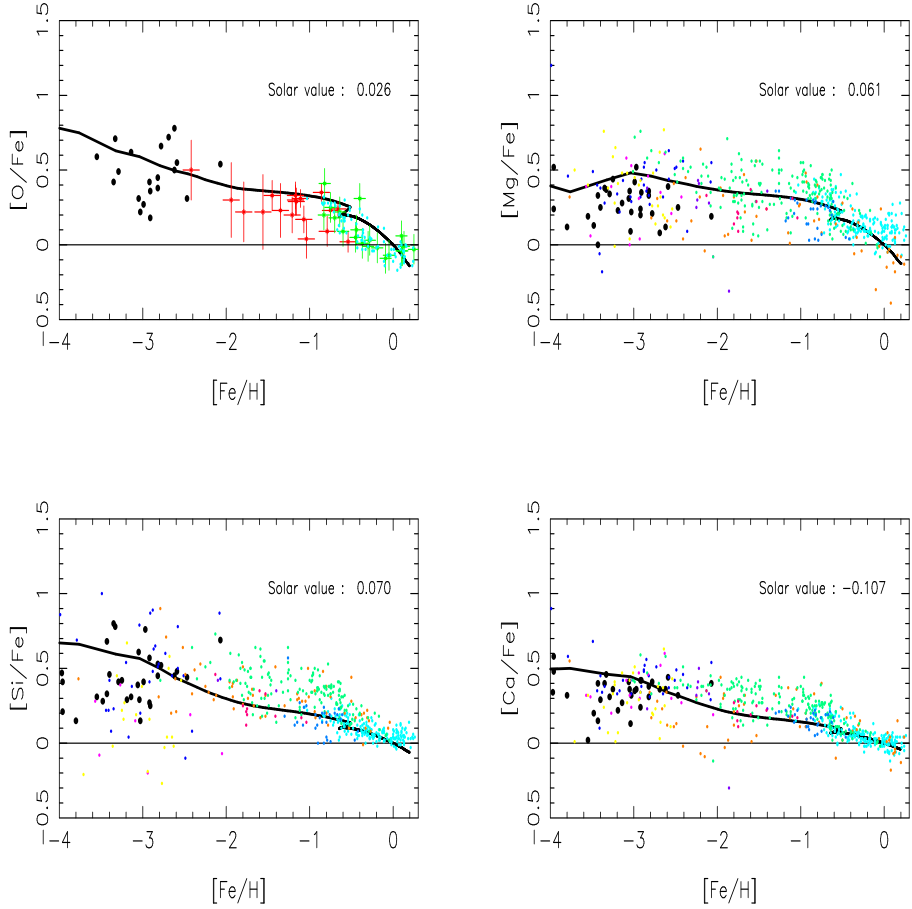}
\caption{Predicted and observed [$\alpha$/Fe] vs. [Fe/H] in the solar neighbourhood.
The models and the data are from Fran\c cois et al. (2004).
The models are normalized to the predicted solar abundances. The predicted abundance ratios at the time of the Sun formation (Solar value) are shown in each panel and indicate a good fit (all the values are close to zero).}
\end{figure}

\begin{figure}
\includegraphics[width=12cm,height=12cm]{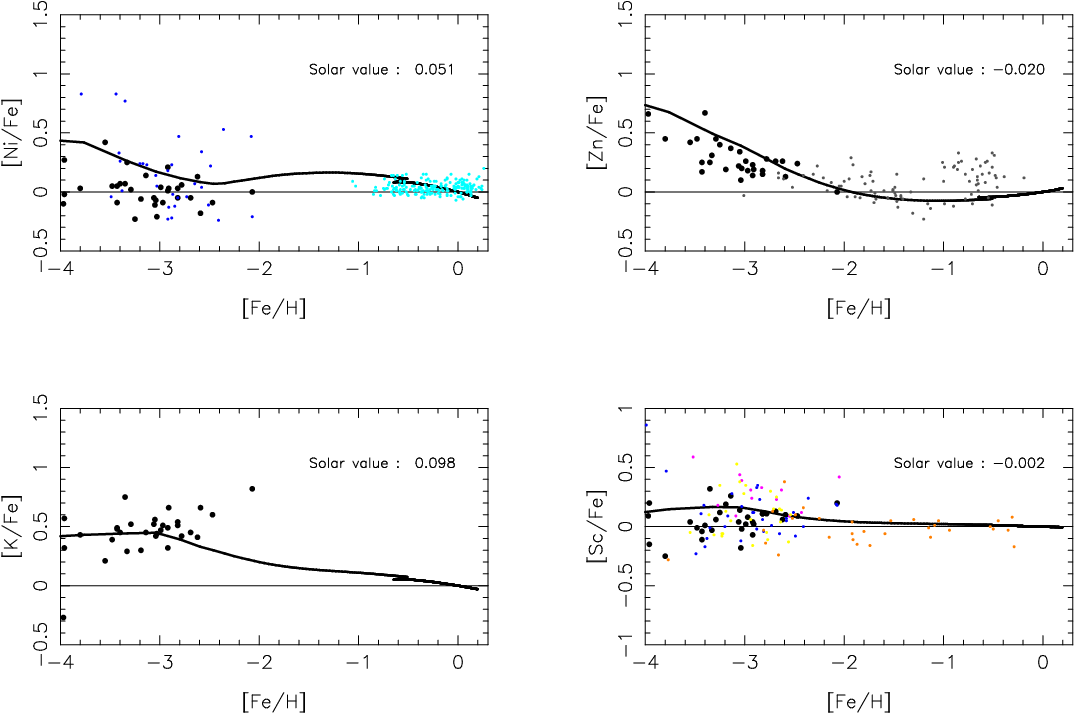}
\caption{The same as Fig.18 for Ni, Zn, K and Sc.
The models and the data are from Fran\c cois et al. (2004).
The models are normalized to the predicted solar abundances. The predicted abundance ratios at the time of the Sun formation are shown in each panel and indicate a good fit.}
\end{figure}

\begin{figure}
\includegraphics[width=12cm,height=12cm]{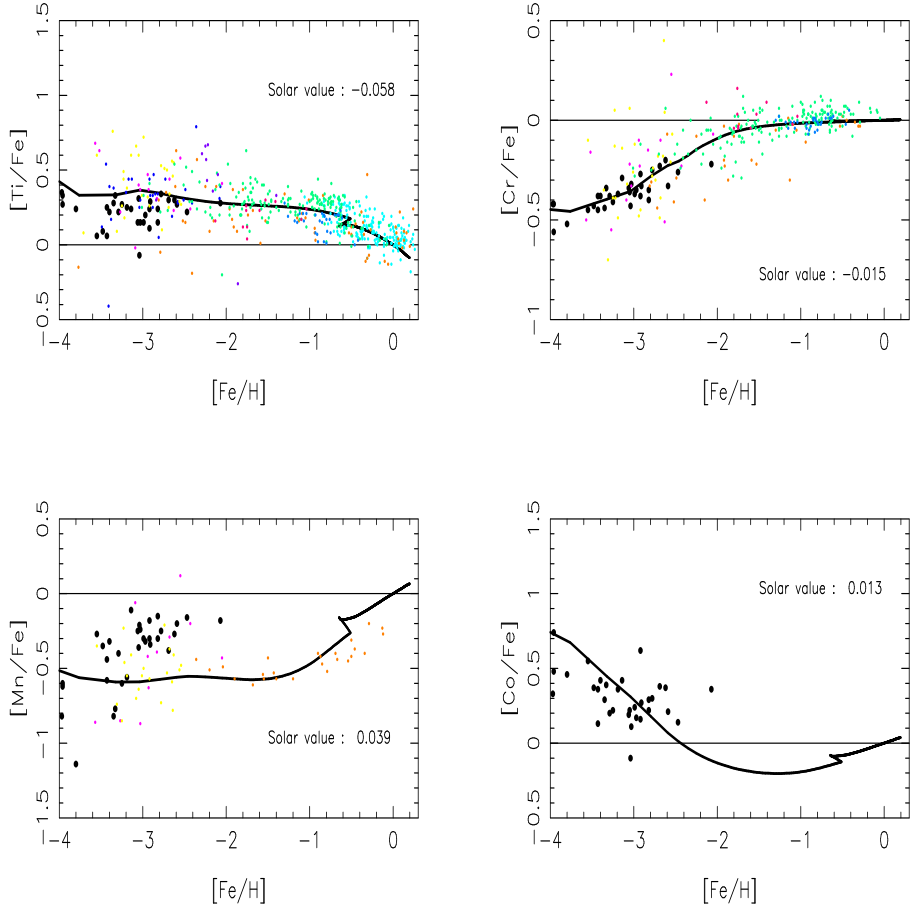}
\caption{The same as in Fig. 18 for Ti, Cr, Mn and Co.
The models and the data are from Fran\c cois et al. (2004).
The models are normalized to the predicted solar abundances. The predicted abundance ratios at the time of the Sun formation are shown in each panel and indicate a good fit.}
\end{figure}

\begin{figure}
\includegraphics[width=10cm,height=8cm]{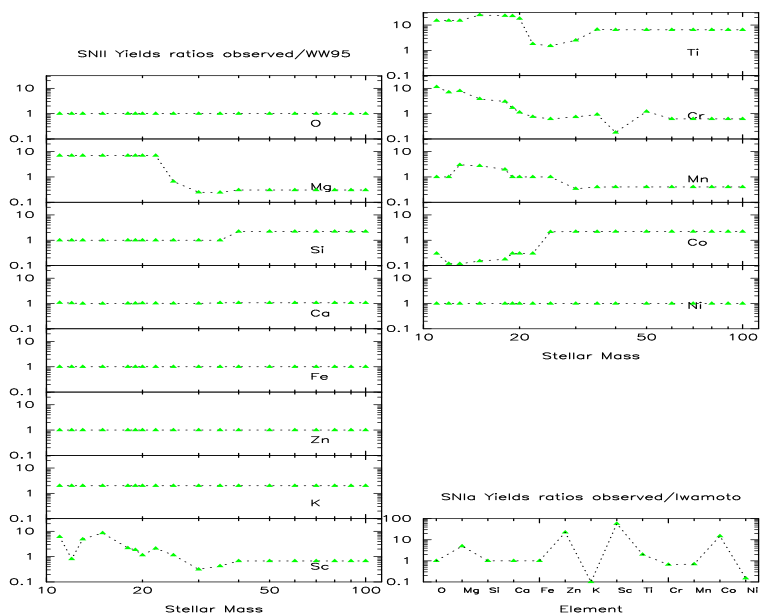}
\caption{Ratios between the empirical yields derived by Fran\c cois et al. (2004) and the yields of WW95 for massive stars. In the small panel at the bottom right we show the same ratios for SNe Ia and the comparison is with the yields of Iwamoto et al. (1999).}
\end{figure}

\begin{figure}
\includegraphics[width=12cm,height=10cm]{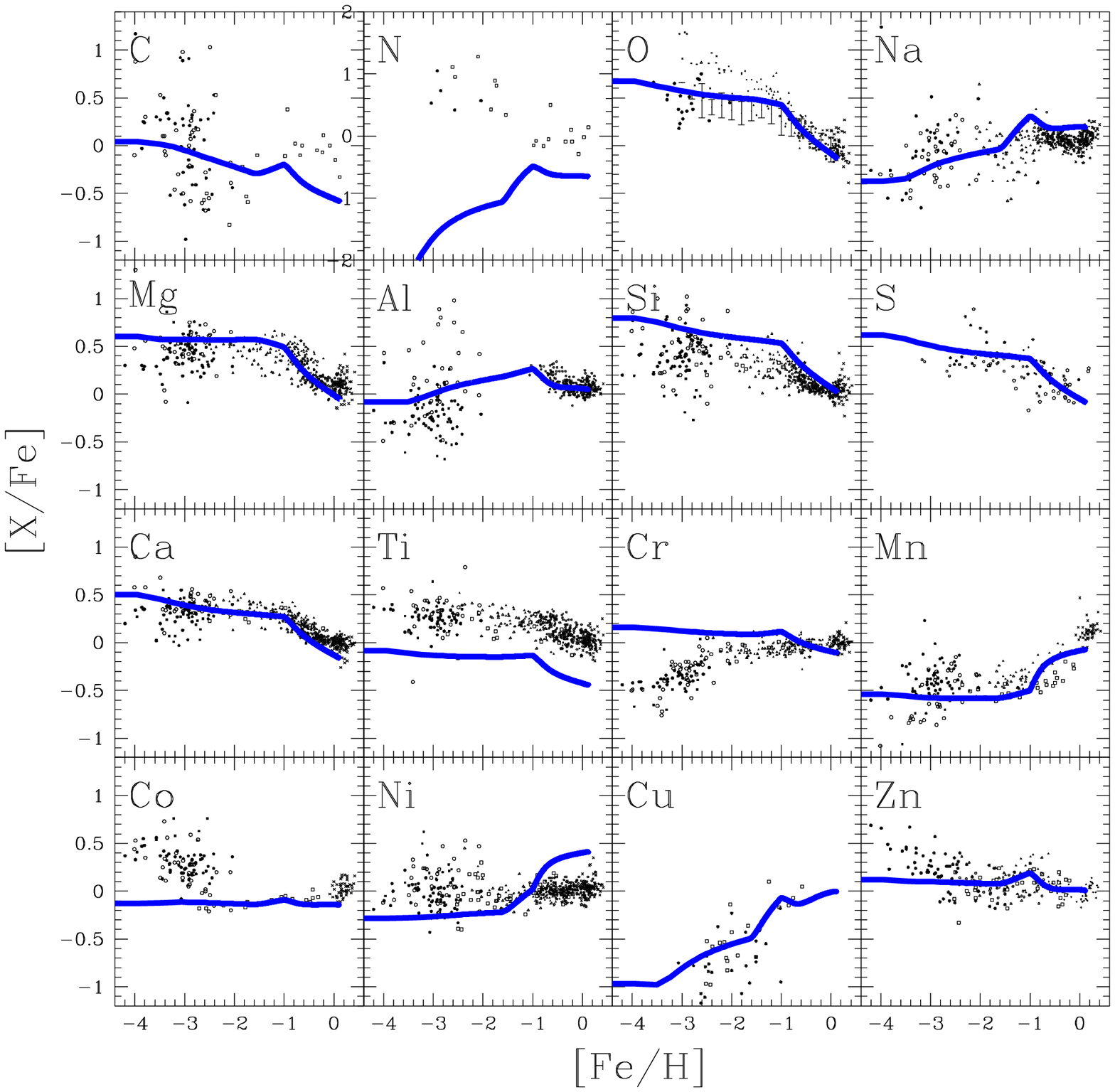}
\caption{Predicted and observed [X/Fe] vs. [Fe/H] in the solar neighbourhood. The predictions are from Nomoto et al. (2006), where all the references to the data can be found, and they have been obtained by means of metal dependent yields. Figure from Nomoto et al. (2006).}
\end{figure}

\subsubsection{The G-dwarf metallicity distribution and constraints on the thin disk formation}
The G-dwarf metallicity distribution is a quite important constraint for the chemical evolution of the solar vicinity. It is the fossil record of the star formation history of the thin disk. If one is able to reproduce such a distribution, then one can have an idea of the SFR and the IMF and, as a consequence, of the gas accretion history. Therefore, to fit the G-dwarf metallicity distribution means to obtain constraints on the mechanism of formation of the thin disk.
Originally, there was the ``G-dwarf problem'' which means that the Simple Model of galactic chemical evolution could not reproduce the distribution of the G-dwarfs. 
It has been since long demonstrated that relaxing the closed-box assumption and allowing for the solar region to form gradually by accretion of gas can solve the problem (Tinsley, 1980; Pagel 1997). Also a variable IMF could solve the problem but it would create other problems (see Martinelli \& Matteucci, 2000). Assuming that the disk forms from pre-enriched gas can also solve the problem but still the gas infall is necessary to have a realistic picture of the 
disk formation. 
The two-infall model can reproduce very well the G-dwarf distribution and also that of K-dwarfs (see Figures 23 and 24),  as long as a timescale for the formation of the disk in the solar vicinity of 7-8 Gyr is assumed. This conclusion is shared by other authors (Alib\'es et al. 2001; Prantzos \& Boissier 1999)

\begin{figure}
\includegraphics[width=10cm,height=8cm]{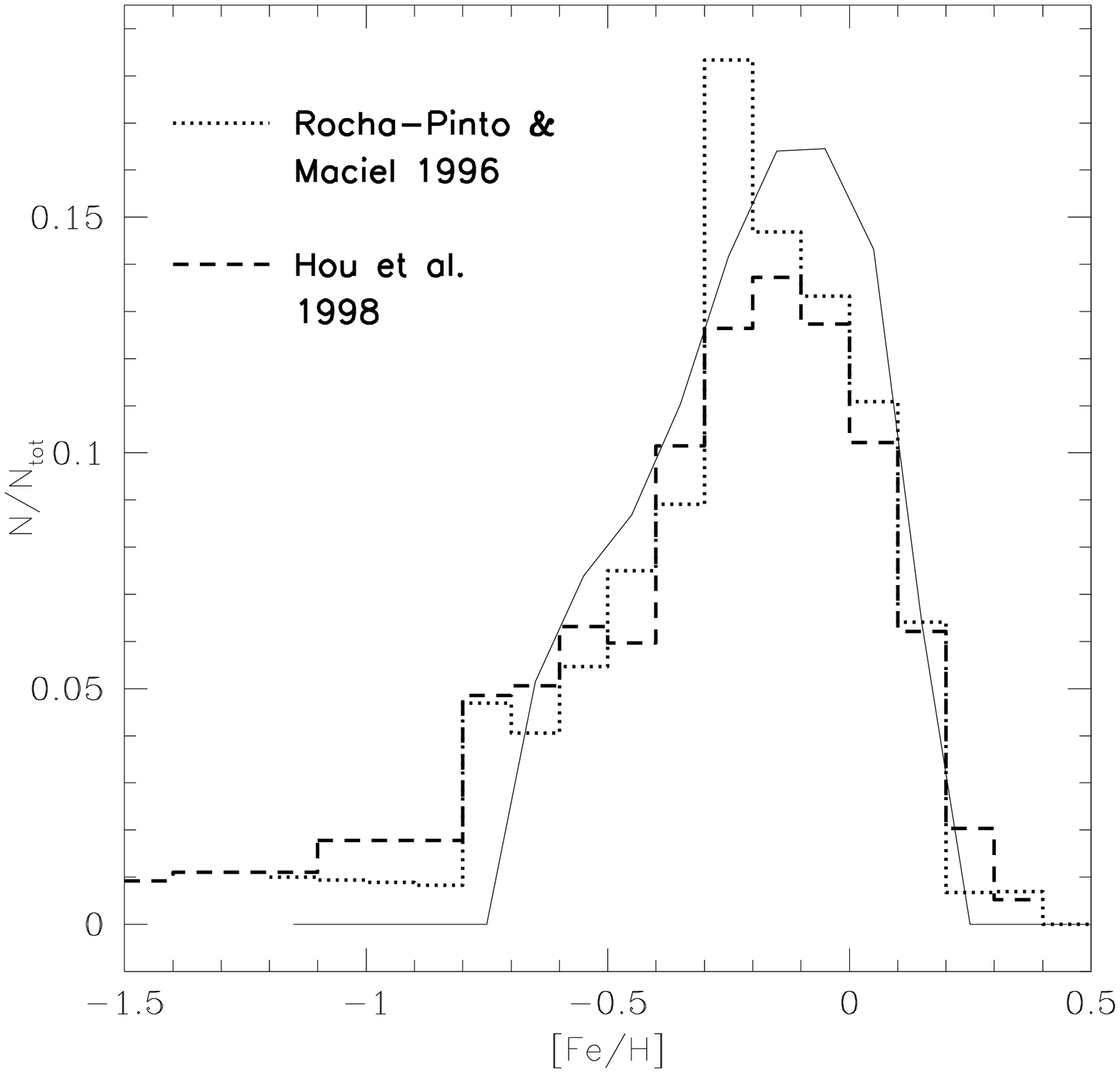}
\caption{The G-dwarf metallicity distribution. The model prediction is from Chiappini et al. (1997) and assumes a timescale for the formation of the local disk of 8 Gyr. The data are represented by the histograms.}
\end{figure}

\begin{figure}
\includegraphics[width=10cm,height=8cm]{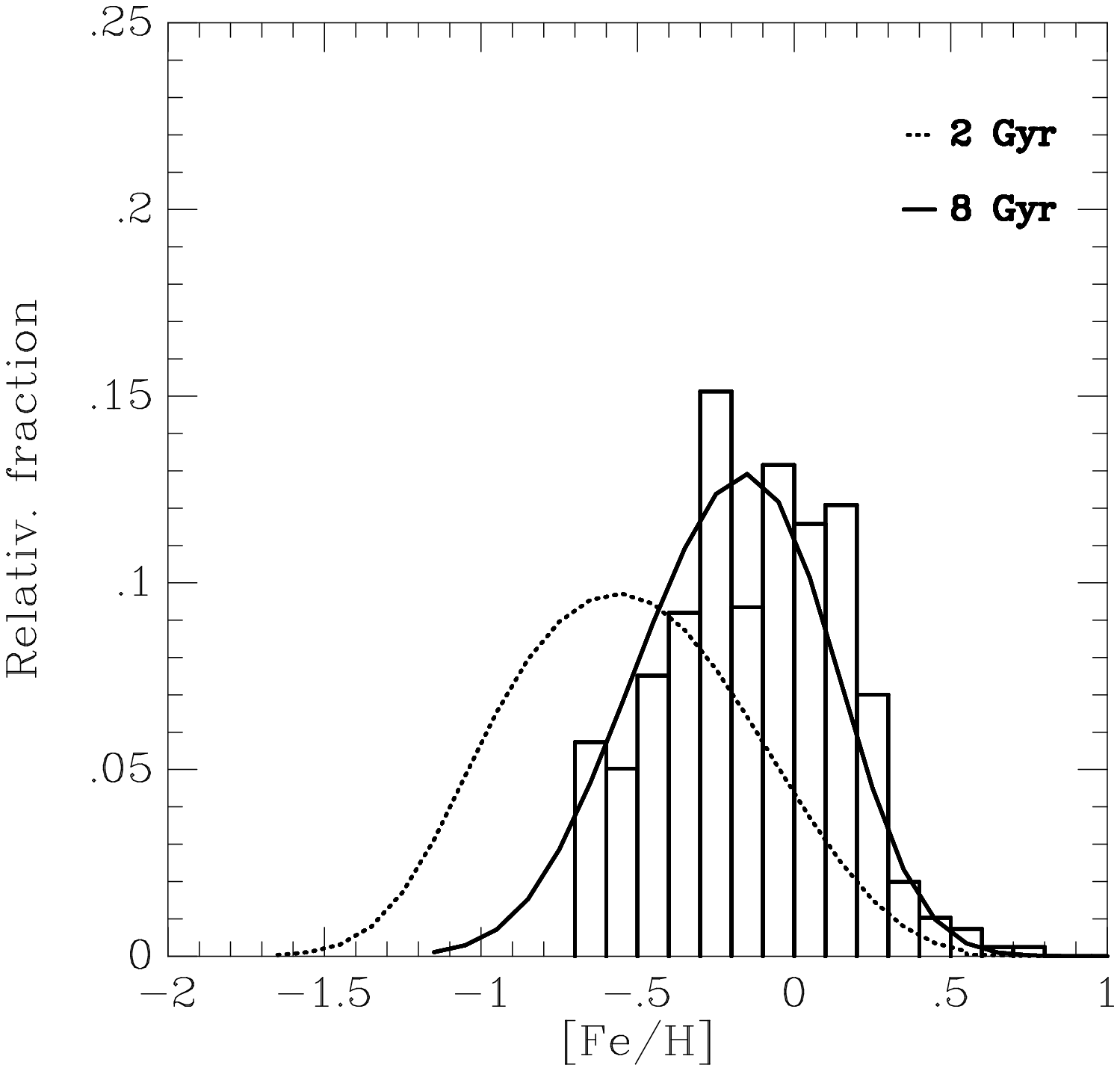}
\caption{The figure is from Kotoneva et al. (2002) and shows the comparison between a sample of K-dwarfs and model predictions in the solar neighbourhood.
The dotted curve refers to the two-infall model with a timescale $\tau=2$ Gyr, whereas the continuous line refers to $\tau= 8$ Gyr, as in Fig.23.}
\end{figure}

\subsubsection{Carbon and Nitrogen evolution}
Carbon and nitrogen deserve a separate discussion from the other elements, in particular $^{14}N$ whose observational behaviour is difficult to reconcile with the theory.
First of all, we should distinguish between {\it primary} and {\it secondary} elements: primary elements are those synthesized directly from H and He, whereas secondary elements are those deriving from metals already present in the star at birth. In the framework of the Simple Model of galactic chemical evolution, the abundance of a secondary element evolves like the square of the abundance of the progenitor metal, whereas the evolution of the abundance of a primary element does not depend on the metallicity.

In Figure 25 we show the predictions of the Simple Model for the ratio N/O, together with data for extragalactic HII regions and Damped Lyman-$\alpha$ systems (DLAs).

It is worth noting that the solutions of the Simple Model for a primary and a secondary element are oversemplifications since the Simple Model does not take into account stellar lifetimes which are very important in $^{14}N$ production, which arises mainly from low and intermediate mass stars, both as a secondary and primary element (e.g. Renzini \& Voli, 1981; van den Hoeck \& Groenewegen 1997). Also $^{12}$C originates mainly from low and intermediate mass stars. 
The contribution to $^{12}C$ from massive stars becomes very important only for metallicities oversolar, if the metallicity dependent mass loss is adopted (e.g. Maeder 1992).
\begin{figure}
\includegraphics[width=16cm,height=12cm,angle=-90]{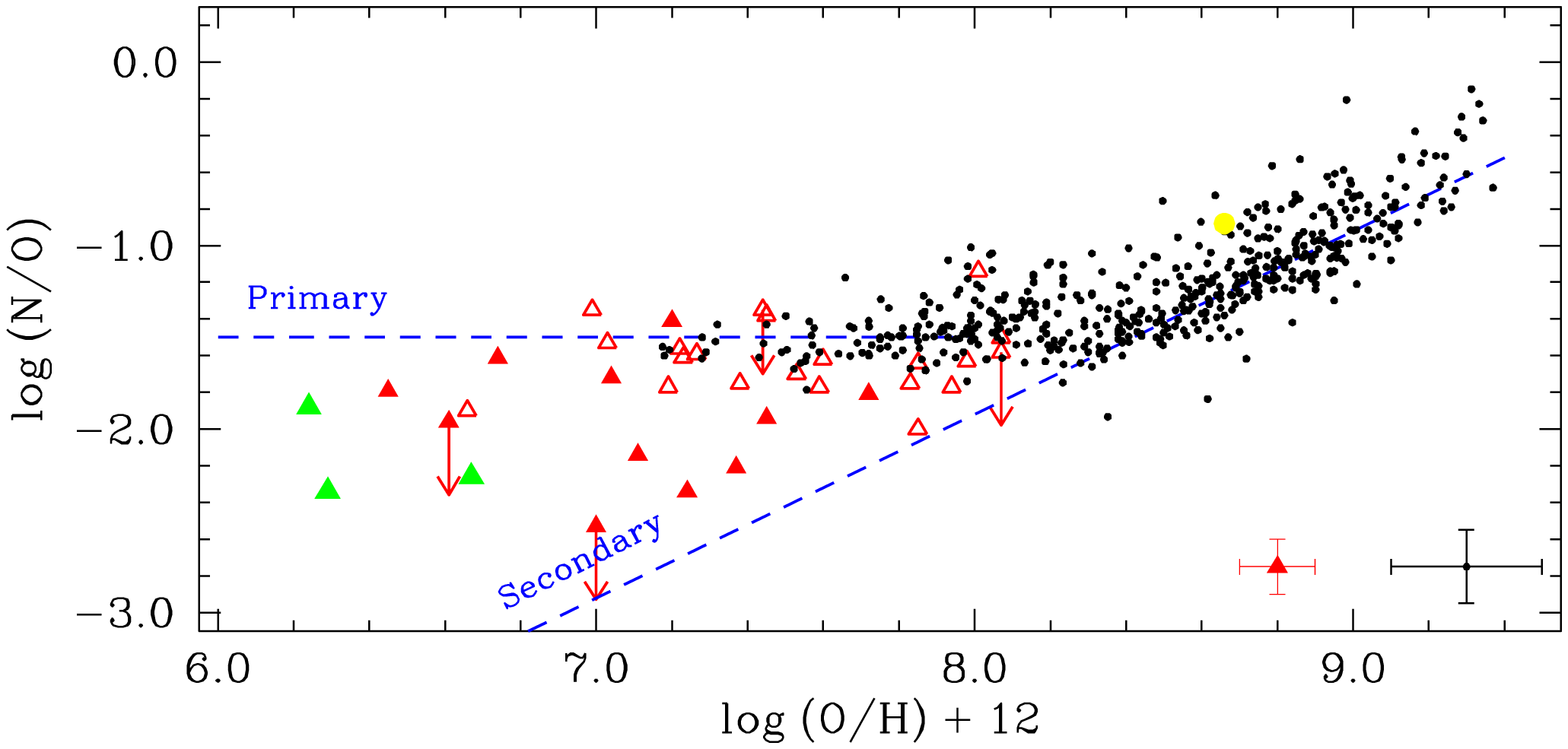}
\caption{The plot of log (N/O) vs. log(O/H)+12: small dots represent extragalactic HII regions, red triangles are Damped-Lyman $\alpha$ systems (DLA), which are high redshift objects (Pettini et al. 2002). The three green triangles are the most recent determinations for DLAs from Pettini et al. (2008).
Dashed lines mark the solution of the 
Simple Model for a primary and a secondary element. Figure from Pettini et al. (2008).}
\end{figure}
The interpretation of the diagram of Figure 25 is not so straightforward since 
extragalactic HII regions and DLAs are galaxies, and not necessarily that 
diagram is an evolutionary one, in the sense that O/H  does not trace the 
time unlike 
[Fe/H] in the Galactic stars.
Galaxies, in fact, may have started forming stars at different cosmic epochs and with different SF histories. However, if we interpret the diagram of Figure 25 as an evolutionary one, then the DLAs and the extragalactic HII regions of low metallicity should be younger and reflect the nucleosynthesis in massive stars and perhaps in intermediate mass stars. The observed plateau for N/O at low metallicity  then would indicate a primary production of N in massive stars. Nitrogen, in fact, is also produced in massive stars: until a few years ago, the N production in massive stars was considered only a secondary process, until Meynet \& Maeder (2001, 2003, 2005) showed that stellar rotation in massive stars can produce primary N.
A better test for the primary/secondary nature of N is represented by  the Galactic stars, since they really represent an evolutionary sequence. In Figures 26 and 27 we show the most recent data on C and N compared with chemical evolution models including N from rotating massive stars.

\begin{figure}
\includegraphics[width=12cm,height=12cm]{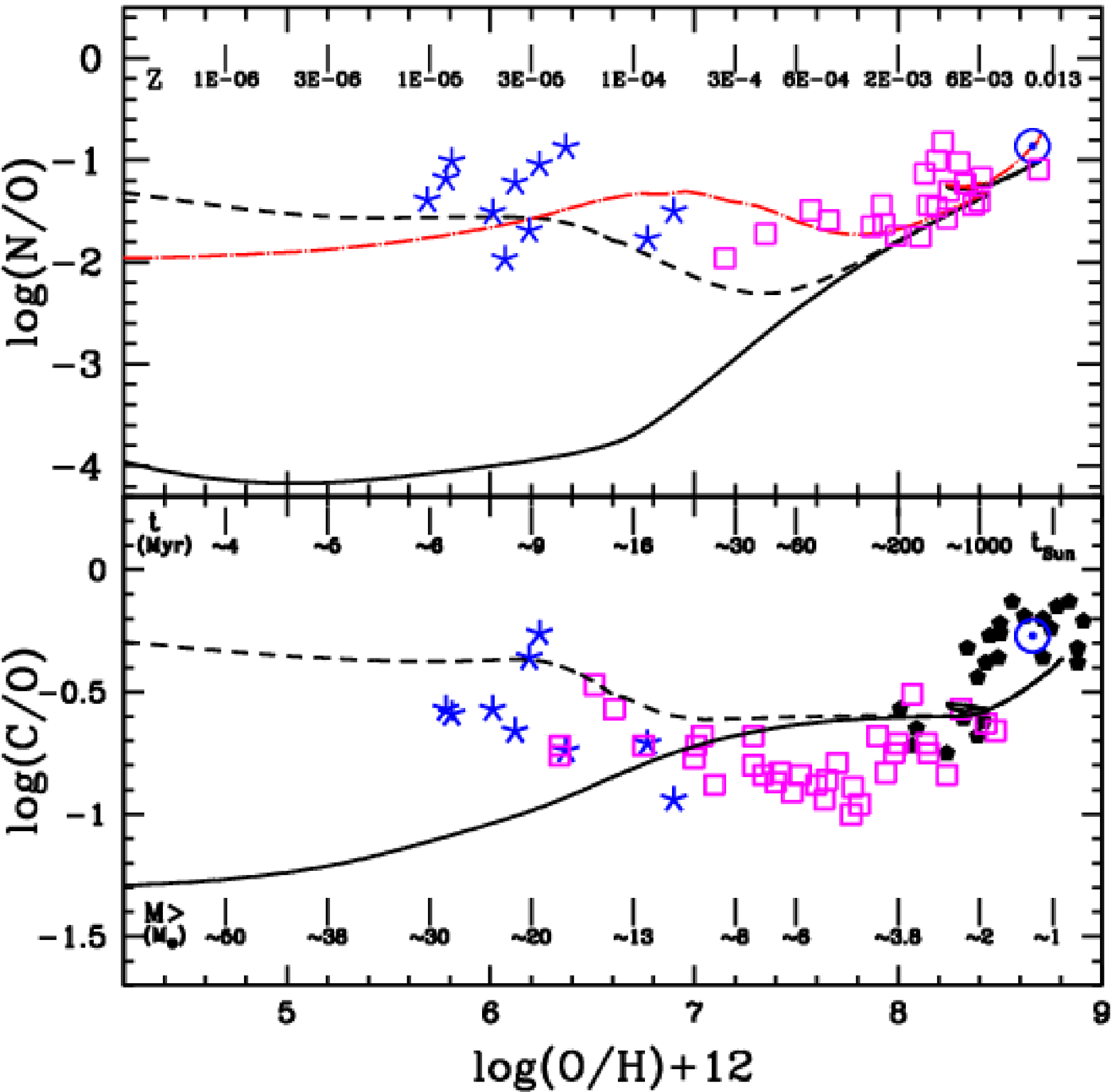}
\caption{Upper panel: solar vicinity diagram log(N/O) vs.
log(O/H) + 12. The data points are from Israelian et al. (2004)
(large squares) and  Spite et al. (2005)(asterisks). Models: the dashed line represents a model with substantial primary N production from massive stars. This was obtained by means of stellar models (Meynet et al. 2006; Hirschi 2007) with faster rotation relative to the work of Meynet \& Maeder (2002) for $Z=10^{-8}$. Lower panel: solar vicinity diagram log(C/O) vs. log(O/H) + 12. The data
are from Spite et al. (2005) (asterisks), Israelian et al. (2004) (squares) and
Nissen (2004) (filled pentagons). Solar abundances (Asplund et al. 2005,
and references therein) are also shown. Figure from Chiappini et al. (2006).}

\end{figure}

As one can see in Figures 26 and 27, the fit with data is good when primary N from massive stars is included. However, there are a few warnings, first of all the measurements of N abundance in stars of low metallicity are still uncertain and then the fact that  the N measurement in the gas in DLAs at high redshift show that at low O abundances there are systems with a log(N/O) $< -2.0$, below the plateau shown by Galactic stars. A plateau in [N/Fe]is also observed in Galactic stars for [Fe/H]$<-3.0$ dex, as shown in Figure 27.
In Figure 27 we show also the [C/Fe] values for Galactic stars but only for low metallicity stars: they indicate a roughly solar ratio like the stars with higher metallicities. Therefore, both [N/Fe] and [C/Fe] seem to show roughly constant solar values over the total [Fe/H] range. In the framework of the time-delay model, this means that C, N and Fe are all formed in the same stars (or with the same lifetimes) and that N is 
mainly a primary element. However, more data are necessary to assess this point and to reconcile the Galactic star data with high redshift DLAs.

\begin{figure}
\includegraphics[width=12cm, height=12cm]{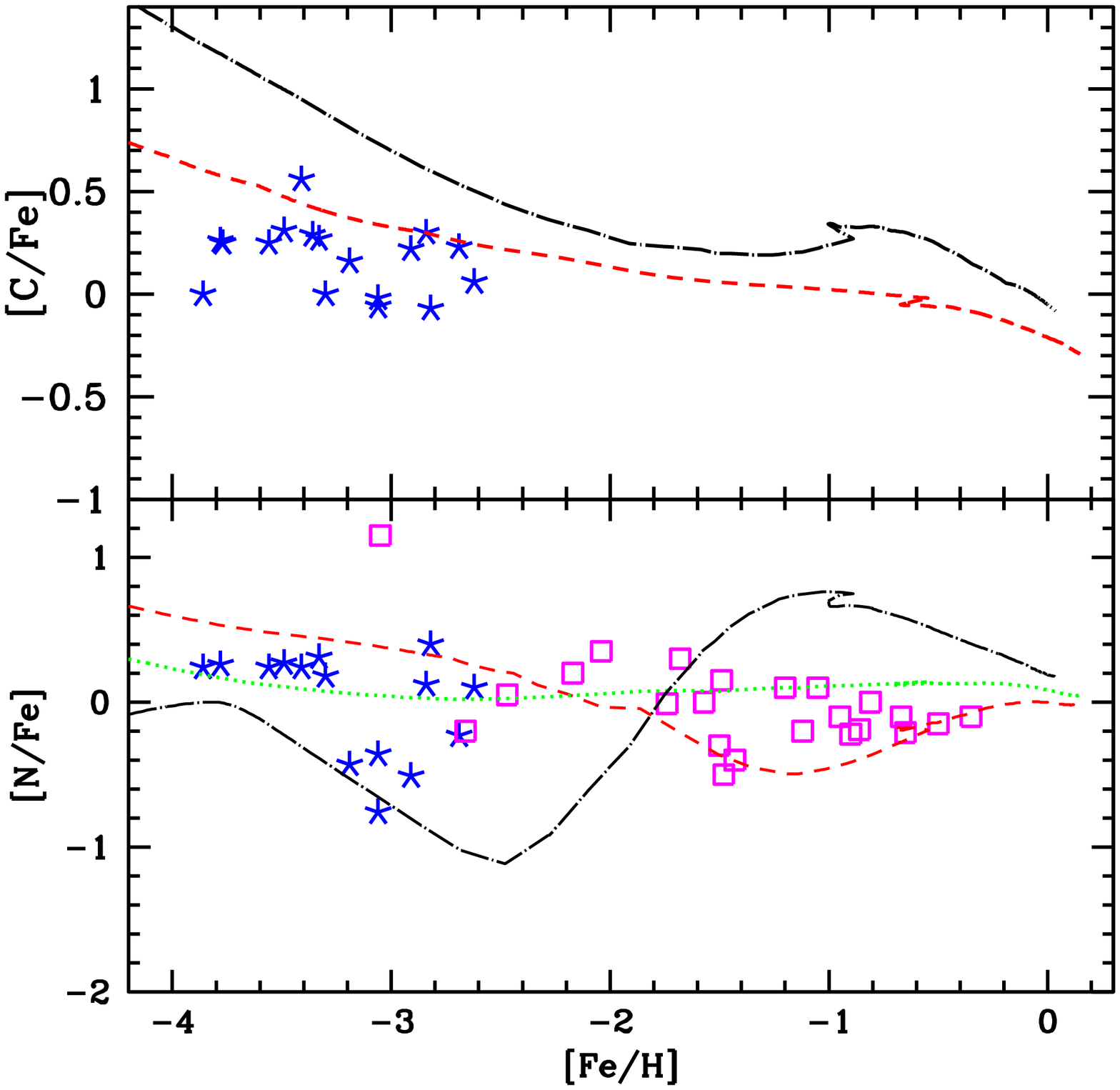}
\caption{Observed and predicted [C/Fe] vs. [Fe/H] (upper panel) and [N/Fe] vs. [Fe/H] (bottom panel) in the solar neighbourhood. The data points are from Cayrel et al. (2004), Spite et al. (2005) (asterisks) and Israelian et al. (2004) (squares). The dot-dashed line represents a model with yields from Chieffi \& Limongi (2002, 2004) for a metallicity  $Z=10^{-6}$ connected to the Pop III stars (only massive stars for that metallicity). The dashed line and the dotted lines represent heuristic models where the yields of C and N have been assumed ``ad hoc''. In particular, the fraction of primary N from massive stars is obtained by the fit to the data at low metallicity. Figure from Chiappini, Matteucci \& Ballero (2005).}
 \end{figure}

\subsubsection{S- and r- process elements}
The s- and r- process elements are generally 
produced by neutron capture on Fe seed nuclei. The former 
are formed during the He-burning phase both in low and massive stars, whereas the latter occur in explosive events such as Type II SNe. Recently, Fran\c cois et al. (2006) have measured the abundances of several very heavy elements (e.g. Ba and Eu) in extremely metal poor stars of the Milky Way. Previous work on the subject had shown a large spread in the abundance ratios of these elements to iron, especially at low metallicities. This spread is confirmed by this more recent study although is less than before, and is at variance with the lack of spread observed in the other elements shown before (e.g. $\alpha$-elements). 
Apart from this problem, not yet solved, these diagrams can be very useful to place constraints on the nucleosynthetic origin of these elements. In particular, Cescutti et al. (2006) by adopting the two-infall model predicted the evolution of [Ba/Fe] and [Eu/Fe] versus [Fe/H], as shown in Figures 28 and 29. They can well fit the average trend but not the spread at very low metallicities since the model assumes instantaneous mixing. In order to fit the Ba evolution, they assumed that Ba is mainly produced as s-process element in low mass stars (1-3$M_{\odot}$) but that a fraction of Ba is also produced as an r-process element in stars with masses 12-30$M_{\odot}$. Europium is assumed to be only an r-process element  produced in the range 12-30$M_{\odot}$.

 \begin{figure}
 \includegraphics[width=12cm,height=10cm]{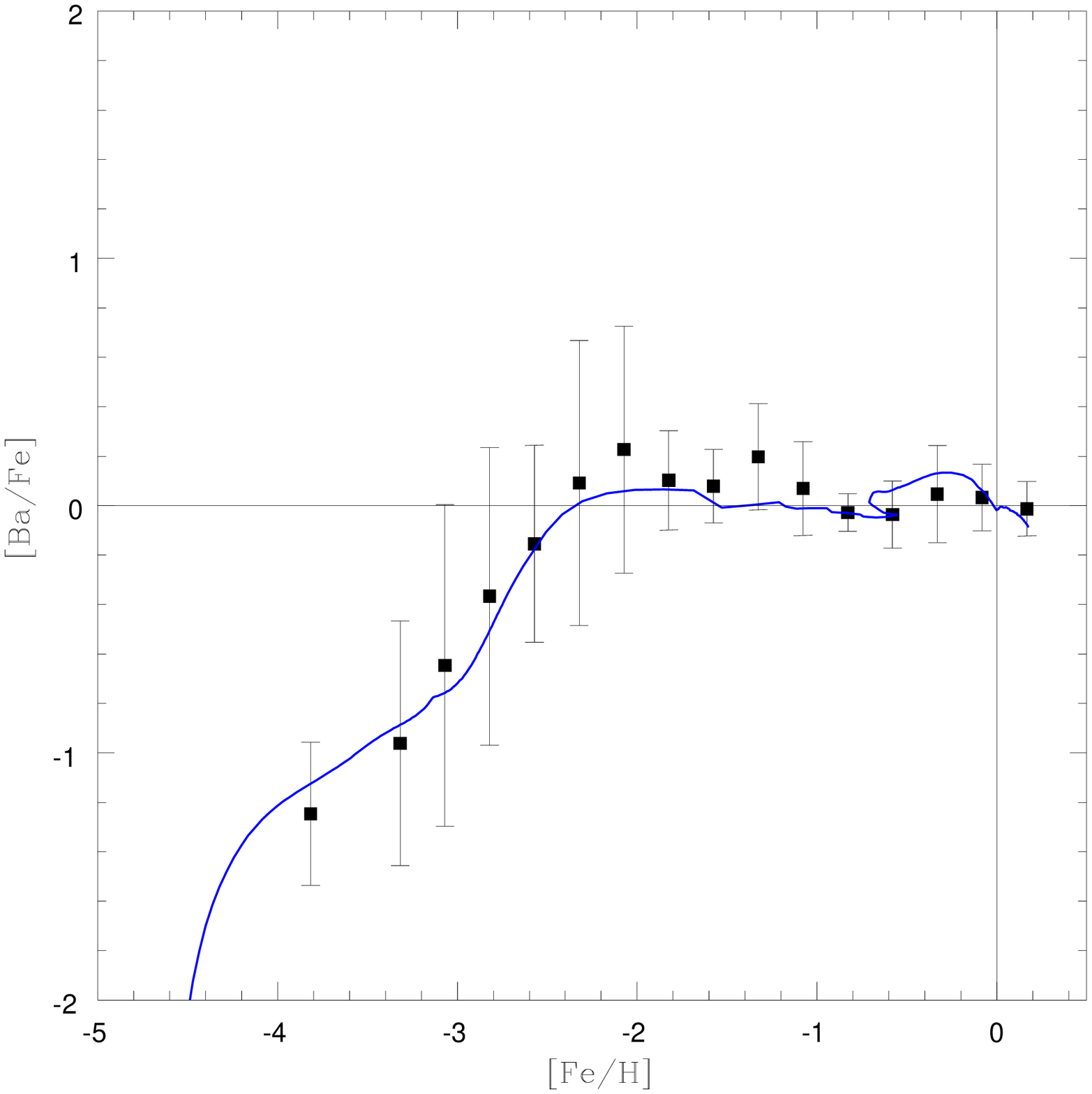}
 \caption{The evolution of Barium in the solar vicinity as predicted by the 
two-infall model (Cescutti et al. 2006). Data are from 
Fran\c cois et al. (2006).}
\end{figure}

\begin{figure}
 \includegraphics[width=12cm,height=10cm]{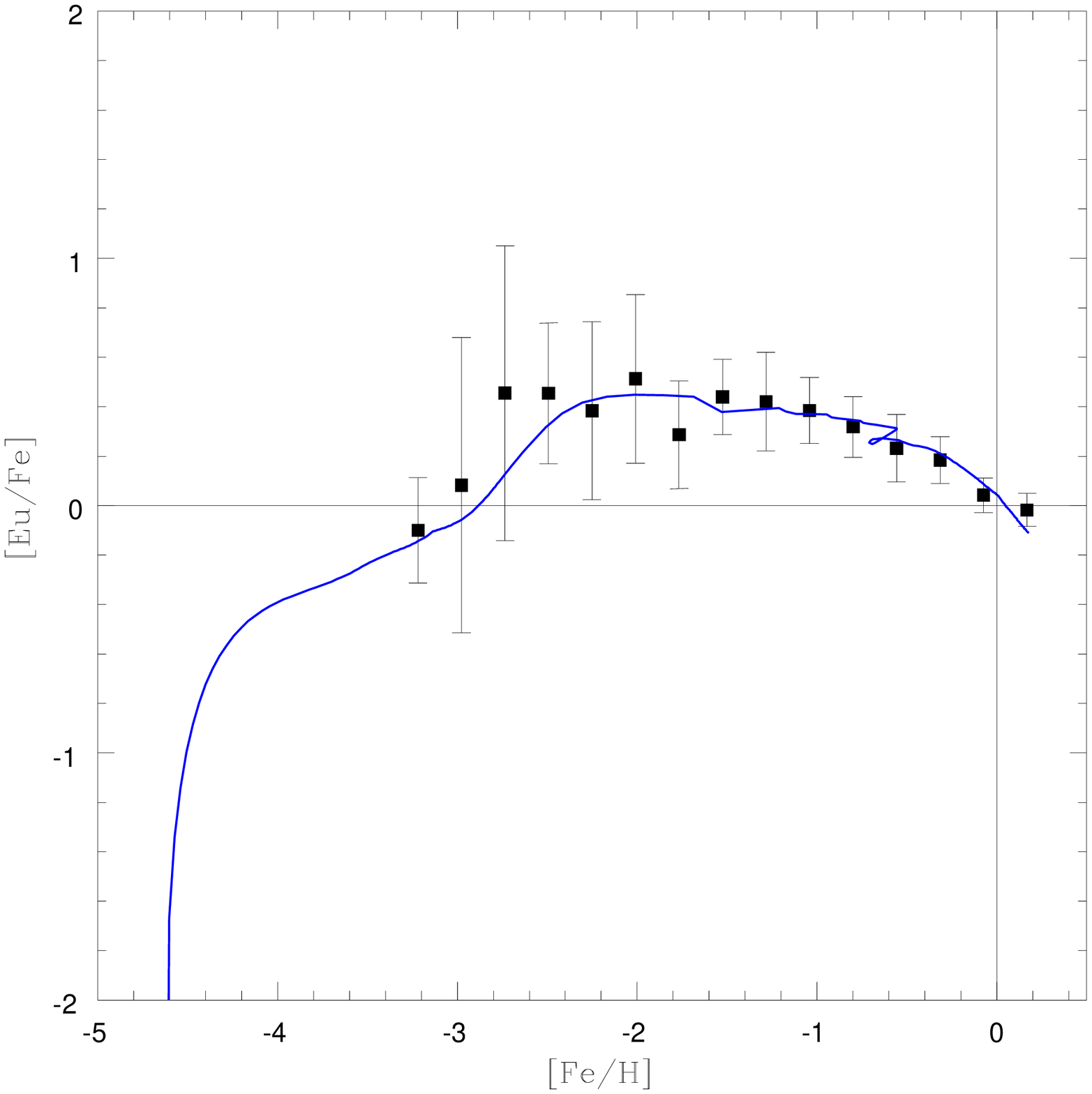}
\caption{The evolution of Europium in the solar vicinity (Cescutti et al. 2006). Data are from Fran\c cois et al. (2006).}
\end{figure}

In order to explain why the s- and r- process elements show a large and probably real spread at very low metallicities, whereas elements such as the $\alpha$-elements show only a little spread, one  could think of a moderately inhomogeneous model coupled with differences in the nucleosynthesis between s- and r- process elements on one side and $\alpha$-elements on the other side (see Cescutti 2008).
Highly inhomogeneous models for the halo evolution, in fact, predict a too large spread for the $\alpha$-elements at low metallicity (e.g. Argast et al. 2000).
It is worth noting the typical secondary behaviour of Ba, whose main production is by means of the s-process, which needs Fe seed nuclei already present in the star, and neutrons which are accreted on these nuclei. The production of neutrons is also dependent on the original metal content, therefore it would be even more precise to speak of Ba as a {\it tertiary} element.

\subsection{The Galactic disk}
A good model of chemical evolution for the Milky Way should reproduce also the features of the Galactic disk. In  particular: abundance gradients, gas and SFR distribution with the galactocentric distance.

\subsubsection{Abundance gradients}
The chemical abundances measured along the disk of the Galaxy suggest that the metal content decreases from the innermost to the outermost regions, in other words there is a negative gradient in metals.
Abundance gradients can be derived from HII regions, planetary nebulae (PNe), open clusters and stars (O,B stars and Cepheids). 
There are two types of abundance determinations in HII regions:
one is based on recombination lines which should have  a weak dependence on the temperature of the nebula (He, C, N, O),
the other is based on collisionally excited lines where a strong dependence is intrinsic to the method 
(C, N, O, Ne, Si, S, Cl, Ar, Fe and Ni). 
This second method has predominated until now.
A direct determination of the abundance gradients from HII regions in the Galaxy from optical lines is difficult because 
of extinction, so usually the abundances for distances larger than 3 Kpc from the Sun are obtained from radio and infrared 
emission lines.

Abundance gradients can also be derived from optical emission lines in PNe. However, the abundances of He, C and N in PNe
are giving only information on the internal nucleosynthesis of the star. So, to derive gradients one should look at the abundances of O, S and Ne, unaffected
by stellar processes. 
Abundance gradients are derived also from measuring the Fe abundance in open clusters (e.g. Carraro et al. 2004; Yong et al. 2005) or from abundances in Cepheids (e.g. Andrievsky et al. 2002 a,b,c, 2004; Luck et al. 2003; Yong et al. 2006) or from abundances in O, B stars (e.g. Daflon \& Cunha, 2004).

In Figure 30 we show theoretical predictions of abundance gradients along the disk of the Milky Way compared with data from HII regions, B stars and PNe. The adopted model is from Chiappini et al. (2001) and is based on an 
inside-out formation of the thin disk. The assumed model does not allow for exchange of gas between different regions of the disk. The disk is, in fact, divided in several concentric shells 2 Kpc wide with no interaction between them.

\begin{figure}
\includegraphics[width=12cm,height=10cm]{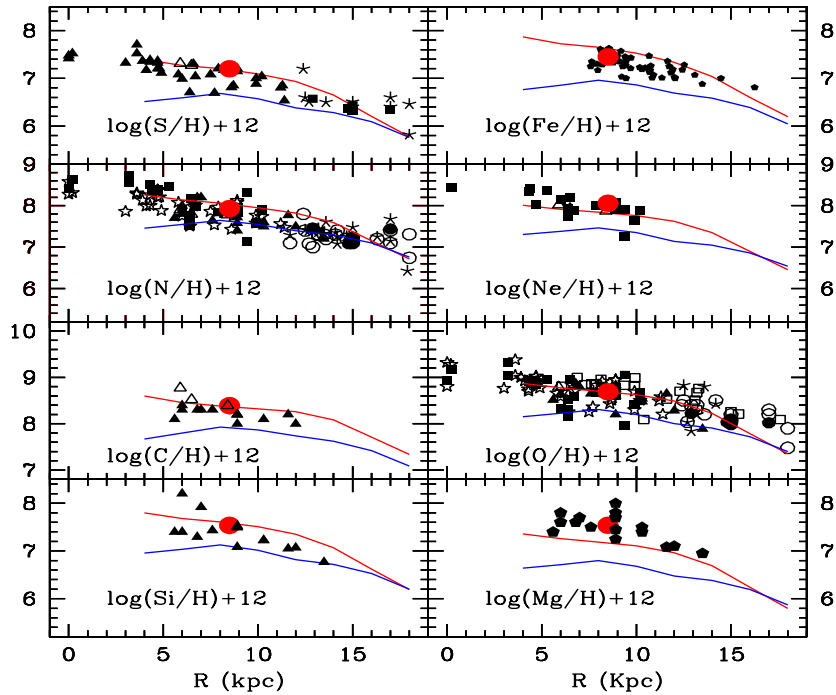}
\hfill
\caption{Spatial and temporal behaviour of abundance 
gradients along the Galactic disk as predicted by the best model of Chiappini et al. (2001). The upper lines in each panel represent the present time gradient, whereas the lower ones represent the gradient a few Gyr ago. It is clear that the gradients tend to stepeen in time, a still controversial result. The data are from HII regions, B stars and PNe (see Chiappini et al. 2001).} 
\end{figure}

As already mentioned, most of the current models agree on the inside-out scenario for the 
disk formation, however not all models agree on the evolution of the gradients 
with time. In fact, some models, although assuming an inside-out formation of the disk, predict a gradient flattening with time (Boissier \&
Prantzos 1999; Alib\`es et al. 2001), whereas others such as that of Chiappini 
et al. (2001) predict a steepening, as shown in Figure 30. 
The reason for the steepening is that in 
the model of Chiappini et al. there is included a threshold density for SF, which 
induces the SF to stop when the density decreases below the threshold. 
This effect is particularly strong in the external 
regions of the Galactic disk, thus
contributing to a slower evolution and therefore to a steepening of the 
gradients with time.
In Figure 31 we show models and some more recent data including Cepheids.

\begin{figure}
\includegraphics[width=12cm,height=13cm]{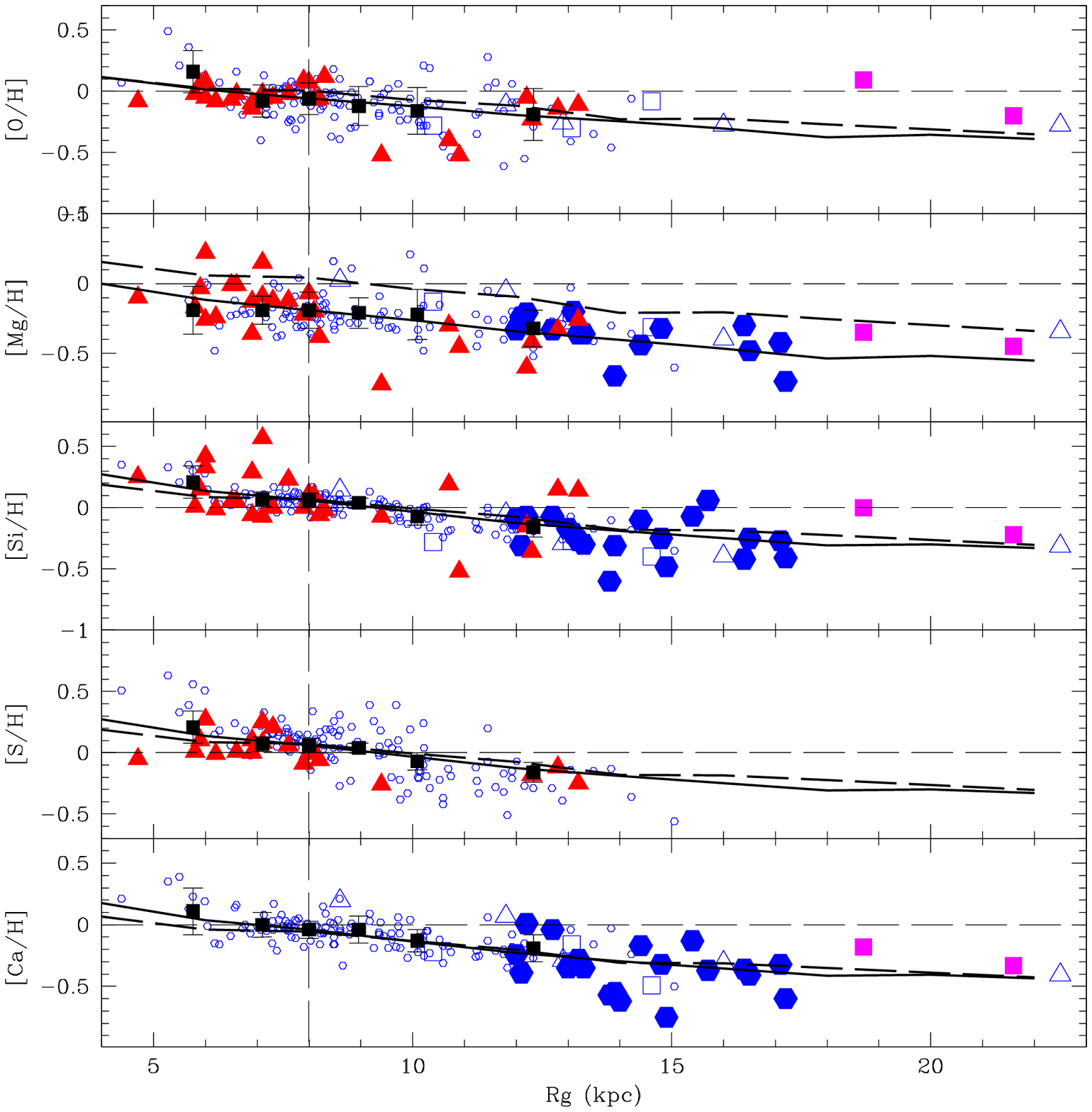}
\hfill
\caption{ Gradients of the $\alpha$-elements along the disk. The predicted gradients for O, Mg, Si, S and Ca are compared with different
sets of data. The small open circles are the data of the Cepheids by Andrievsky et al. (2002a,b,c; 2004) and Luck et al. (2003). The solid triangles
are the data by Daflon \& Cunha (2004) (OB stars), the open 
squares are the data by Carney et al. (2005) (red giants), the 
solid hexagons are the data by Yong et al. (2006) (Cepheids), the
open triangles are the data by Yong et al. (2005) (open clusters)
and the solid squares are the data by Carraro et al. (2004) (open
clusters). The most distant value for Carraro et al. (2004) and Yong
et al. (2005) refers to the same object: the open cluster Berkeley 29.
The thin solid line represents the model predictions 
at the present time normalized to the
mean value of the Cepheids at 8 Kpc;
the dashed line represents the predictions of the model at the epoch of
the formation of the solar system normalized to the observed solar abundances
by Asplund et al. (2005). This prediction should be compared
with the data for red giant stars and open clusters (Carraro et al. 2004;
Carney et al. 2005; Yong et al. 2005). The models and the Figure are from Cescutti et al. (2007).} 
\end{figure}

In the Chiappini et al. model, the fit to the gradients is obtained by means of the inside-out formation of the Galactic disk. Numerical simulations of abundance gradients show that no gradient arises if one assumes the same timescale of disk formation at any galactocentric distance. The different timescale of accretion influences the SFR, thus creating a gradient in the SFR and therefore in the resulting metal content. However, it should be said that the effect of the threshold is also important and tends to steepen the gradients.

In Figure 32  we show the results of Boissier \& Prantzos (1999) for abundance gradients and also 
for the gas and SFR distribution along the disk.Note the gradients are flattening with time. 
\begin{figure}
\includegraphics[width=14cm,height=12cm]{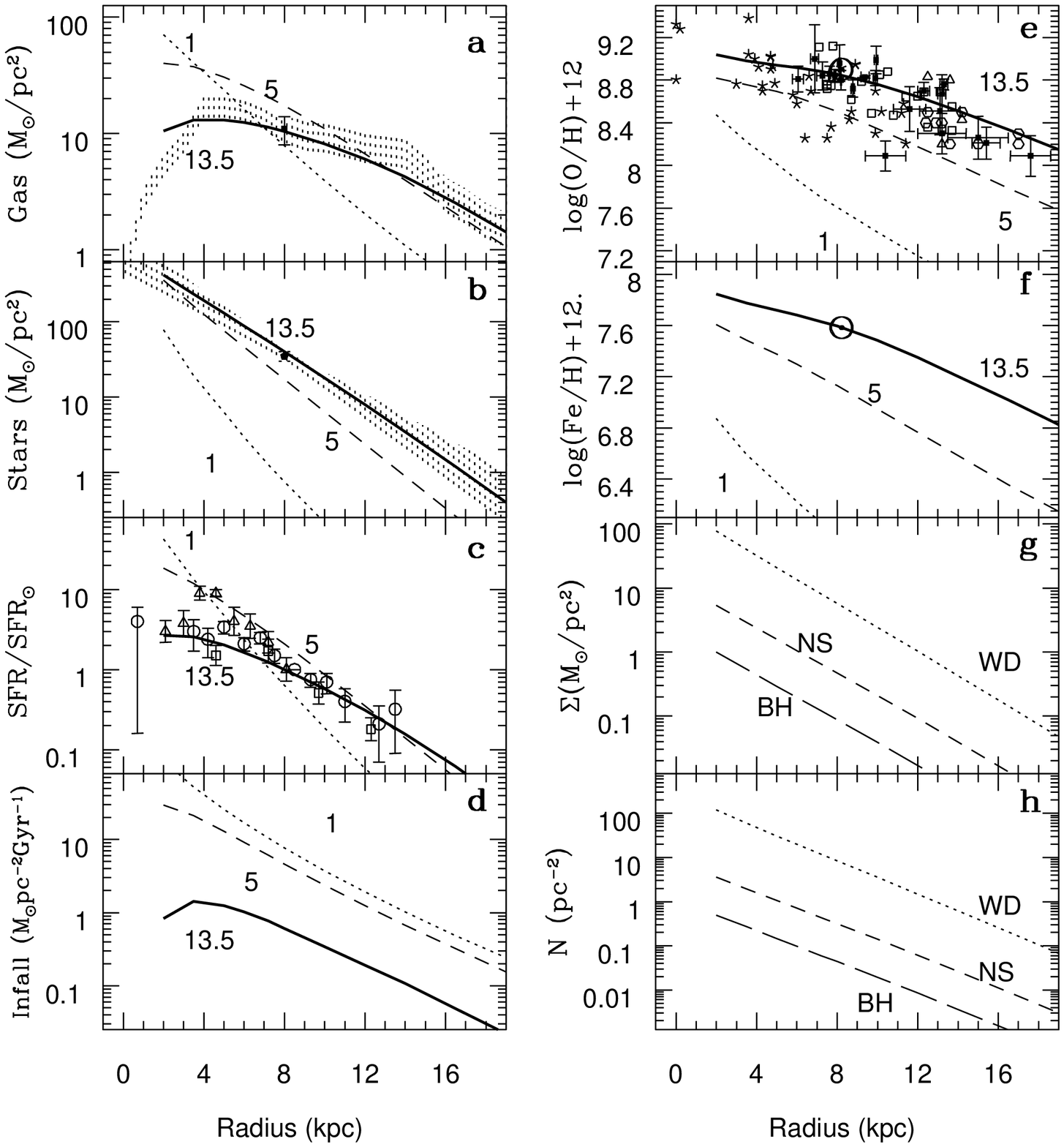}
\hfill
\caption{Comparison between model predictions and observations for the disk of the Milky Way. The figure is from Boissier \& Prantzos (1999). Top left panel:
gas distribution along the disk. Top right panel: the O gradient at the present time (curve with label 13.5) and at two other different cosmic epochs (5 Gyr and 1 Gyr from the beginning). Second left panel: the surface mass density of living stars. Second right panel: the Fe gradient. Third left panel: the gradient of the SFR normalized to the value at the solar ring. Third right panel: the predicted distribution of the current surface mass densities of stellar remnants 
(WDs), black holes (BH) and neutron stars (NS). Fourth left panel: the predicted infall rate along the disk at three different cosmic epochs. Fourth right panel: the predicted distributions of surface densities by number of the stellar remnants. } 
\end{figure}

\subsection{The Galactic bulge}
\subsubsection{Bulge formation}

The bulges of spiral galaxies are generally distinguished in true bulges, hosted by S0-Sb galaxies and ``pseudobulges'' hosted in later type galaxies
(see Renzini 2006 for references). Generally, the properties (luminosity, colors, line strenghts) of true bulges are very similar to those of elliptical galaxies.
In the following, we will refer only to true bulges and in particular to the bulge of the Milky Way.
The bulge of the Milky Way is, in fact, the best studied bulge and several scenarios for its formation have been put forward in past years.
As summarized by Wyse \& Gilmore (1992) the proposed scenarios are:
\begin{itemize}

\item the bulge formed by accretion of extant stellar systems which eventually 
settle in the center of the Galaxy.

\item The bulge was formed by accumulation of gas at the center of the Galaxy and subsequent evolution with either fast or slow star formation.

\item  The bulge was formed by accumulation of 
metal enriched gas from the halo, thick disk or thin disk
in the Galaxy center.

\end{itemize}

In the context of chemical evolution, the Galactic bulge was first
modeled by Matteucci \& Brocato (1990) who predicted that the
[$\alpha$/Fe] ratio for some elements (O, Si and Mg) should be 
supersolar over almost the whole metallicity range, in analogy with
the halo stars, as a consequence of assuming a fast bulge evolution
which involved rapid gas enrichment in Fe mainly by Type II SNe. 
At that time, no data were available for chemical abundances; the
predictions of Matteucci \& Brocato (1990)
were confirmed for a few $\alpha$-elements (Mg,
Ti) by the observations of McWilliam \& Rich (1994, hereafter MR94), whereas 
for other $\alpha$-elements (e.g. Ca, Si) the observed trend was different.
Other discrepancies regarding the Mg overabundance came from Sadler et
al. (1996).
In order to better assess these points, Matteucci et al. (1999)
studied a larger set of abundance ratios, by means of a detailed
chemical evolution model whose parameters were calibrated so that the
metallicity distribution observed by MR94 could be fitted.
They concluded that an evolution much faster than that in the solar
neighbourhood and even faster than that of the halo (see also Renzini,
1993) is necessary for the MR94 metallicity distribution to be
reproduced, and that an IMF index flatter ($x=1.1-1.35$) than that of
the solar neighbourhood is needed as well. 
They also made predictions about the evolution of several abundance
ratios which were meant to be confirmed or disproved by subsequent
observations, namely that $\alpha$-elements should in general be
overabundant with respect to Fe, but some (e.g. Si, Ca) less than
others (e.g. O, Mg), and that the [$^{12}$C/Fe] ratio should be solar
at all metallicities. 

Samland et al. (1997) developed a self-consistent chemo-dynamical
model for the evolution of the Milky Way components starting from a
rotating protogalactic gas cloud in virial equilibrium, which
collapses owing to dissipative cloud-cloud collisions. They found that
self-regulation due to a bursting star formation and subsequent
injection of energy from Type II supernovae led to the development of
``contrary flows'', i.e. alternate collapse and outflow episodes in
the bulge. This caused a prolonged star formation episode lasting over
$\sim4\times10^9$ yr. They included stellar nucleosynthesis of O, N and
Fe, but claimed that gas outflows prevent any clear correlation
between local star formation rate and chemical enrichment.
With their model, they could reproduce the oxygen gradient of HII
regions in the equatorial plane of the Galactic disk and
the metallicity distribution of K giants in the bulge (Rich, 1988),
field stars in the halo and G dwarfs in the disk, but they did not
make predictions about abundance ratios in the bulge.
In general, hierarchical clustering models of galaxy formation do not support
the conclusion of a fast formation and evolution of the bulge. In Kauffmann (1996) the bulges 
form through violent relaxation and destruction of disks in major mergers. The stars of the 
destroyed disk build the bulge, and subsequently the bulge has to be rebuilt. This implies 
that late type spirals should have older bulges than early type ones, since the build-up of a 
large disk needs a long time during which the galaxy has to evolve undisturbed. This is not 
confirmed by observations, since the high metallicity and the the narrow age distribution 
observed in bulges of local spirals are not compatible with their merger origin (see Wyse, 1999).

Moll\'a et al. (2000) proposed a multiphase model in the context of
the dissipative collapse scenario of the Eggen et al. (1962) picture. 
They supposed that the bulge formation occurred in two main infall
episodes, the first from the halo to the bulge, on a timescale
$\tau_H=0.7$ Gyr, and the
second from the bulge to a so-called core population in the very
nuclear region of the Galaxy, on a timescale $\tau_B \gg \tau_H$. 
The three zones (halo, bulge, core) interact via supernova winds and
gas infall. 
They concluded that there is no need for accretion of external
material to reproduce the main properties of bulges and that the
analogy to ellipticals is not justified.
Because of their rather long timescale for the bulge formation, these
authors did not predict a noticeable difference in the trend of the
[$\alpha$/Fe] ratios but rather suggested that they behave more akin to that
in the solar neighbourhood (contrary to several indications from
abundance data, e.g. MR94).

Immeli et al. (2004)
investigated the role of cloud dissipation in the formation and dynamical evolution of star forming gas rich disks by means of a 3D chemodynamical model.
They found that the galaxy evolution proceeds very differently depending on whether the gas disk or the stellar disk first become unstable. This in turn depends on how efficiently the cold cloud medium can dissipate energy. If the gas cools efficiently, a starburst takes place which gives rise to enhanced [$\alpha$/Fe] ratios, thus in agreement with a fast bulge formation.

A more recent model was proposed by Costa et al. (2005), in which the
best fit to observations is achieved by means of a double infall
model. 
An initial fast (0.1 Gyr) collapse of primordial gas is followed by a
supernova-driven mass loss and then by a second slower (2 Gyr) infall
episode, enriched by the material ejected by the bulge during the
first collapse. 
Costa et al. (2005) claimed that the mass loss is necessary to reproduce
the abundance distribution observed in PNe, and because
the predicted abundances would otherwise be higher than observed.
However, it should be noted again that the abundances derived from PNe can be affected by internal 
stellar processes and therefore are meaningless for studying galactic chemical evolution.
With their model, they are able to reproduce the trend of [O/Fe]
abundance ratio observed by Pomp\'eia et al. (2003) and the data of
nitrogen versus oxygen abundance observed by Escudero \& Costa (2001)
and Escudero et al. (2004). It must be noted however that 
Pomp\'eia et al. (2003) obtained abundances for ``bulge-like'' dwarf
stars. This ``bulge-like'' population consists of old ($\sim 10-11$
Gyr) metal-rich nearby stars whose kinematics and metallicity suggest
an inner disk or bulge origin and a mechanism of radial migration,
perhaps caused by the action of a Galactic bar, but the birthplace of
these stars is undoubtedly not certain.

\subsubsection{Interpretation of bulge data and other galaxies}

In summary, MR94 first measured the metallicity distribution and the 
[$\alpha$/Fe] ratios in the Galactic bulge and confirmed partly 
the predictions of 
Matteucci \& Brocato (1990) that all of the $\alpha$-elements should be 
enhanced relative to Fe for a large range of [Fe/H]. In fact, MR94 found that 
not all the $\alpha$-elements were enhanced, in particular oxygen.
Very recently, medium- and high-resolution spectroscopy of bulge stars
was performed (Rich \& McWilliam, 2000; Fulbright et al., 2006, 2007;
Zoccali et al., 2006; Lecureur et al. 2007), and it seems to indicate that 
also O is enhanced, thus supporting the suggestion of a 
fast formation of the bulge. 
The metallicity distribution of stars in the bulge and the [$\alpha$/Fe] ratios
greatly help 
in selecting the most probable scenario for the bulge 
formation. In Figure 33 we present the predictions by Matteucci (2003) of the 
[$\alpha$/Fe] ratios as functions of [Fe/H] in galaxies of different morphological type. In particular, for the Galactic bulge or an elliptical galaxy of the same mass, for the solar vicinity region and for an irregular magellanic galaxy (LMC and SMC).
The underlying assumption is that different objects undergo different  histories of star formation, being very fast in the spheroids (bulges and ellipticals), moderate in spiral disks and slow and perhaps gasping in irregular gas rich galaxies. The effect of different star formation histories is evident in Figure 33 where the predicted  [$\alpha$/Fe] ratios in the bulge and ellipticals remain high and almost constant for a large interval of [Fe/H]. This is due to the fact that, since star formation is very intense, the bulge reaches very soon a solar metallicity thanks only to the SNe II; then, when SNe Ia start exploding and ejecting Fe into the ISM, 
the change in the slope occurs at larger [Fe/H] than in the solar vicinity.
In the extreme case of irregular galaxies the situation is opposite: here the star formation is slow and when the SNe Ia start exploding the gas is still very metal poor. 
This scheme is quite useful since it can be used to identify galaxies only by looking at their abundance ratios. 
A model for the bulge behaving as shown in Figure 33 is able to reproduce also the observed metallicity distribution of bulge stars (see Matteucci \& Brocato 1990; Matteucci et al. 1999).
The scenario suggested in these papers 
favors the formation of the bulge by means of a short and strong starburst, in agreement with 
Elmegreen (1999) and Ferreras et al. (2003). A similar model, although updated with the inclusion of the development of a galactic wind and more recent stellar yields,  has been presented by Ballero et al. (2007): it shows how a model with intense star formation (star formation efficiency $\sim 20 Gyr^{-1}$) and rapid assembly of gas (0.1 Gyr)
can best reproduce the most recent accurate data on abundance ratios and metallicity distribution.

\begin{figure}
\includegraphics[width=12cm,height=12cm]{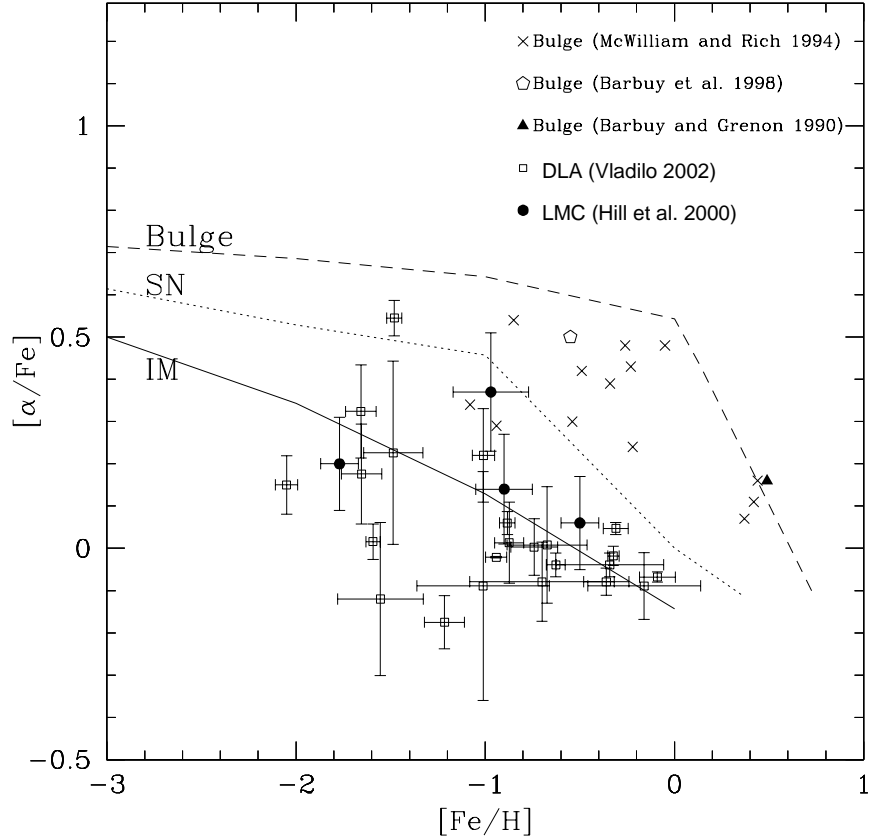}
\hfill
\caption{The predicted [$\alpha$/Fe] vs. [Fe/H] relations for the Galactic bulge (upper curve), the solar vicinity (median curve) and irregular galaxies (low curve). Data for the bulge are reported for comparison. Data for the LMC and DLA systems are also shown for comparison, indicating that DLAs are probably irregular galaxies. Figure and references are from Matteucci 
(2003).}\label{fig}
\end{figure}

In Figures 34 and 35 we show a detailed comparison between model predictions for the Galactic bulge and data on O and Mg.

\begin{figure}
\includegraphics[width=12cm,height=12cm]{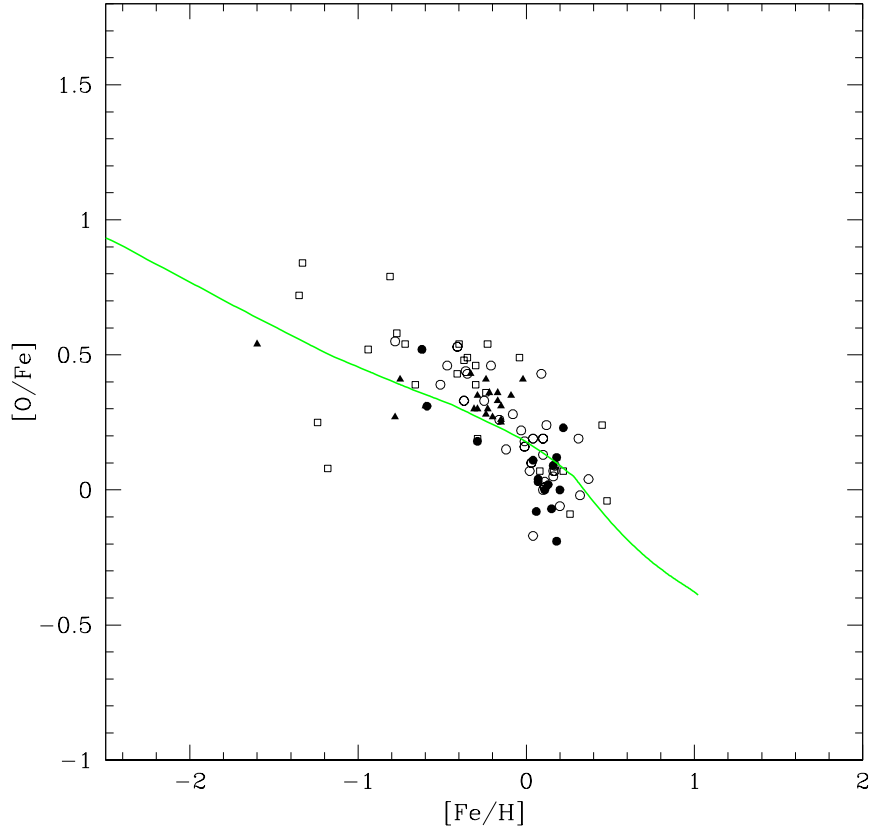}
\hfill
\caption{The predicted [O/Fe] vs. [Fe/H] relation for the Bulge (curve), compared with the most recent data. The chemical evolution model is that of Ballero et al. (2007), where references to the data can be found. }
\end{figure}

\begin{figure}
\includegraphics[width=12cm,height=12cm]{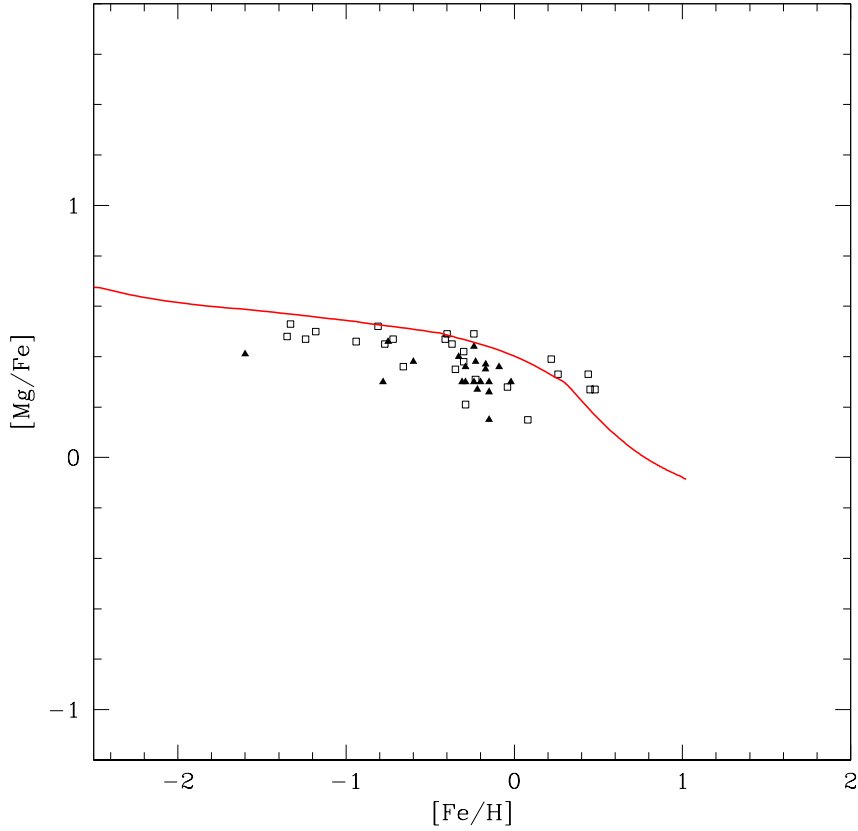}
\hfill
\caption{The predicted [Mg/Fe] vs. [Fe/H] relation for the bulge (curve), compared with the most recent data. The chemical evolution model is that of Ballero et al. (2007), where references to the data can be found. }
\end{figure}

As one can see, the plateau in the [$\alpha$/Fe] is longer than in the solar neighbourhood, since in the bulge the slope of the [$\alpha$/Fe] ratio starts changing drastically only for [Fe/H]$>$ 0.0 dex. The long plateau is well explained by the  model of Ballero et al. (2007) assuming a very fast formation of the bulge. It is worth noting that the [O/Fe] ratio has a steeper slope than [Mg/Fe] and this could be due to differences in the nucleosynthesis of these elements (e.g. McWilliam et al. 2007).

The IMF assumed for the bulge is usually flatter than the IMF of the solar neighbourhood and this is generally dictated by the fit of the bulge metallicity distribution which peaks at a higher [Fe/H] than the G-dwarf metallicity distribution in the solar vicinity. Numerical calculations have indicated that the main parameter influencing the peak of the distribution is the IMF, as clearly shown in  Figures 36 and 37.

\begin{figure}
\includegraphics[width=10cm,height=8cm]{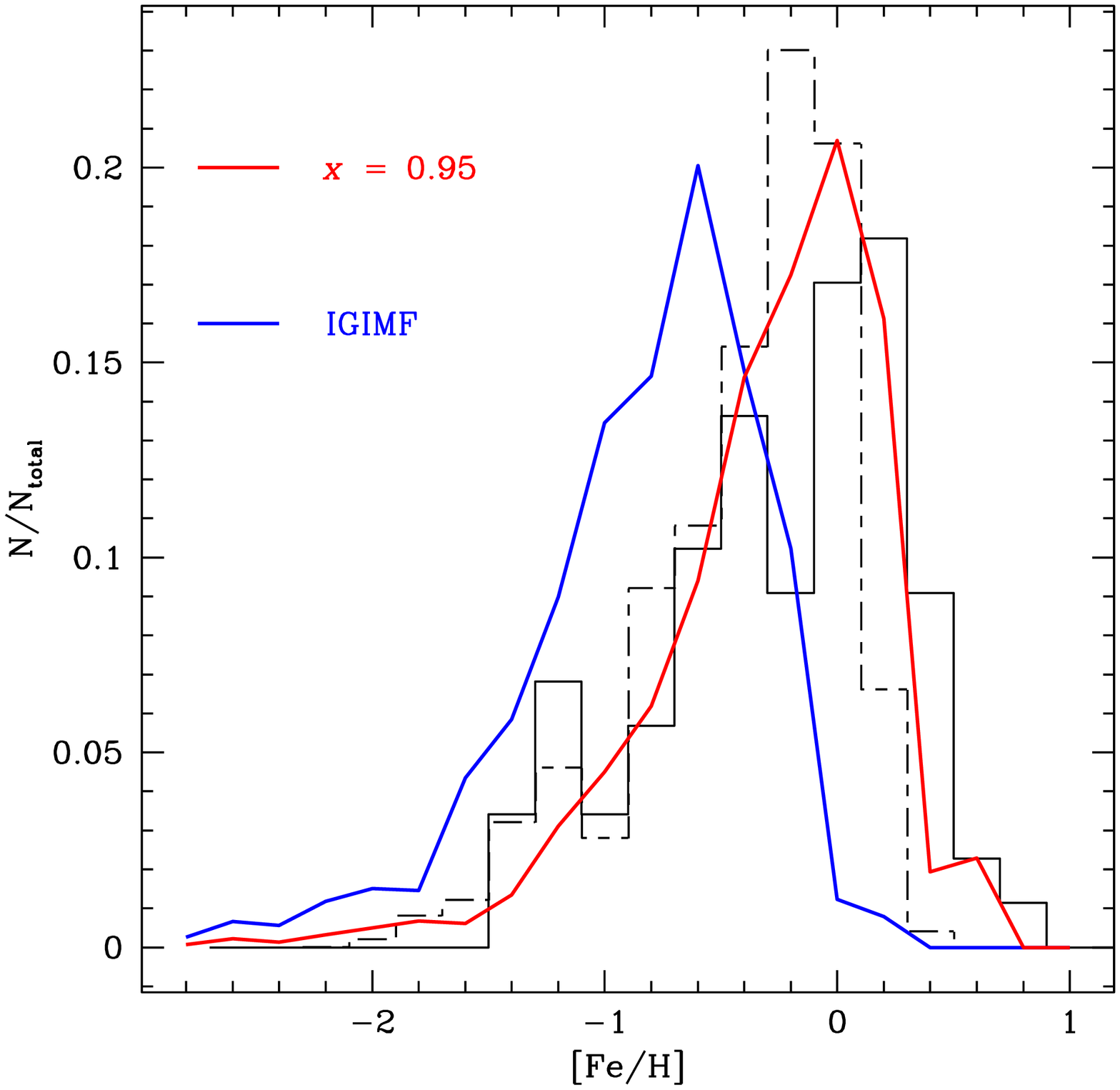}
\hfill
\caption{The predicted and observed metallicity distribution in the Galactic bulge.
The data are from Zoccali et al. (2003)(dashed histogram) and Fulbright et al. 
(2006) (continuous histogram). In particular, the model with the peak at the lower metallicity is computed with an IMF which is similar to that of the solar vicinity and indicated by IGIMF, whereas the distribution which best fits the data is computed with a flat IMF (x=0.95 for $M> 1 M_{\odot}$).}
\end{figure}

\begin{figure}
\includegraphics[width=12cm,height=8cm]{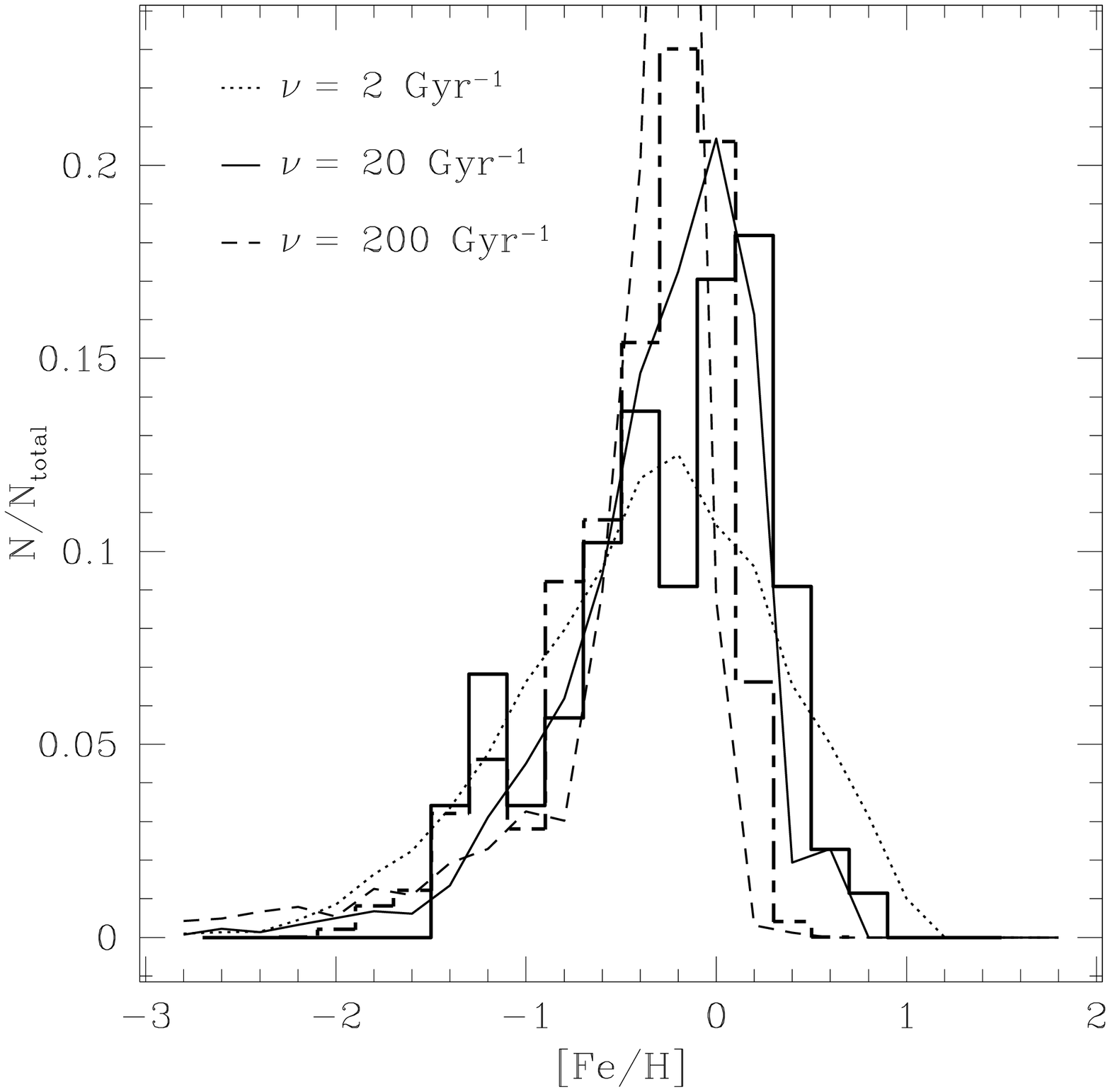}
\hfill
\caption{The predicted and observed metallicity distribution in the Galactic 
bulge.
The data are from Zoccali et al (2003)(dashed histogram) and Fulbright et al. (2006) (continuous histogram). The lines are the predictions of models with the same IMF but different SF efficiencies, as indicated in the Figure. }
\end{figure}

In summary, the comparison between the models (Ballero et al. 2007) on one side, and the 
metallicity distribution and the [$\alpha$/Fe] ratios on the other, strongly indicates that the Galactic bulge is very old and must  have formed very quickly during a strong starburst (with a SF efficiency much higher than in the disk). 
The metallicity distribution in particular, seems to suggest an IMF flatter than in the disk 
with an exponent for massive stars in the range $x=1.35-0.95$. However, to assess more precisely 
this point we need more data: in particular, a flatter IMF predicts that the overabundances of $\alpha$-elements relative to Fe and to the Sun should be higher in the bulge than in the disk. This is not entirely clear from the available data, although Zoccali et al. (2006) conclude that the [O/Fe] ratios in bulge stars are higher than in thick and thin disk stars (see Figure 38). The timescale for the bulge formation by accretion of gas lost from the halo is 0.1 Gyr and certainly no longer than 0.5 Gyr.

\begin{figure}
\includegraphics[width=12cm,height=10cm]{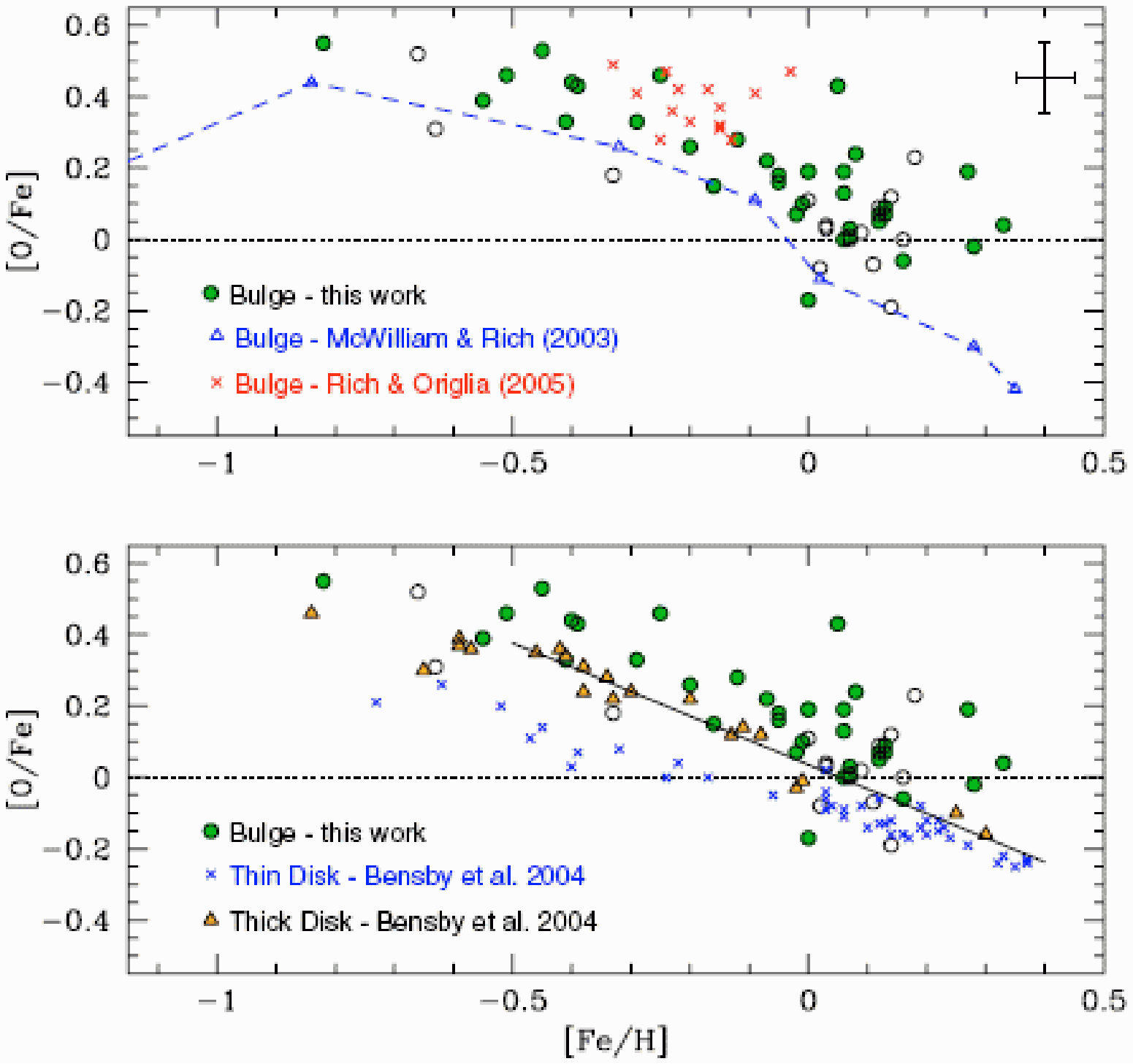}
\hfill
\caption{Upper panel: the [O/Fe] vs. [Fe/H] for bulge stars as measured by Zoccali et al. (2006) (circles), along with previous determinations in other bulge stars from optical (open triangles) and near-IR spectra (crosses). Open symbols refer to spectra with lower S/N or with the O line partially blended with telluric absorption. Lower panel: [O/Fe] trend in the bulge and in thick and thin disk stars (crosses and triangles). The solid line shows a linear fit to the thick data points with [Fe/H]$>$ -0.5 dex and is meant to emphasize that all bulge stars with$-0.4 < [Fe/H] < +0.1$ are more O-enhanced than the thick disk stars. If confirmed, this trend may indicate a flatter IMF for the bulge than for the disk. This plot also suggests that there is a systematic difference between bulge and disk stars, thus excluding that bulge stars were once disk stars migrated in the bulge. Figure and references from Zoccali et al. (2006).}
\end{figure}

\section{What we have learned about the Milky Way}

From the discussions of the previous sections we can extract some important 
conclusions on the formation and evolution of the Milky Way, derived from 
chemical abundances.
In particular:
\begin{itemize}
\item 
The inner halo formed on a timescale of 1-2 Gyr at maximum, the outer halo 
formed on longer timescales perhaps from accretion of satellites or gas. 

\item The disk at the solar ring formed on a timescale not shorter than 7 Gyr.

\item The whole disk formed inside out with timescales of the order of 2 Gyr 
or less in the inner regions and 10 Gyr or more in the outermost regions.

\item The abundance gradients arise naturally from the assumption of the 
inside-out formation of  the disk. A threshold density for the star formation helps in steepening the gradients in the outer disk regions.

\item The bulge is very old and formed very quickly on a timescale smaller 
than even the inner halo and  not larger than 0.5 Gyr.

\item The IMF seems to be different in the bulge and the disk, being flatter 
in the bulge, although more abundance data are necessary before drawing firm 
conclusions.

\end{itemize}

\section{The time-delay model and the Hubble sequence}

In this section we will discuss how different star formation histories affect the evolution of galaxies of different morphological type and in particular how the abundance patterns are expected to change with the star formation.

\subsection{Star formation and Hubble sequence}

Sandage (1986), on the basis of a work of Gallagher, Hunter \& Tutukov (1984) who measured SFR in galaxies, suggested a possible interpretation of the Hubble sequence in terms of different star formation histories. In this picture, ellipticals and bulges must have suffered an intense and strongly declining SFR, whereas late type galaxies must have undergone through a less intense, almost constant  (spirals) and even increasing with time (irregulars) SFR. In Figure 39 is illustrated such a behaviour of the SFR for E, S0 and Sa galaxies: there are two important timescales $t_c$, the gas collapse time, and $t_s$, the star formation time, namely the timescale on which the gas in a galaxy is consumed 
by means of star formation (the inverse of the SF efficiency).
The interplay between these two quantities can be crucial for the formation of the different galactic morphological types. In fact, if $t_s << t_c$, most of the stars form before the collapse is over and the gas does not have time to dissipate energy and settle into a disk. In this case, the resulting galaxy will be a spheroid, whereas if $t_s > t_c$ the gas has time to dissipate energy and form a spiral galaxy. This picture is certainly too simplistic to fully describe the reality of galaxy formation but it seems to work well when we are going to interpret galaxy formation by studying the stellar populations in galaxies.

\begin{figure}
\includegraphics[width=12cm,height=10cm]{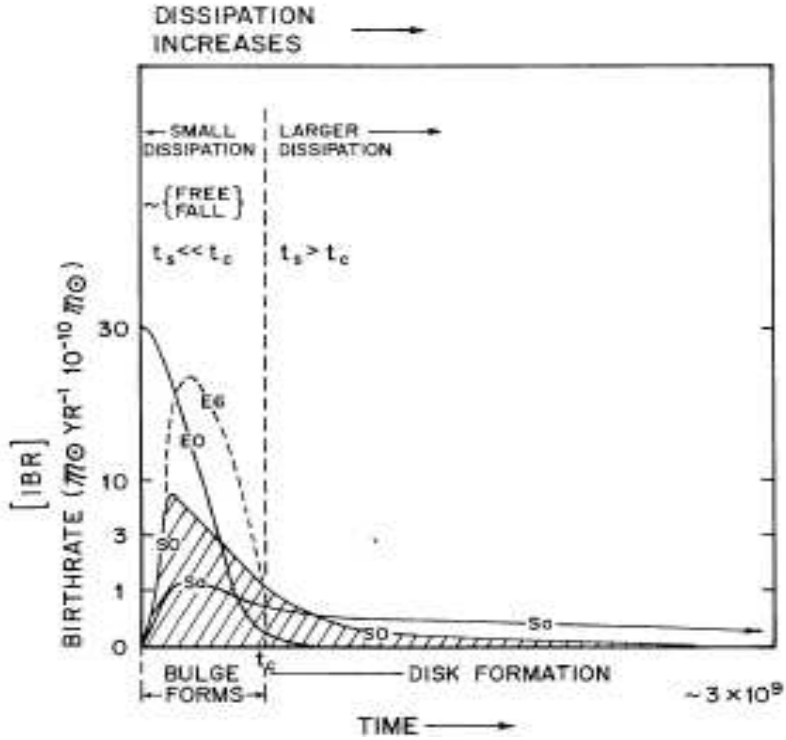}
\hfill
\caption{Schematic representation of the change of the SFR with time for galaxies of types E, S0, and Sa. The dashed vertical line at the collapse time $t_c$ separates regions of low energy dissipation (to the left) from those of high energy dissipation (to the right). Bulges form in the left region, disks in the right. The integral under the curves gives the total number of stars formed (per unit galaxy mass). The integral under the S0 curve is shaded. Figure from Sandage (1986).}
\end{figure}

Later on, Kennicutt (1998a,b) measured again the SFR in star forming galaxies and suggested similar behaviours for the different galactic morphological types, as shown in Figure 40. In this Figure, the behaviour of the SFR in
 spirals is obtained by  fitting the mean value of the parameter $b$, measured by Kennicutt et al. (1994). The parameter $b$ is the ratio between the present time SFR and the average SFR in the past, as defined in eq.(16).
Early type spiral galaxies are characterized by rapidly declining SFRs, with $b \sim 0.01-0.1$, 
whereas  late type spirals have formed stars since a long time at an almost constant rate 
with $b=1$. Finally, ellipticals and S0 have long ago ceased forming stars and have  $b=0$.

\begin{figure}
\includegraphics[width=13cm,height=10cm]{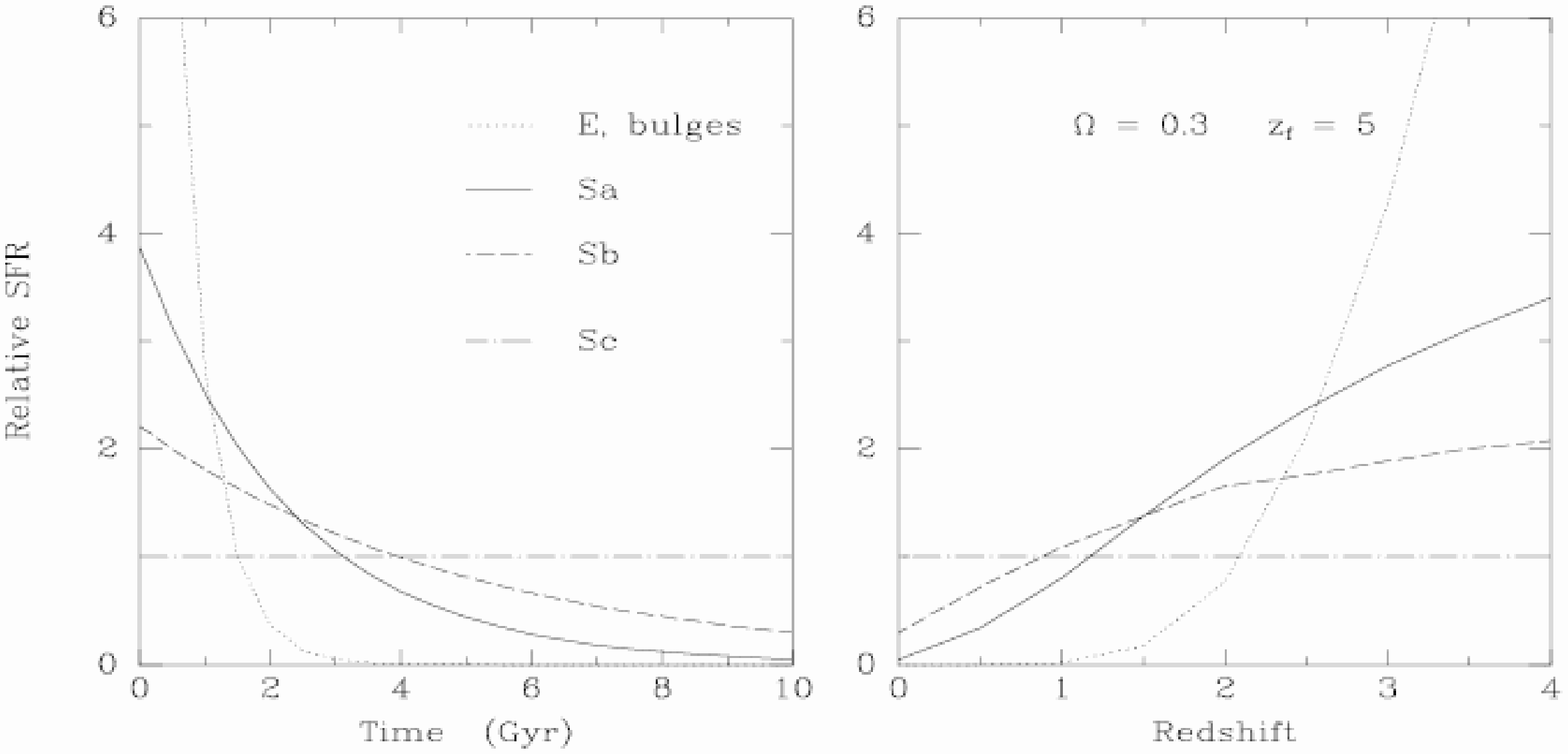}
\hfill
\caption{Schematic illustration of the stellar birthrate for different Hubble types. The left panel shows the evolution of the relative SFR with time following Sandage (1986). The curves for spiral galaxies are exponentially declining SFRs which fit the mean values of the birthrate parameter $b$ as measured by Kennicutt et al. (1994). The curve for elliptical galaxies and bulges is an arbitrary dependence for an e-folding time of 0.5 Gyr, for comparative purposes only. The right panel shows the corresponding evolution in SFR with redshift, for an assumed cosmological density parameter $\Omega=0.3$ and formation redshift 
$z_f=5$. Figure from Kennicutt (1998b).}
\end{figure}

In Figure 41 we show the SF histories which give rise to the [$\alpha$/Fe] vs. [Fe/H] relations of 
Fig. 33.

\begin{figure}
\includegraphics[width=12cm,height=10cm]{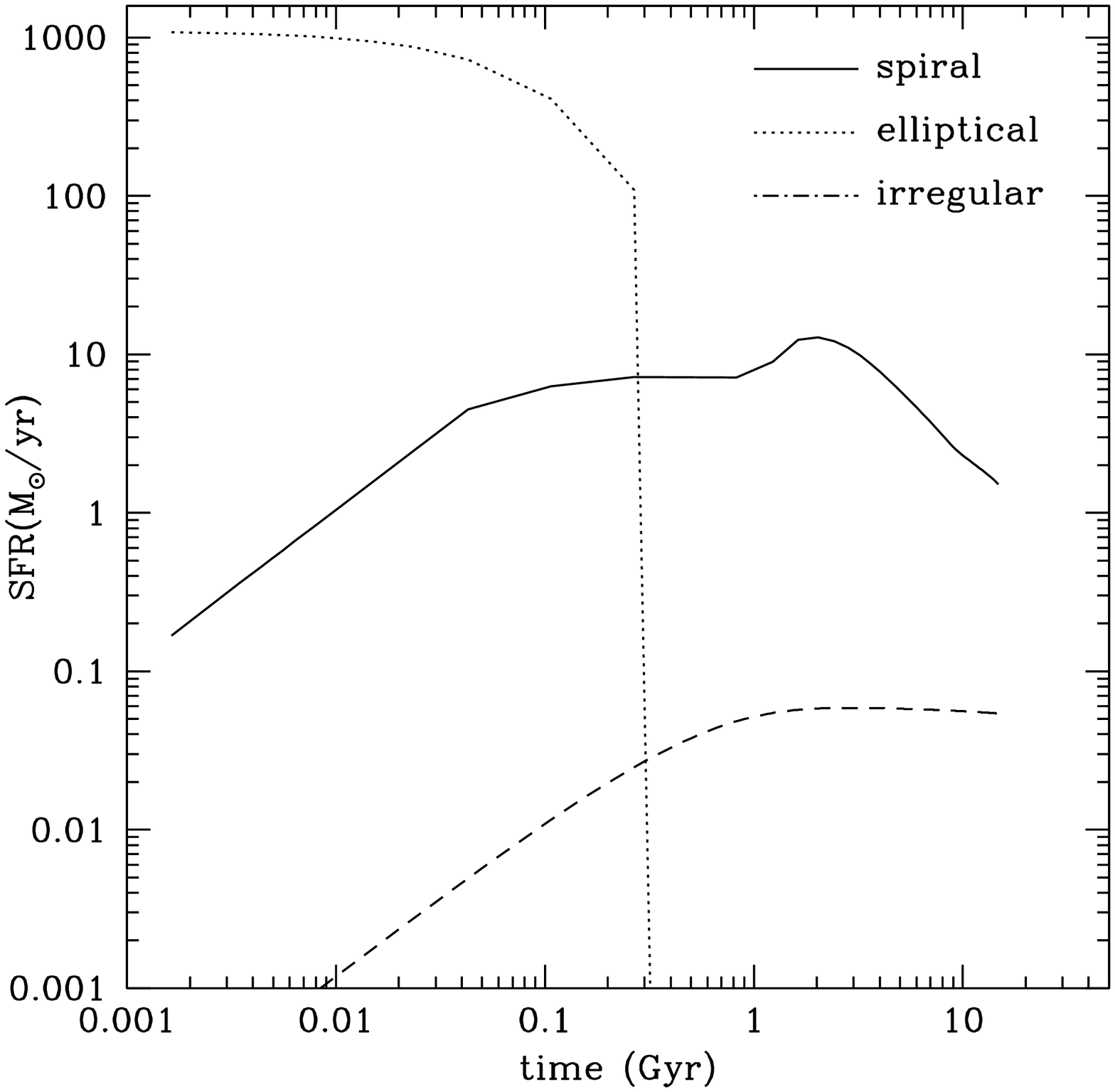}
\hfill
\caption{The predicted histories of SF in galaxies of different 
morphological type, with decreasing efficiency of SFR from  ellipticals to irregulars. 
Figure from Calura (2004).}
\end{figure}

In Figure 42 we show the predicted Type Ia SN rates according to the SFRs of Figure 41. The assumed progenitor model for Type Ia SNe is the single degenerate with the delay time distribution as in Matteucci \& Recchi (2001).

\subsubsection{The typical timescale for SN Ia enrichment}

The predicted Type Ia SN rates for galaxies with different morphologies show a difference in the maximum SN Ia rate which is reached quite early in ellipticals and it occurs later and later moving to late types. Matteucci \& Recchi (2001) suggested to assume the time for the occurrence of the maximum SNIa rate as the typical timescale for the chemical enrichment from these SNe. It depends on the star formation history of each galaxy, on the IMF and on the stellar lifetimes. As we have already shown, the IMF together with stellar lifetimes represent the distribution of the time-delays (DTD) with which SNe explode; therefore, when a DTD is specified, the Type Ia SN rate depends on the SF history only.
In summary, in ellipticals and bulges this timescale is 0.3-0.5 Gyr since the beginning of star formation, in the solar vicinity there is a first peak at 1 Gyr then it decreases slightly and increases again till 3 Gyr (due to the two-infall episodes). In irregular galaxies the maximum is reached at $\sim 4$ Gyr and then the rate remains constant.

\begin{figure}
\includegraphics[width=14cm,height=14cm]{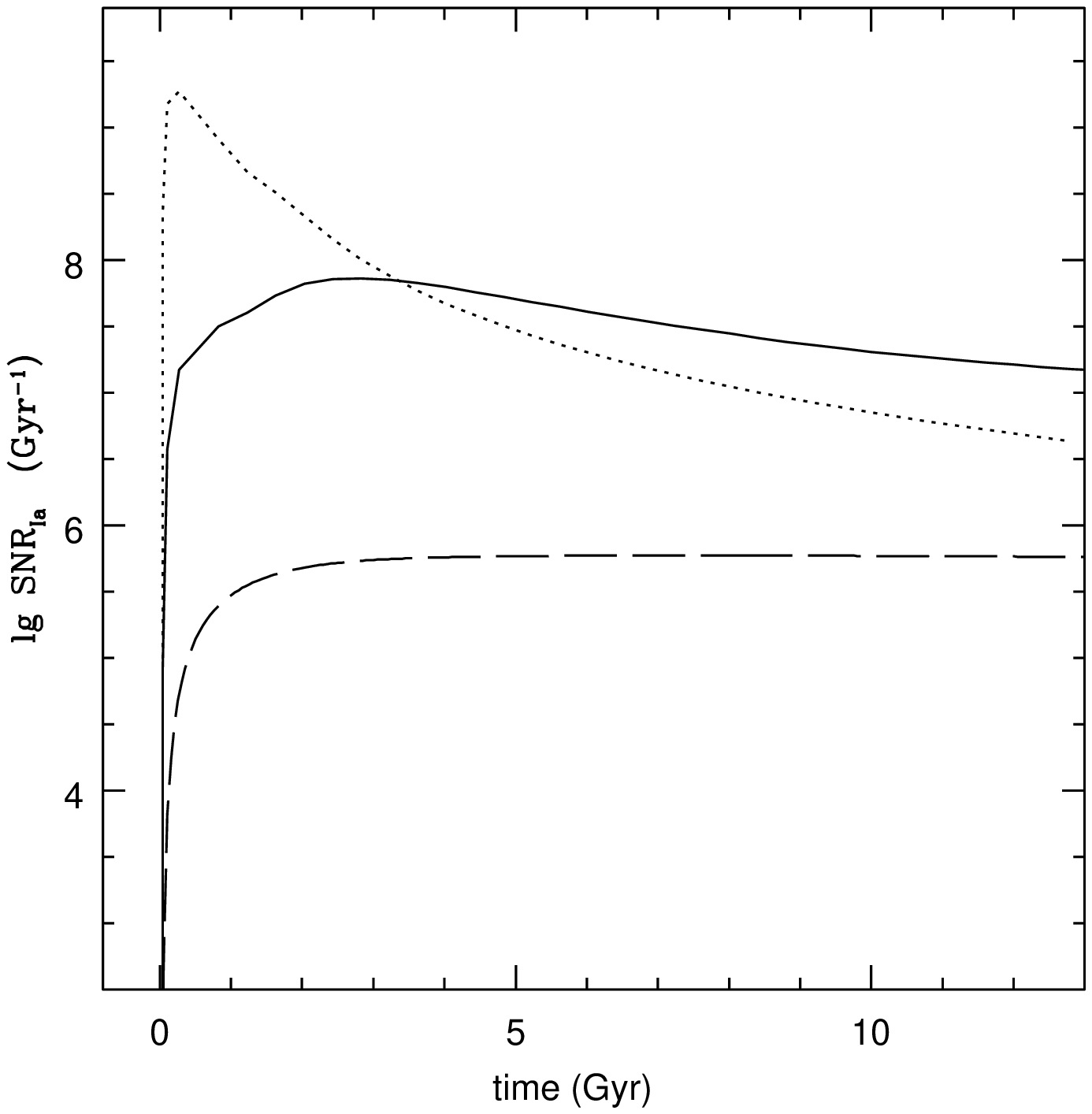}
\hfill
\caption{Predicted Type Ia SN rates (expressed in $Gyr^{-1}$ obtained according to the SF
histories of Figure 41. As one can see the elliptical (dotted line) reaches a maximum at 0.3 Gyr, 
whereas the spiral (continuous line) at $\sim$ 2 Gyr and the irregular (dashed line) has a rate 
increasing to its maximum at around 5 Gyr and then is roughly constant for the rest of the 
galactic lifetime.The minimum time delay for the Type Ia SNe to appear is 30 Myr, hardly visible in this plot.}
\end{figure}

\section{Dwarf Spheroidals of the Local Group}
A different pattern for the [$\alpha$/Fe] vs. [Fe/H] relation compared to the solar vicinity is observed in dwarf spheroidal galaxies (dSphs) of the Local Group, as shown in Figure 43, and this can be easily interpreted in the framework of the time-delay model coupled with different star formation histories.

\begin{figure}
\includegraphics[width=12cm,height=10cm]{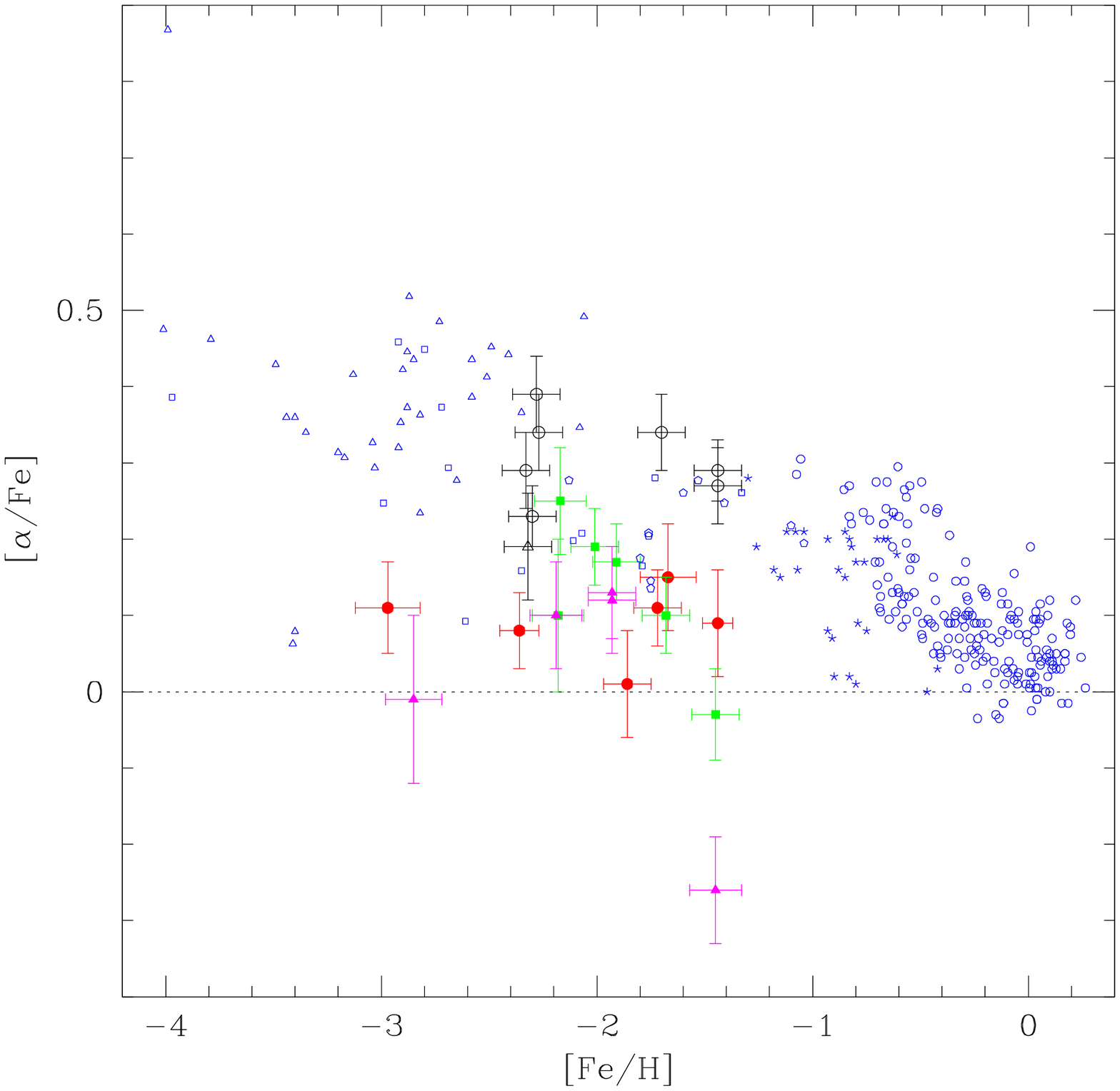}
\hfill
\caption{Observed [$\alpha$/Fe] vs. [Fe/H] in the Milky Way (small points) and 
in dSphs
(points with error bars). Figure from Shetrone et al. (2001).}
\end{figure}

Before interpreting the [$\alpha$/Fe] diagram, we recall the current ideas about the formation of the dSphs.

\subsection{How do dSphs form?}

Cold dark matter (CDM) models for galaxy formation predict that the dSphs, systems with luminous masses of the order of $10^{7}M_{\odot}$, are the first objects to form stars and that all stars in these systems should form on a timescale $<1$ Gyr, since the heating and gas loss, due to reionization, must have halted the SF soon.
However, observationally all dSph satellites of the Milky Way contain old stars indistinguishable from those of Galactic globular clusters and they seem to have experienced SF for long periods ($> 2$ Gyr, Grebel \& Gallagher 2004).
The histories of SF for these galaxies are generally derived from the color-magnitude diagram (e. g. Mateo  1998).
By looking at the [$\alpha$/Fe] vs. [Fe/H] relations for dSphs, as shown in Figure 43,  one can immediately suggest, on the basis of the time-delay model, that their evolution should have been characterized by a slow and protracted SF, at variance with the suggestion of a fast episode of SF truncated by the heating due to reionization.

\subsubsection{Dark matter in dSphs}
The dSph satellites of the Milky Way are considered the smallest dark matter dominated systems in the universe.
In the past years there have been a few attempts at deriving the amount of dark matter in dSphs, in particular by measuring the mass to light ratios versus magnitude for these galaxies (e.g. Mateo 1998; Gilmore et al. 2007).
Gilmore et al. (2007) suggested that the dSphs have a shallow central dark matter distribution and no galaxy is found with a dark mass halo less massive than $5 \cdot 10^{7}M_{\odot}$, as shown in Figure 44.

\begin{figure}
\includegraphics[width=12cm,height=10cm]{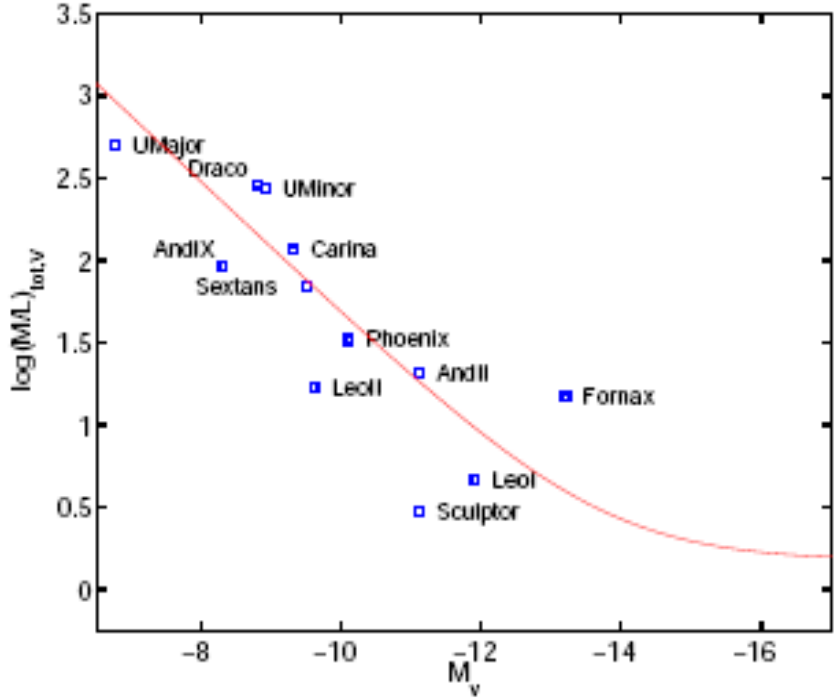}
\hfill
\caption{Dark matter in dSphs: mass to light ratios versus absolute 
V magnitude for some Local Group dSphs. The solid curve shows the relation expected if all the dSphs contain about $4 \cdot 10^{7}M_{\odot}$ of dark matter interior 
to their stellar distributions. Figure from Gilmore et al. (2007).}
\end{figure}

\subsubsection{Observations of dSphs}
In recent years there  has been a fast development in the field of chemical evolution of dSphs of the Local Group due to the increasing amount of data on chemical abundances derived from high resolution spectra (e.g. Smecker-Hane \& McWilliam, 1999; Bonifacio el al. 2000, 2004; Shetrone et al. 2001, 2003; Tolstoy et al.2003; Bonifacio et al. 2004; Venn et al.2004; Sadakane et al. 2004; Fulbright et al. 2004; McWilliam \& Smecker-Hane 2005a,b; Monaco et al. 2005; Geisler et al. 2005). The abundances of $\alpha$-elements (O, Mg, Ca, Si,) plus the abundances of s- and r- process elements (Ba, Y, Sr, La and Eu) were measured with unprecedented accuracy. Besides these high-resolution studies we recall also the measure of the metallicities of many red giant stars in several dSphs by Tolstoy et al. (2004), Koch et al. (2006), Helmi et al. (2006) and Battaglia et al. (2006) obtained from the low-resolution Ca triplet. An interesting result of Helmi et al. (2006) is that they did not find stars with [Fe/H]$<$ -3.0 dex and that the metallicity distribution of the stars in dSphs is different from that of halo stars in the Milky Way.
Other important information, as already mentioned, comes from the photometry of dSphs of the Local Group and in particular from the color-magnitude diagrams. From these diagrams one can infer the history of SF of these galaxies. We recall the studies of Hernandez et al. (2000), Dolphin (2002), Bellazzini et al. (2002), Rizzi et al. (2003), Monelli et al. (2003), Dolphin et al. (2005). The color-magnitude diagrams seem
to indicate that the majority of dSphs had one rather long episode of SF with the exception of Carina for which four episodes of SF have been suggested (Rizzi et al. 2003).

\subsubsection{Chemical evolution of dSphs}
Several papers have appeared in the last few
years concerning the chemical 
evolution of dSphs.
For example Carigi et al. (2002) computed models for the chemical evolution of 
four dSphs by adopting the SF histories derived, from color-magnitude diagrams, by 
Hernandez et al. (2000). In their model they assumed gas infall
and computed the gas thermal energy heated by SNe in order to study galactic 
winds. In fact, the dSphs must have lost their gas in one way or another 
(galactic winds and/or ram pressure stripping) since they appear completely 
without gas. They assumed that the wind is sudden and devoids the galaxy of 
gas instantaneously.
The adopted IMF is the Kroupa et al. (1993) IMF as in the solar vicinity.
They predicted a too high metallicity for Ursa Minor and did not match the 
correct slope for the observed [$\alpha$/Fe] ratios, unless the history of SF in this galaxy was assumed to be different than suggested by its color-magnitude diagram, as shown in Figure 45.

\begin{figure}
\includegraphics[width=12cm,height=10cm]{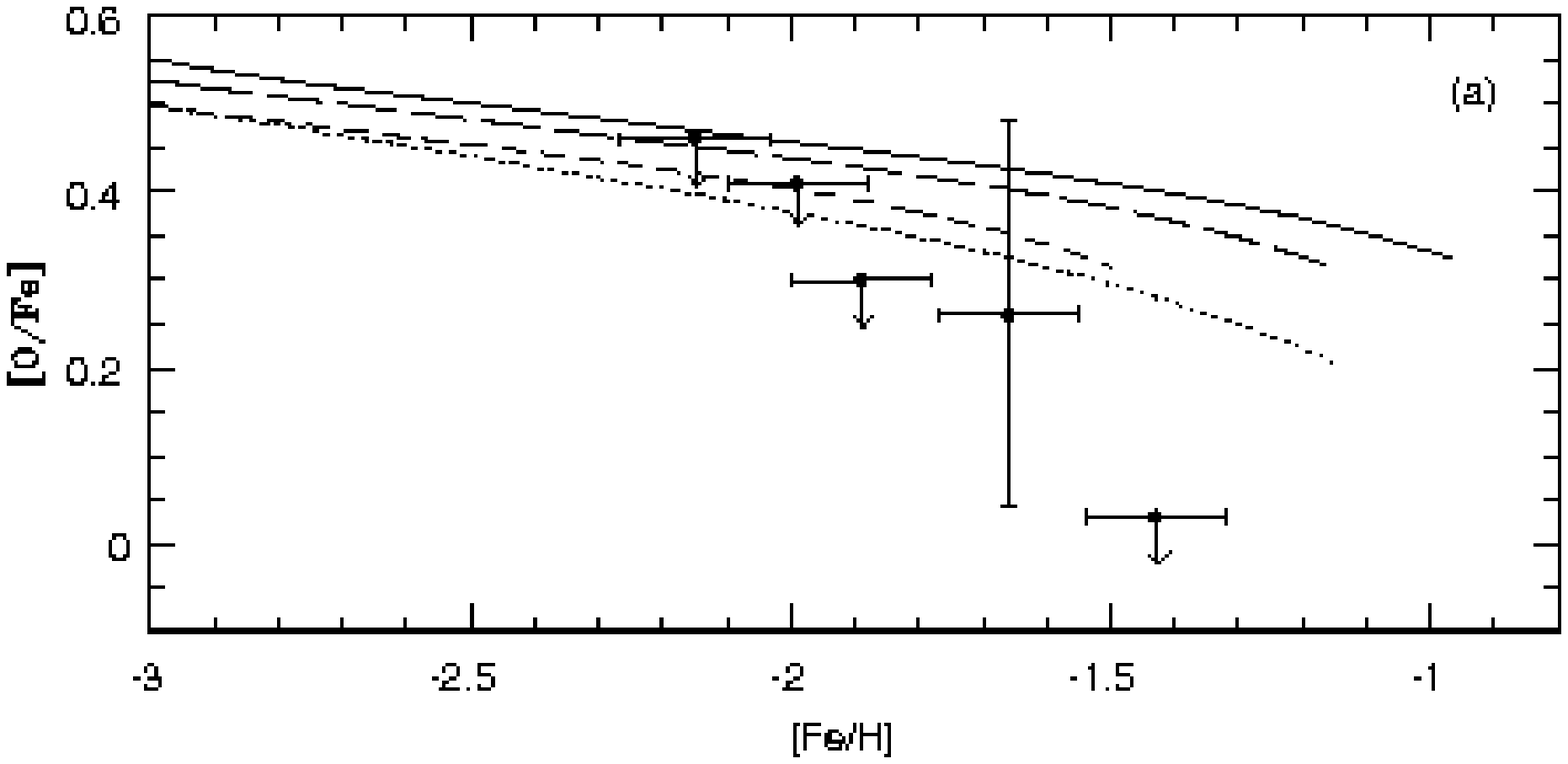}
\includegraphics[width=12cm,height=10cm]{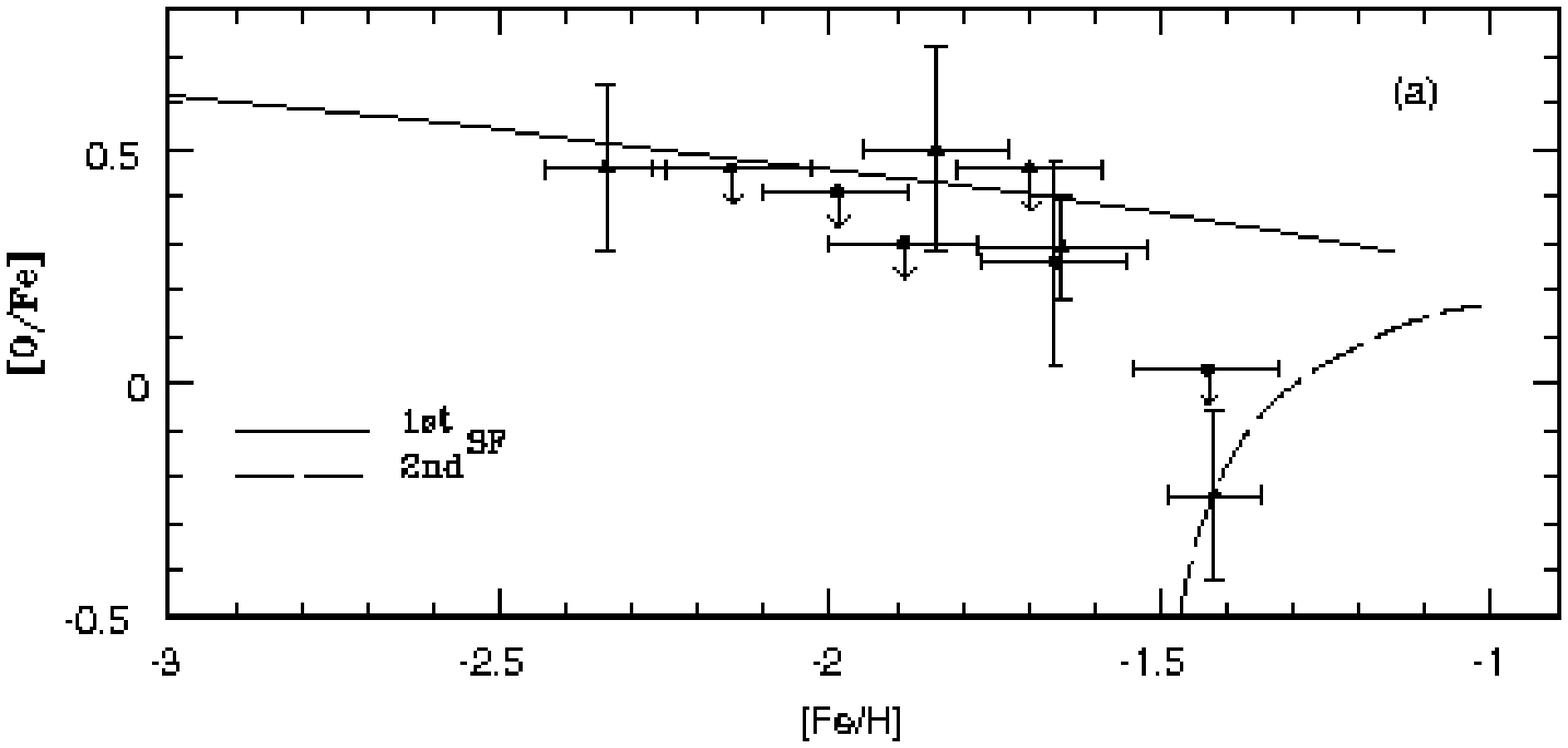}
\hfill
\caption{Observed and predicted [O/Fe] vs. [Fe/H] relation for the galaxy Ursa 
Minor. Models and Figures by Carigi et al. (2002). The upper Figure presents models assuming only one burst of SF at high redshift, as suggested by the color-magnitude diagram of this galaxy. The lower Figure  shows the predictions of a model with two separate bursts which seems to fit better the data.}
\end{figure}

Then, Ikuta \& Arimoto (2002) proposed 
a closed box model (no infall nor outflow) for dSphs. In this Simple Model they had to assume some external cause to stop star formation, such as ram pressure stripping. They tested different IMFs and suggested that these galaxies had suffered very low SFRs (1-5\% of that in the solar neighbourhood) and that the SF had a long duration ($> $ 3.9-6.5Gyr).
In Figure 46 are shown their predictions for [Mg/Fe] in dSphs. Also here, the predicted slope of the [Mg/Fe] ratio is flatter than observed.

\begin{figure}
\includegraphics[width=12cm,height=10cm]{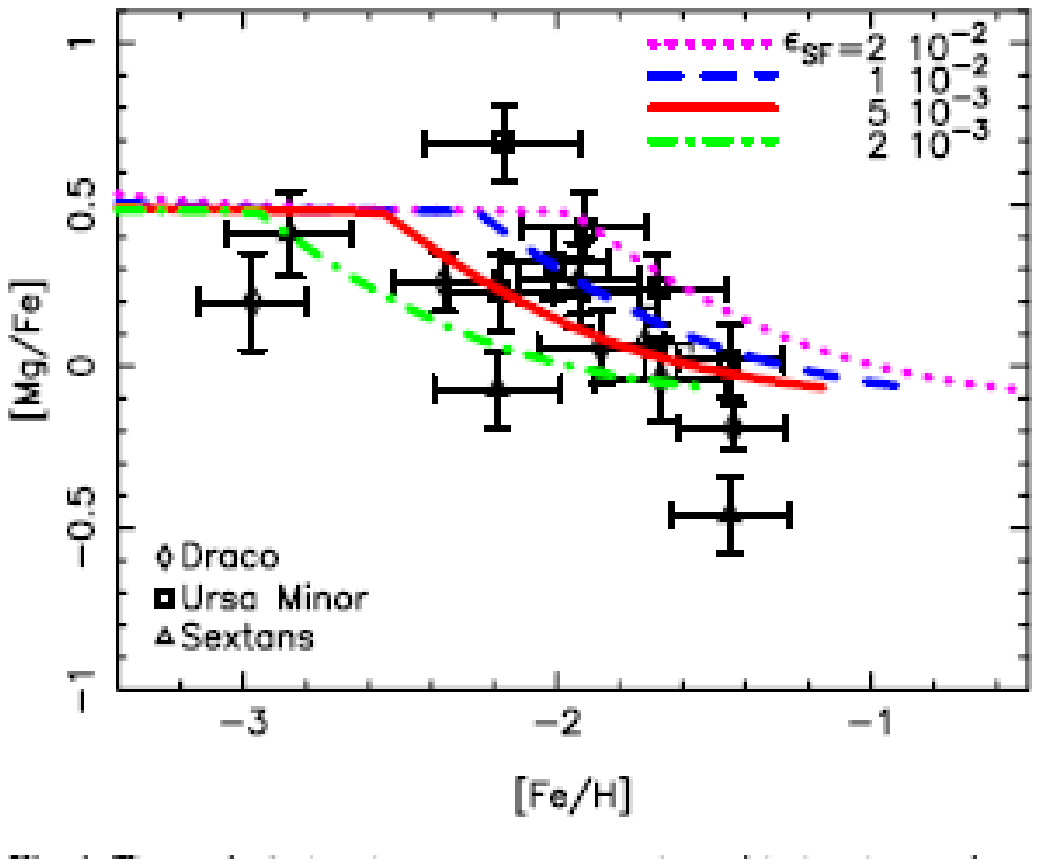}
\hfill
\caption{Observed and predicted [Mg/Fe] vs. [Fe/H] relation for the dSph 
Draco, Ursa Minor and Sextans. The different curves refer to different 
SF efficiencies ($\epsilon_{SF}$)
expressed in $Gyr^{-1}$, which are equivalent to the quantity $\nu$. 
Figure from Ikuta \& Arimoto (2002).}
\end{figure}
More recently, Fenner et al. (2006) suggested a model with a galactic wind for 
Sculptor: they indicated an efficiency of SF of $0.05  Gyr^{-1}$. 
They concluded, from the study of the [Ba/Y] ratio,
that chemical evolution in dSphs is inconsistent with the SF being truncated after reionization (at redshift z=8). In fact, the high value of this ratio measured in stars indicates strong s-process production from low mass stars which have very long lifetimes.

\subsubsection{The results of Lanfranchi \& Matteucci}

Lanfranchi \& Matteucci (2003; 2004 hereafter LM04) developed models for dSphs of the Local 
Group. First they tested a ``standard model'' devised for describing an 
average dSph galaxy. This model was based on the following assumptions:

\begin{itemize}
\item one long star formation episode of duration $\sim$ 8 Gyr,

\item a small star formation efficiency, namely the star formation rate per 
unit mass is 1-10\% of that in the solar vicinity,

\item a strong galactic wind develops when the thermal energy of the gas 
equates the binding energy of the gas. The rate of gas loss is assumed to be 
several times the SFR, as in eq. (27),
with typical values of $\lambda$ (wind parameter) between 5 and 15.

The condition for the onset of the wind is written as:
\begin{equation}
(E_{th})_{ISM} \ge E_{Bgas}
\end{equation}
namely, that the thermal energy of the gas is larger or equal to its 
binding energy.
The thermal energy of gas due to SN and stellar wind heating is:
\begin{equation}
(E_{th})_{ISM}=E_{th_{SN}}+ E_{th_{w}}
\end{equation}

with the contribution of SNe being:
\begin{equation}
E_{th_{SN}}= \int^{t}_{0}{\epsilon_{SN}R_{SN}(t^{`})dt^{`}},
\end{equation}

while the contribution of stellar winds is:
\begin{equation}
E_{th_{w}}=\int^{t}_{0}\int^{100}_{12}{ \varphi(m) \psi(t^{`}) \epsilon_{w}dm
dt^{`}}
\end{equation}

with
$\epsilon_{SN}= \eta_{SN}\epsilon_{o}$ and $\epsilon_o=10^{51}$erg 
(typical SN energy),
and
$\epsilon_{w}= \eta_{w}E_{w}$ with $E_{w}= 10^{49}$erg 
(typical energy injected by the stellar wind of a $20M_{\odot}$ star taken as representative).
$\eta_{w}$ and $\eta_{SN}$ are two free parameters and indicate the 
efficiency of
energy transfer from stellar winds and SNe into the ISM, respectively,  
quantities still largely unknown.
It is assumed that $\epsilon_w=0.03$ for the stellar winds, and that $\epsilon_{SN}=0.03$ for Type II SNe and $\epsilon_{SN}=1.0$ for Type Ia SNe, as suggested by Recchi et al. (2001).
The total mass of the galaxy is expressed as 
$M_{tot}(t)=M_{*}(t)+M_{gas}(t)+M_{dark}(t)$
with $M_L(t)=M_{*}(t)+M_{gas}(t)$
and the binding energy of gas is:

\begin{equation}
E_{Bgas}(t)=W_L(t)+W_{LD}(t)
\end{equation}

with:
\begin{equation}
W_L(t)=-0.5G{ M_{gas}(t) M_L(t) \over r_L}
\end{equation}
which is the potential well due to the luminous matter and
with:
\begin{equation}
W_{LD}(t)= -Gw_{LD}{M_{gas}(t) M_{dark} \over r_L}
\end{equation}
which represents the potential well due to the interaction between dark and 
luminous matter,
where $w_{LD} \sim {1 \over 2\pi} S(1+1.37S)$,
with $S= r_L/r_{D}$, being the ratio between the galaxy effective radius and 
the radius of the dark matter core. 
The typical model for a dSph starts with an initial baryonic mass of 
$10^{8} M_{\odot}$ and ends up, after the wind, with a luminous mass of 
$\sim 10^{7}M_{\odot}$. 
The dark matter halo is assumed to be ten times larger than the luminous mass 
but diffuse ($S=0.1$). The galactic wind in these galaxies develops 
after several Gyr from the start of SF, according to the different assumed 
SF efficiency. Soon after, but not immediately, the wind has started, the SF 
decreases very strongly until it halts completely.

\item The IMF is that of Salpeter (1955) for all galaxies.

\item Each galaxy is supposed to have formed by infall of gas clouds of 
primordial chemical composition, on a timescale  not longer than 0.5 Gyr.

\end{itemize}

In Figure 47 we show the [$\alpha$/Fe] ratios for different 
$\alpha$-elements and for different efficiencies of SF, as predicted by the standard model of Lanfranchi \& Matteucci (2003).

 \begin{figure}
\includegraphics[width=12cm,height=10cm]{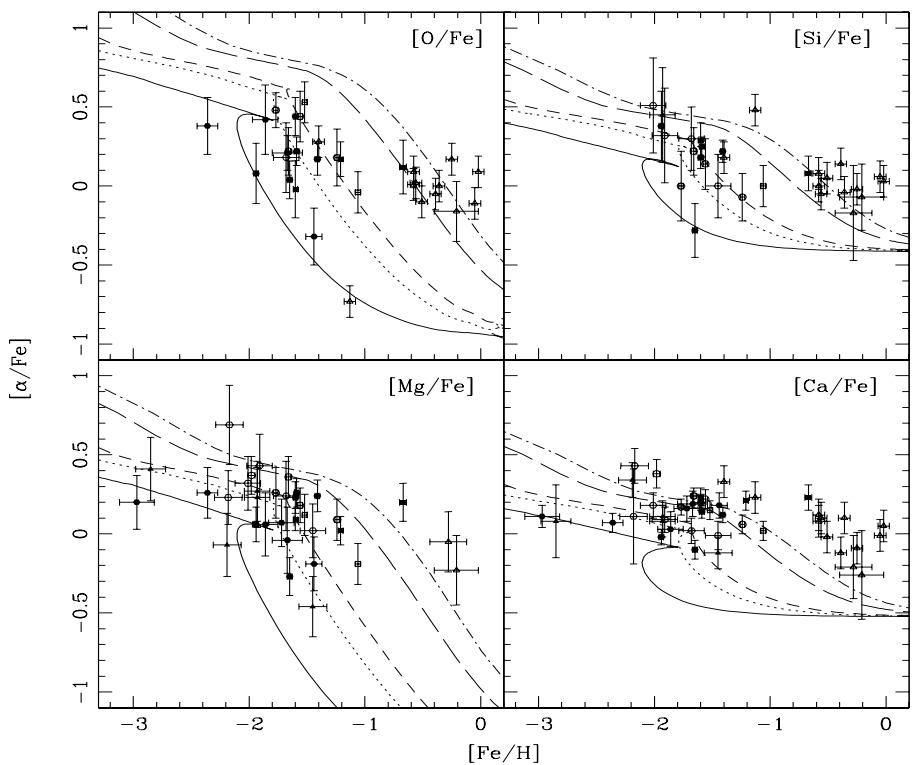}
\hfill
\caption{Observed and predicted [$\alpha$/Fe] versus [Fe/H]. The different lines refer to the ``standard model'' with different SF efficiency $\nu$ going from 1 (dashed-dotted lines) to 0.01 $Gyr^{-1}$ (continuous lines). The points represent stars in different dSphs: Sagittarius (open triangles), Draco (filled hexagons), Carina (filled circles), Ursa Minor (open hexagons), Sculptor (open circles), Sextans (filled triangles), Leo I (open squares) and Fornax (filled squares). 
Figure and  references from  Lanfranchi \& Matteucci (2003).}
\end{figure}

As one can see, the [$\alpha$/Fe] ratios show a clear change in slope followed by a steep decline, in agreement with the data. The change in slope corresponds to the occurrence of the galactic wind which starts emptying the galaxy of gas. In such a situation the SF starts to decrease as does therefore the production of the $\alpha$-elements from massive stars, whereas Fe continues to be produced since its progenitors have long lifetimes. This produces the steep slope: the low SF efficiency and the wind, which decreases furtherly the SF. In this situation, the time-delay model predicts an earlier and steeper decline of the [$\alpha$/Fe] ratios, as we have already discussed.

In LM04, the histories of star formation of 
specific galaxie, as suggested by their color-magnitude diagrams, 
were taken into account, and they developed models for 
six dSphs: Carina, Ursa Minor, Sculptor, Draco, Sextans and Sagittarius.
In Table 1 we show the assumed SF formation histories and model 
parameters. In particular, in column 1 are the galaxy names, in column 2 are 
the initial baryonic masses, in column 3 are the SF efficiencies, in column 4 are the wind parameters, in column 5 are the numbers of SF episodes, in column 6 are the times at which the SF episodes start, 
in column 7 are the durations, in Gyr, of the SF episodes and in column 8 the 
assumed IMF.

%

\begin{table*}
\begin{center}\scriptsize
\caption[]{Models for dSph galaxies. $M_{tot}^{initial}$
is the baryonic initial mass of the galaxy, $\nu$ is the star-formation
efficiency, $\lambda$ is the wind efficiency, and $n$, $t$ and $d$
are the number, time of occurrence and duration of the SF
episodes, respectively.}
\begin{tabular}{lccccccc}
\hline\hline\noalign{\smallskip}
galaxy &$M_{tot}^{initial} (M_{\odot})$ &$\nu(Gyr^{-1})$ &$\lambda$
&n &t($Gyr$) &d($Gyr$) &$IMF$\\
\noalign{\smallskip}
\hline
Sextan  &$5*10^{8}$ &0.01-0.3 &9-13 &1 &0 &8 &Salpeter\\
Sculptor &$5*10^{8}$ &0.05-0.5 &11-15 &1 &0 &7 &Salpeter\\
Sagittarius &$5*10^{8}$ &1.0-5.0 &9-13 &1 &0 &13 &Salpeter\\
Draco  &$5*10^{8}$ &0.005-0.1 &6-10 &1 &6 &4 &Salpeter\\
Ursa Minor &$5*10^{8}$ &0.05-0.5 &8-12 &1 &0 &3 &Salpeter\\
Carina &$5*10^{8}$ &0.02-0.4 &7-11 &2 &6/10 &3/3 &Salpeter\\
\hline\hline
\end{tabular}
\end{center}
\end{table*} 

%

\begin{figure}
\includegraphics[width=12cm,height=10cm]{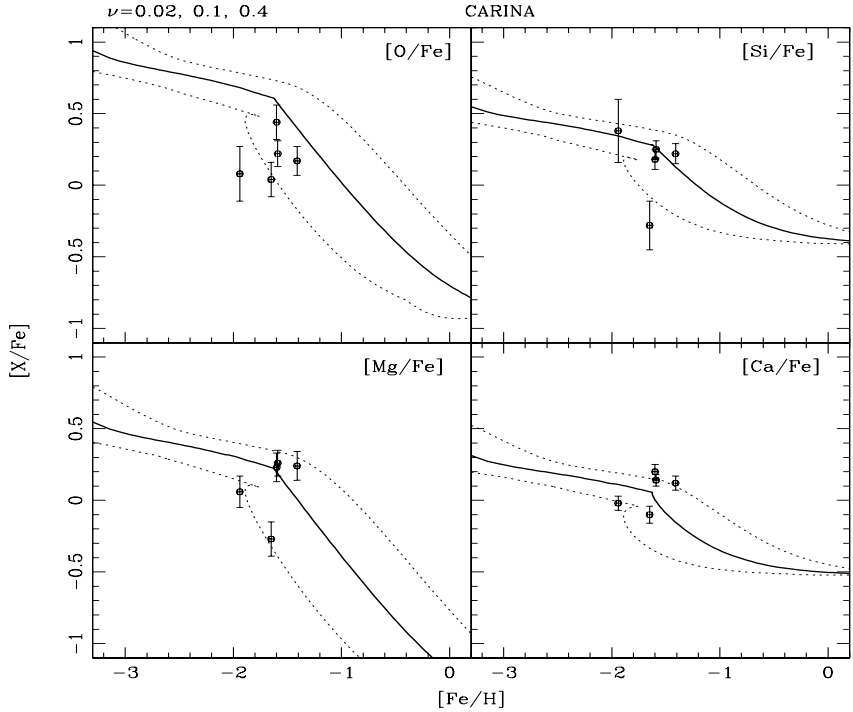}
\hfill
\caption{Observed and predicted [$\alpha$/Fe] vs. [Fe/H] relation for the galaxy Carina. The different lines represent models with different SF efficiency. The continuous line represents the best model and corresponds to the efficiency $\nu=0.1 Gyr^{-1}$. Figure from LM04.}
\end{figure}

\begin{figure}
\includegraphics[width=12cm,height=10cm]{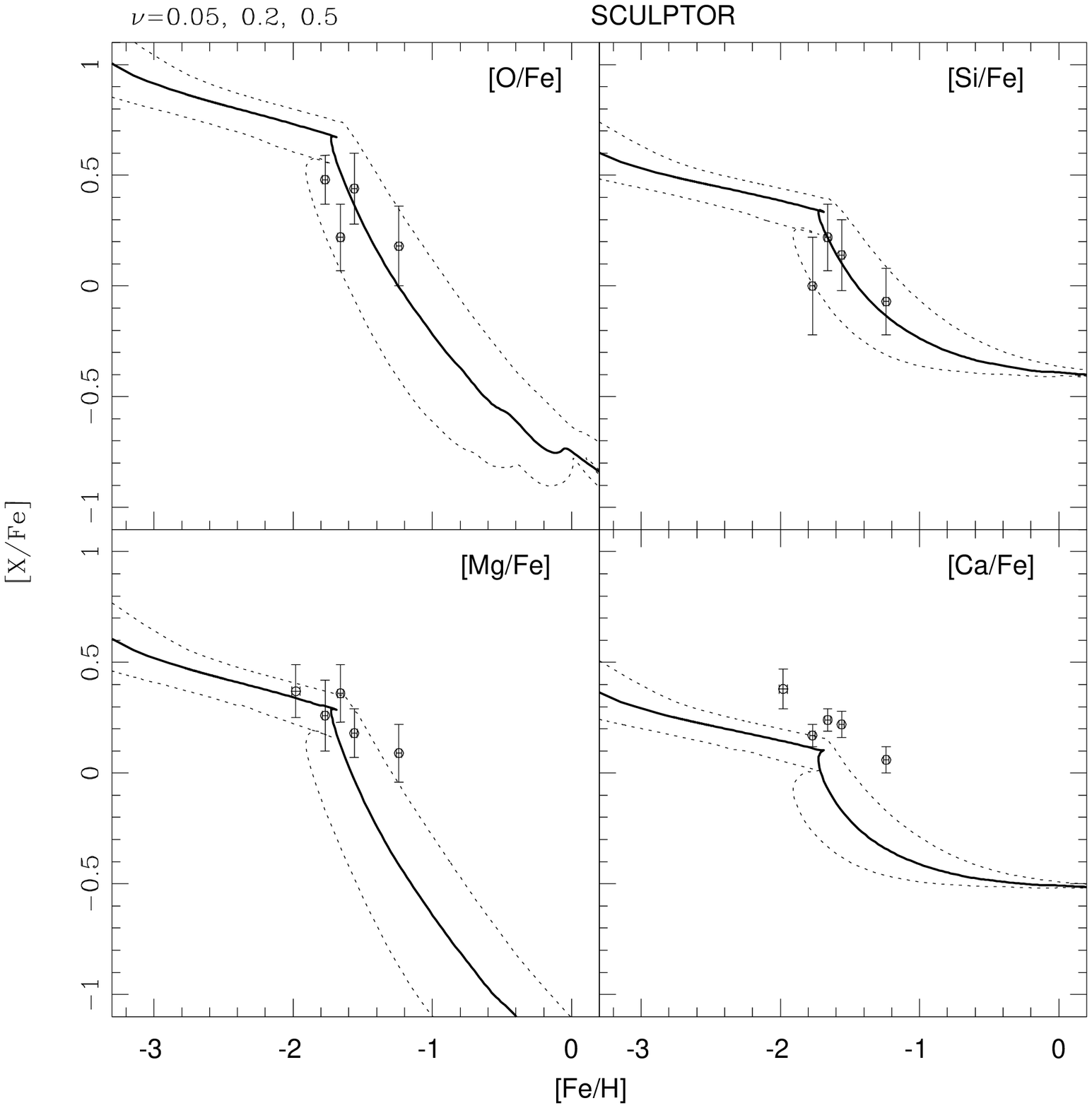}
\hfill
\caption{Observed and predicted [$\alpha$/Fe] vs. [Fe/H] relation for the galaxy Sculptor. The different lines represent models with different SF efficiencies 
($ \nu=0.05, 0.2, 0.5 Gyr^{-1}$). The continuous line represents the best model ($\nu= 0.2 Gyr^{-1}$). Figure from LM04.}
\end{figure}

\begin{figure}
\includegraphics[width=12cm,height=10cm]{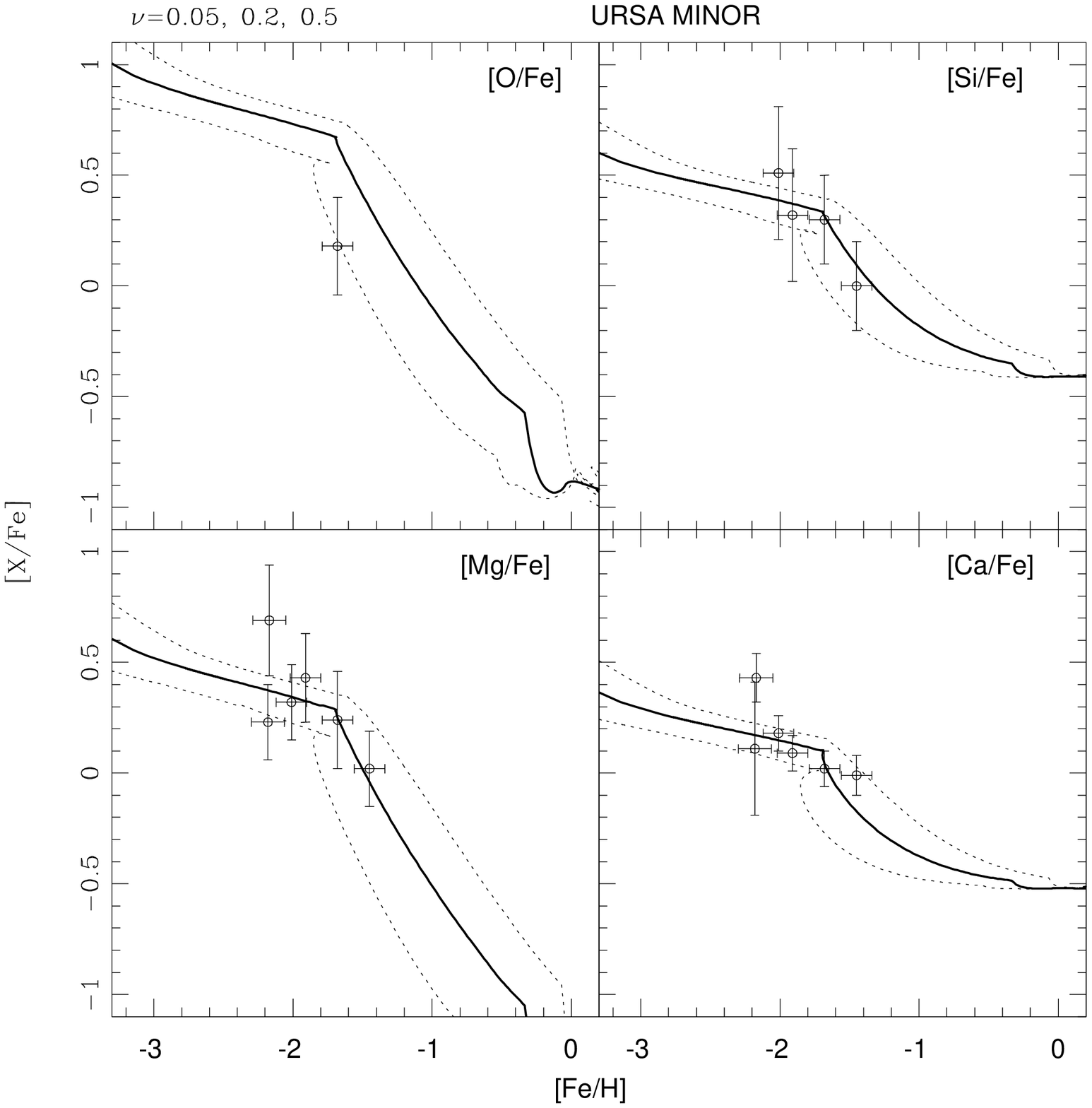}
\hfill
\caption{Observed and predicted [$\alpha$/Fe] vs. [Fe/H] relation for the galaxy Ursa Minor. The different lines represent models with different SF efficiencies. The continuous line represents the best model ($\nu= 0.2 Gyr^{-1}$). Figure from LM04.}
\end{figure}

\begin{figure}
\includegraphics[width=12cm,height=10cm]{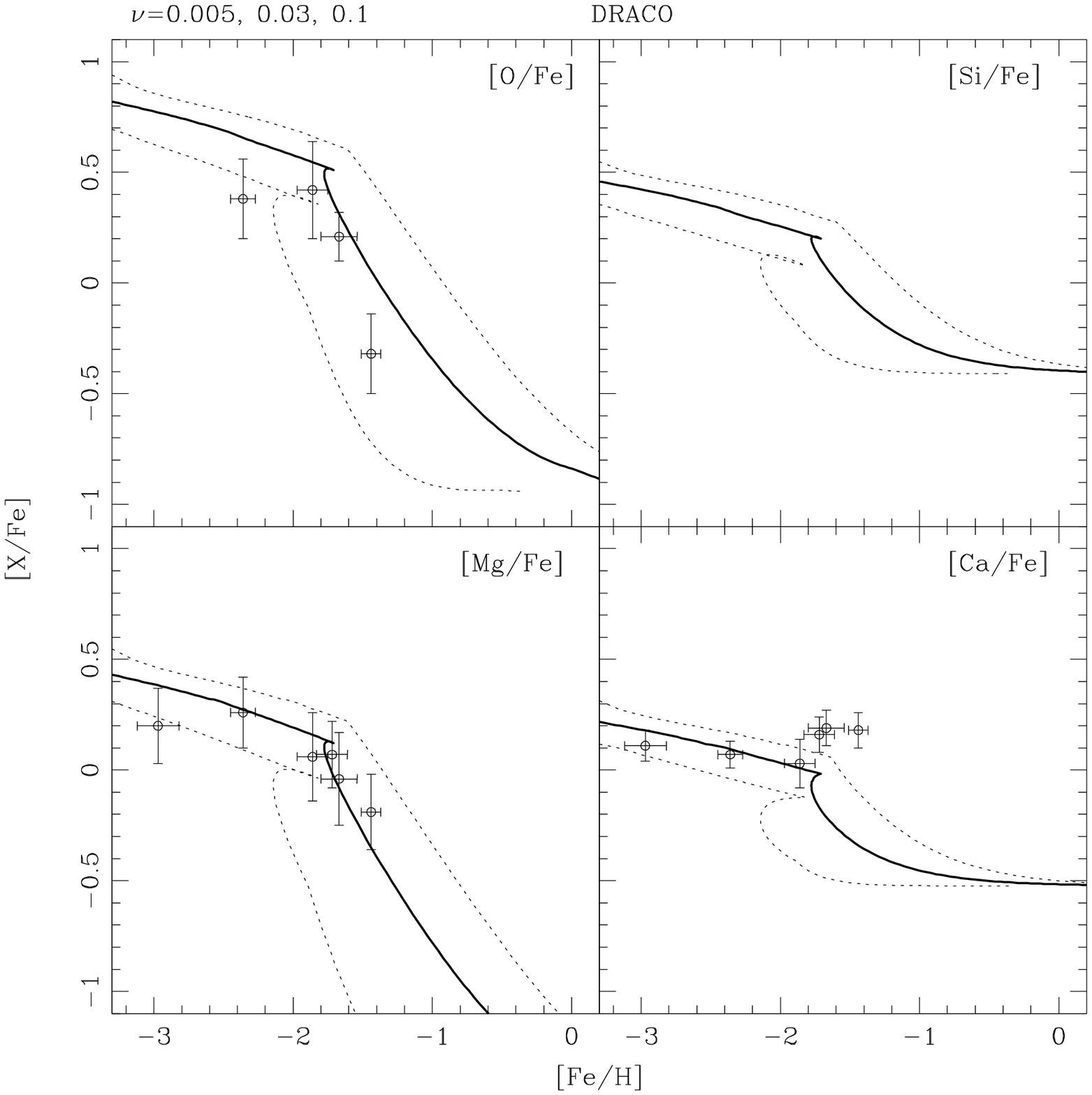}
\hfill
\caption{Observed and predicted [$\alpha$/Fe] vs. [Fe/H] relation for the galaxy Draco. The different lines represent models with different SF efficiencies. The continuous line represents the best model ($\nu= 0.03 Gyr^{-1}$). Figure from LM04.}
\end{figure}

\begin{figure}
\includegraphics[width=12cm,height=10cm]{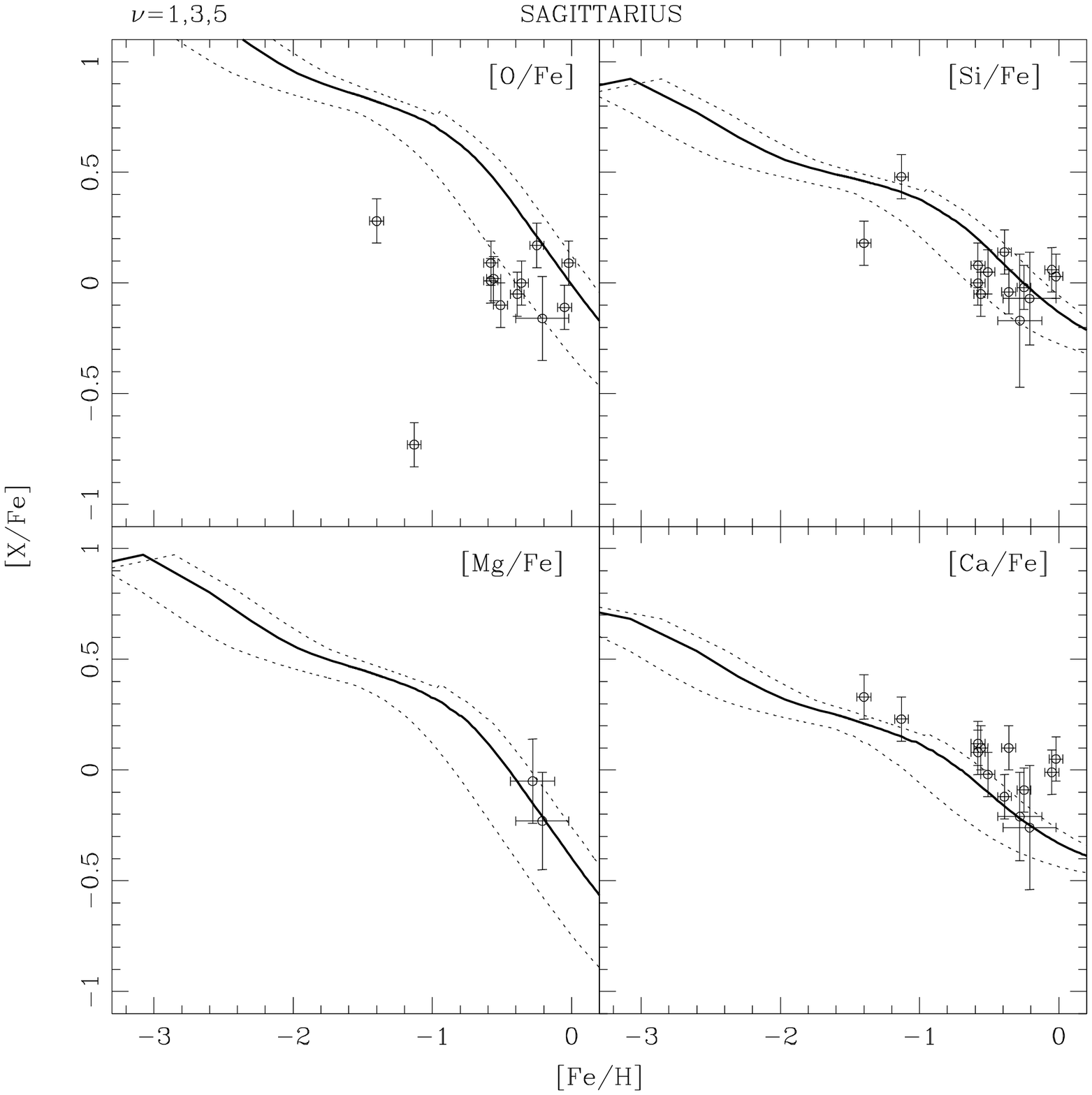}
\hfill
\caption{Observed and predicted [$\alpha$/Fe] vs. [Fe/H] relation for the galaxy Sagittarius. The different lines represent models with different SF efficiencies. The continuous line represents the best model($\nu= 3.0 Gyr^{-1}$) . Here the efficiency of SF is relatively high but the strong galactic wind makes it effectively much lower. Figure from LM04.}
\end{figure}

In Figures 48, 49, 50, 51 and 52we show the predictions for specific dSphs by LM04.
As one can see, the [$\alpha$/Fe] data in the dSphs are well reproduced and in particular the steep decline of the [$\alpha$/Fe] ratio is well reproduced.
This steep decline is due again to the low efficiency SFR, a feature common also to the other 
models, coupled with a strong and continuous galactic wind which gradually empties the galaxies of gas. In the previous models either the galactic wind was not present or it was assumed instantaneous or not as strong as in LM04, thus predicting a flatter slope for the descent of the [$\alpha$/Fe] vs. [Fe/H], as we have already seen. 

Lanfranchi et al. (2006) computed also the expected abundances of s- and r- process elements in dSphs, by adopting the same nucleosynthesis prescriptions used for the chemical evolution of the Milky Way. In particular, they adopted the prescriptions of Cescutti et al. (2006) for Ba, Y, La, Sr and Eu: Ba, Sr , La and Y are mainly s-process elements produced on long timescales by low mass stars ($1-3M_{\odot}$), but they have also a small r-process component originating in stars in the mass range 12-30$M_{\odot}$. The Eu instead is considered as a pure r-process element produced in the stellar mass range  12-30$M_{\odot}$.

In Figures 53, 54, 55 and 56 we show the predictions of Lanfranchi et al. (2006) for s-and r-process elements in dSphs  compared with the available data. Also in this case the agreement looks good, although more data are necessary before 
drawing firm conclusions.
The general tendency for the $\alpha$-elements in dSphs is to be less 
overabundant relative to Fe and the Sun than the stars of the solar vicinity 
with the same [Fe/H].
This is due to the lower SFR in dSphs (the effect is increased by the galactic 
wind)  which acts to shift the curve [$\alpha$/Fe] vs. [Fe/H] for the solar 
vicinity towards left in the diagram, whereas a stronger SF than in the solar 
neighbourhood moves the solar vicinity curve toward right in the diagram 
(see Figure 33).
The same holds for s- and r- process elements: in this case, since [s/Fe] vs. [Fe/H] first increases sharply at low metallicities and then it flattens at higher ones (the opposite of what happens for the $\alpha$-elements), the dSphs show a higher [s/Fe] than the stars in the solar vicinity at the same [Fe/H]. 
This shows again the effect of the time-delay model.

The differences between the [s/Fe] vs. [Fe/H] in dSphs and the Milky Way are shown in Figures 57, where data and predictions of Ba and Eu for Sculptor are compared with data and predictions of Ba and Eu in the solar vicinity.

\begin{figure}
\includegraphics[width=12cm,height=10cm]{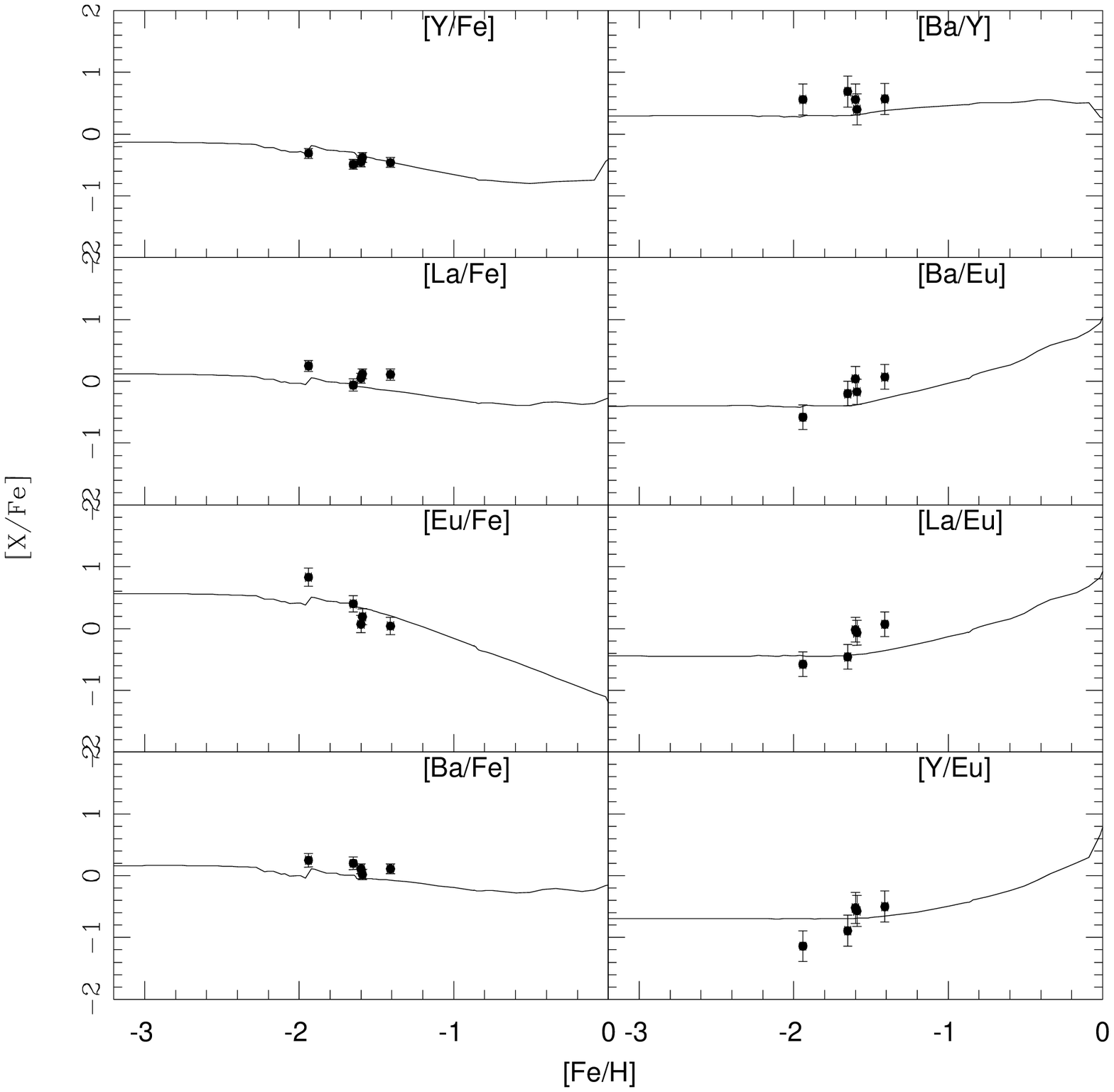}
\hfill
\caption{Predicted and observed [s,r/Fe] vs. [Fe/H] for the galaxy Carina. The model is from Lanfranchi, Matteucci \& Cescutti
 (2006a), where the references for the data can be found.}
\end{figure}

\begin{figure}
\includegraphics[width=12cm,height=10cm]{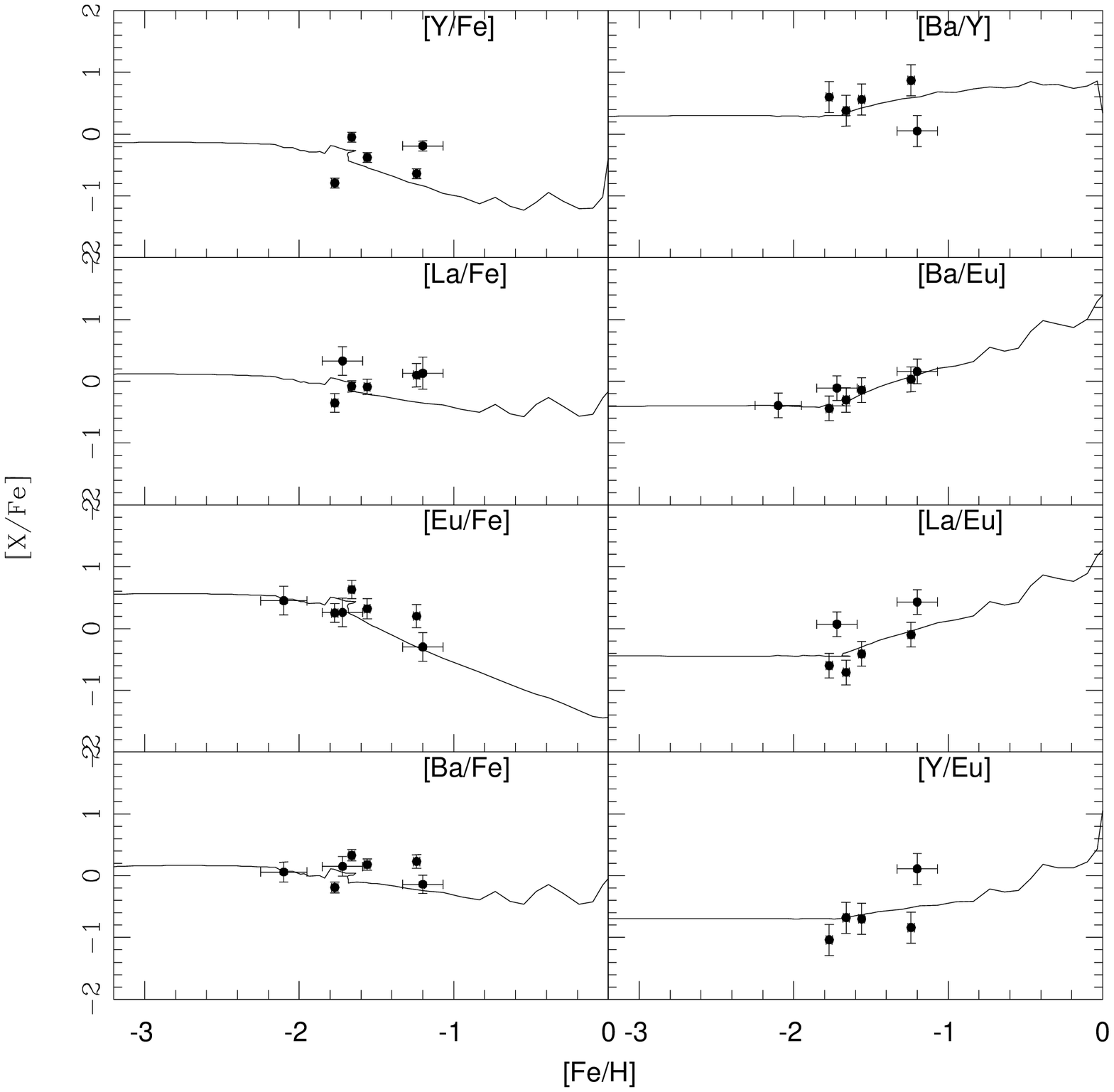}
\hfill
\caption{Predicted and observed [s,r/Fe] vs. [Fe/H] for the galaxy Sculptor.
The model is from Lanfranchi, Matteucci \& Cescutti (2006a), where the references for the the data can be found.}
\end{figure}

\begin{figure}
\includegraphics[width=12cm,height=10cm]{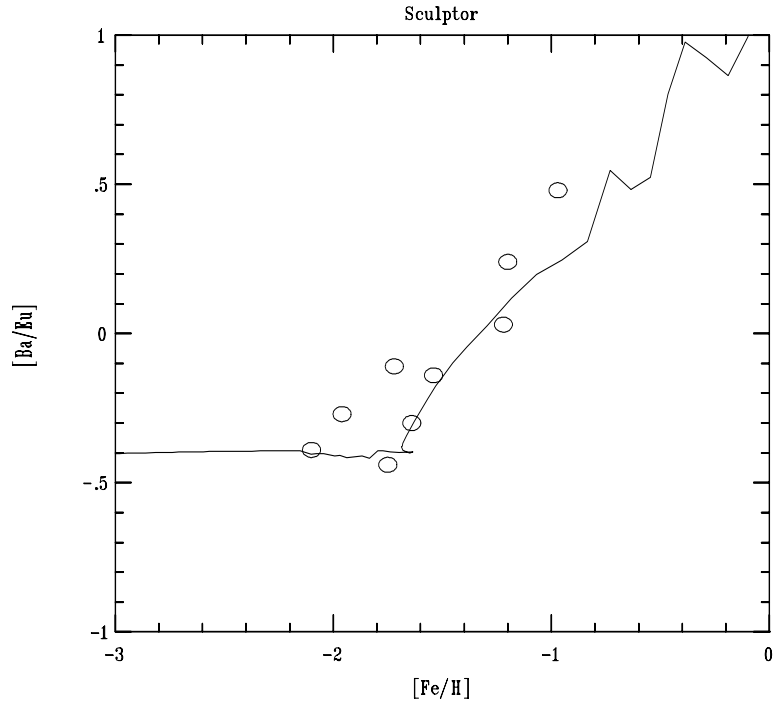}
\hfill
\caption{Predicted and observed [Ba/Eu] vs. [Fe/H] for the galaxy Sculptor.The model is from Lanfranchi, Matteucci \& Cescutti (2006a), the data and the figure are from Geisler et al. (2007).}
\end{figure}

\begin{figure}
\includegraphics[width=12cm,height=10cm]{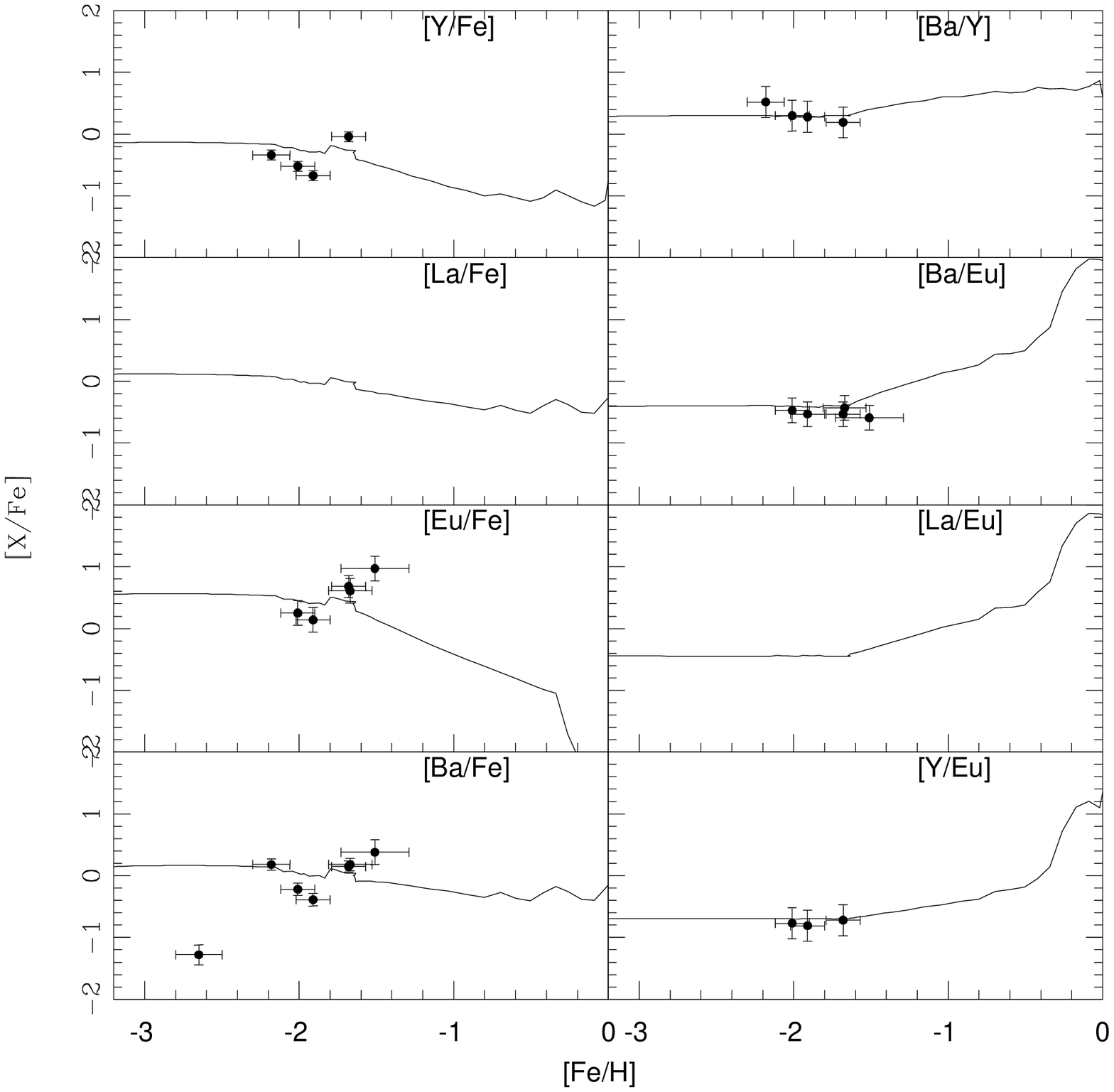}
\hfill
\caption{Predicted and observed [s,r/Fe] vs. [Fe/H] for the galaxy Ursa Minor. The model is from Lanfranchi, Matteucci \& Cescutti (2006a), where the references for the the data can be found.}
\end{figure}

\begin{figure}
\includegraphics[width=12cm,height=9cm]{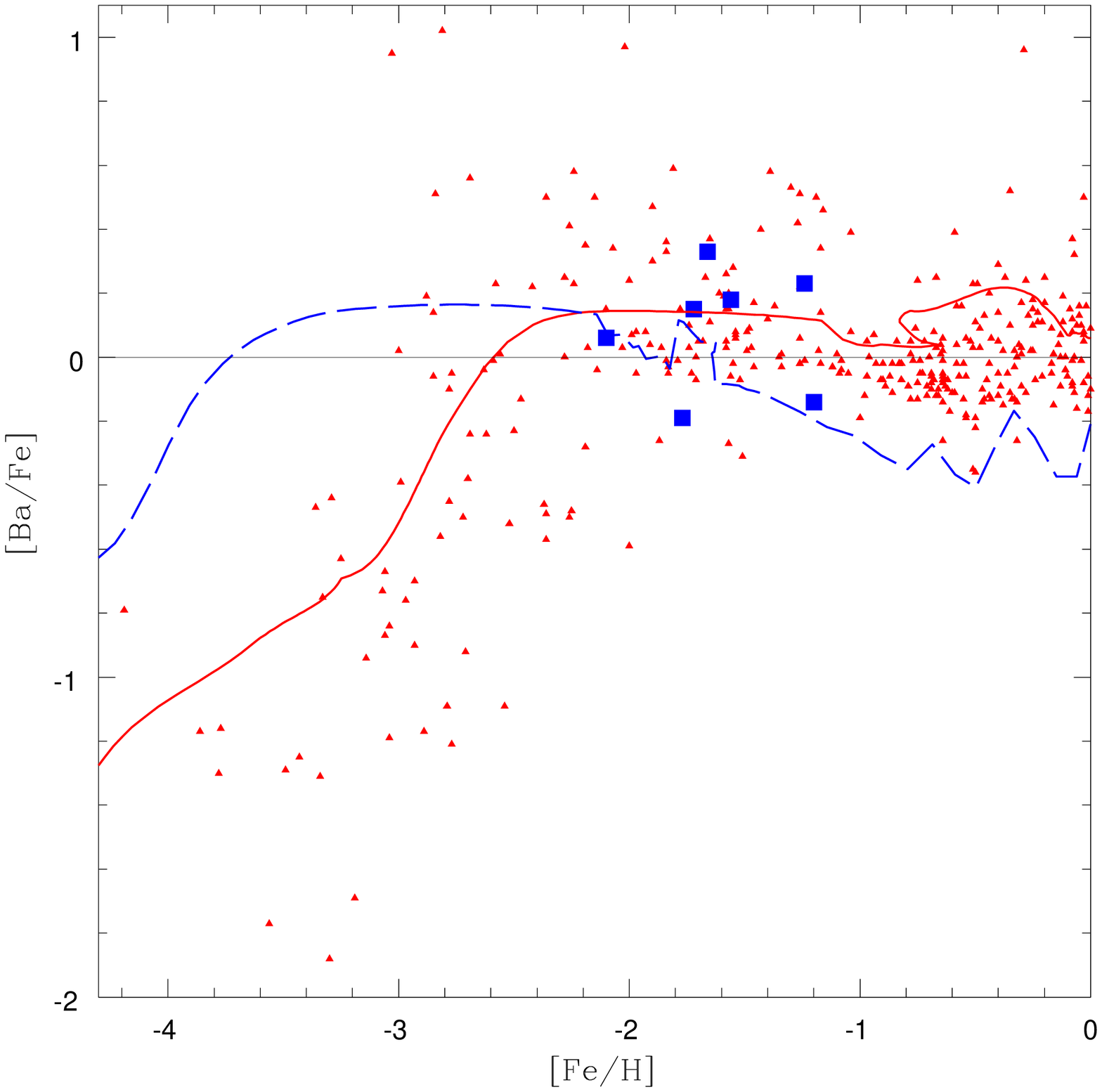}
\includegraphics[width=12cm,height=9cm]{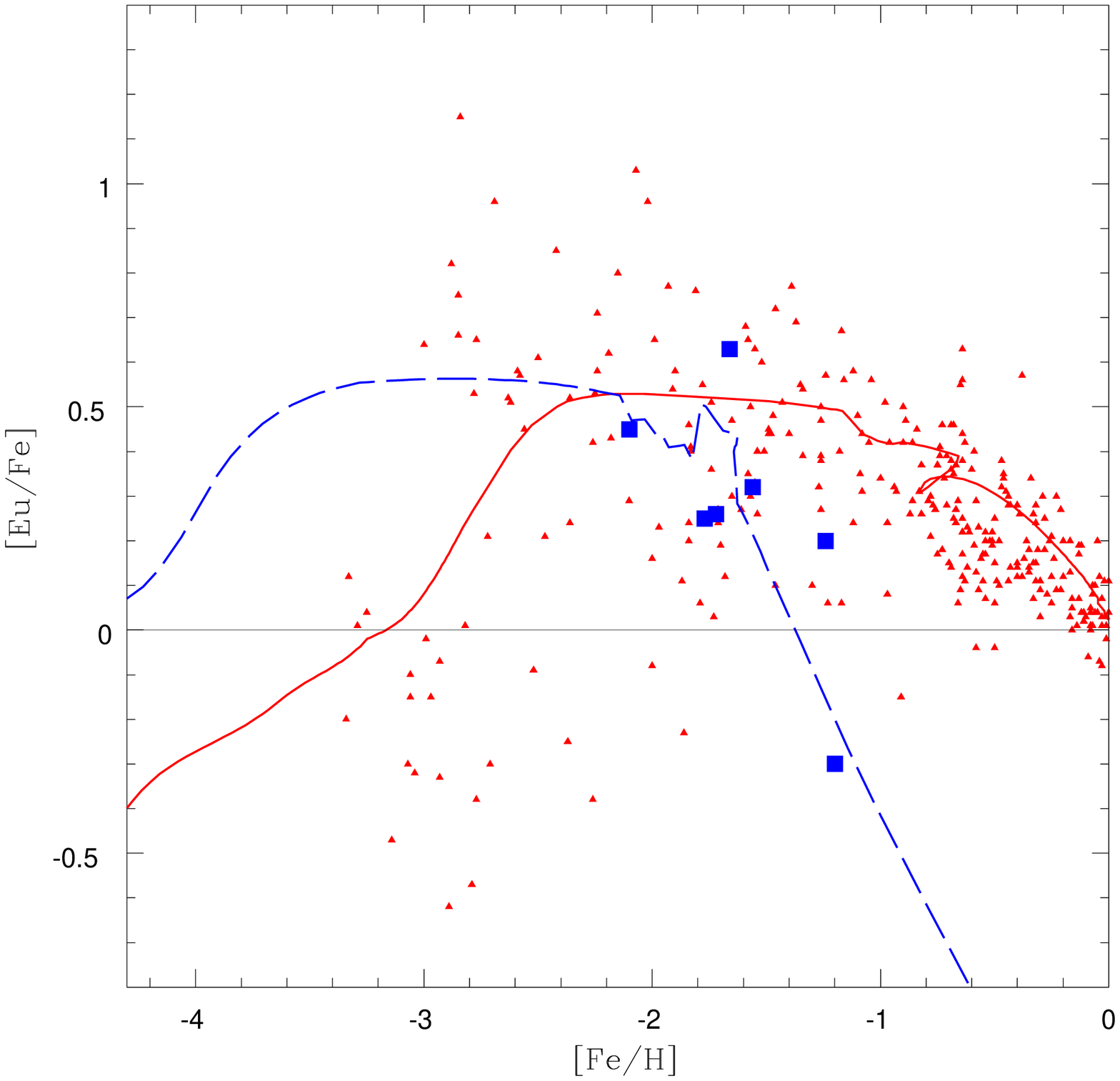}
\hfill
\caption{Comparison between data and predictions for Ba and Eu in Sculptor and in the Milky Way. Model for the Milky Way: continuous lines. Model for Sculptor: dashed lines. Data for the Milky Way: triangles. Data for Sculptor: full squares. Models from Lanfranchi \& al. (2007) where the references to the data can be found.}
\end{figure}

Finally, another important constraint for models of galactic chemical evolution is represented by the stellar metallicity distribution. In Figures 58 we show the predictions for the stellar metallicity distribution of Carina compared with the observed one and the agreement is very good. The observed distribution is from Koch et al. (2006) who measured the metallicity of 437 giants in Carina by means of Ca triplet and then transformed it into [Fe/H] through a suitable calibration.
In Figures 58 we also show the comparison between the stellar metallicity distribution in Carina and the G-dwarf metallicity distribution in the solar vicinity. As 
one can see, the Carina distribution lies in a range of smaller metallicities due to the lower efficiency of SF assumed for this galaxy.

To this purpose, owing a word of caution is appropriate: in fact, the Ca triplet in principle traces the abundance of Ca and not that of Fe and we know that Ca and Fe evolve in a different way since Ca is mainly produced in Type II SNe, whereas Fe is produced mainly in Type Ia SNe. This different evolution of Ca and Fe leads, in the Koch paper, to an uncertainty of 0.2 dex. Besides that, the globular clusters which serve as calibrators for obtaining [Fe/H] lie in  the range -2.0 - -1.0 dex, whereas Koch's data extend down to lower metallicities.

\begin{figure}
\includegraphics[width=12cm,height=10cm]{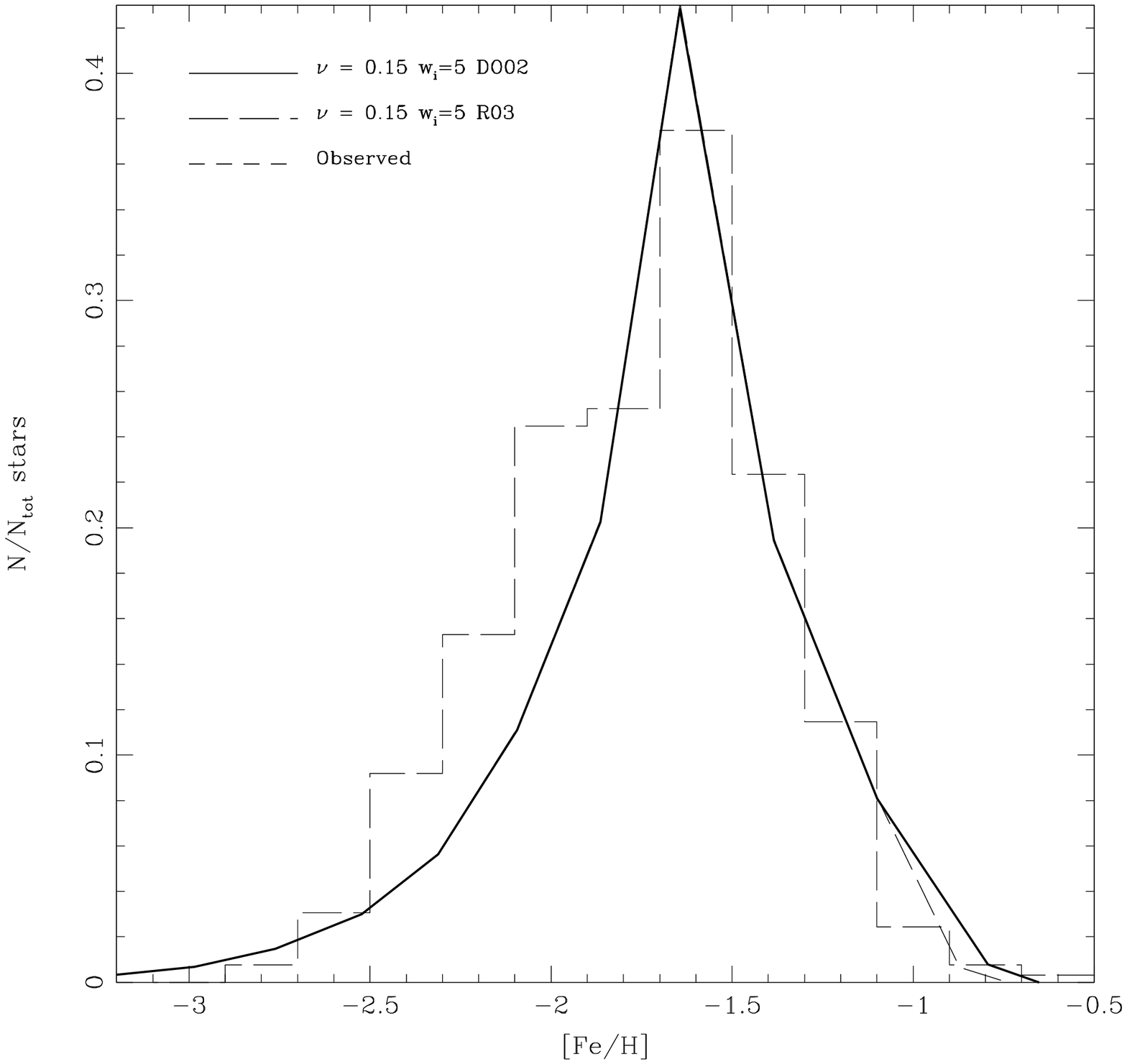}
\includegraphics[width=12cm,height=10cm]{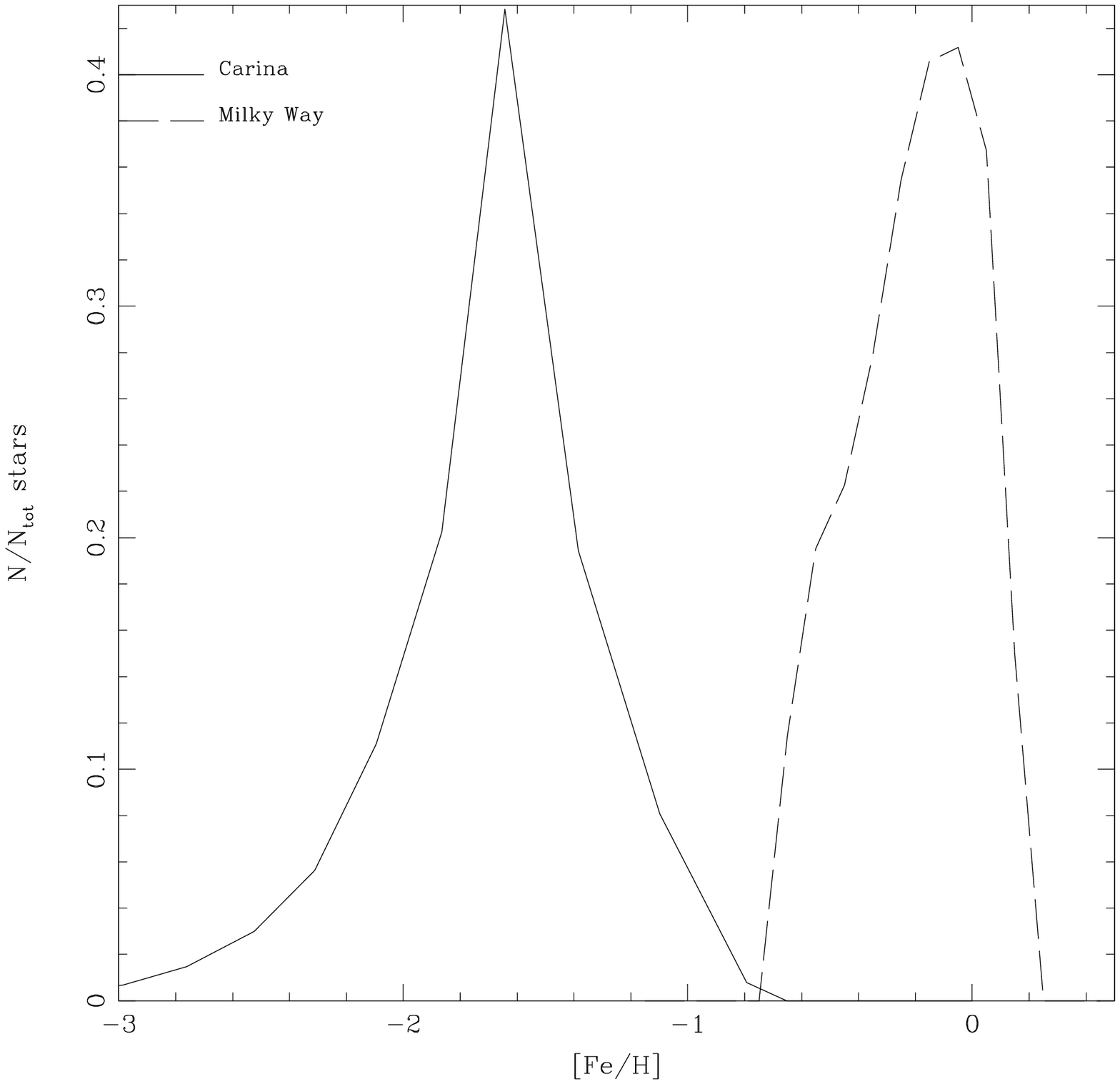}
\hfill
\caption{Upper panel: stellar metallicity distribution for Carina. Model from Lanfranchi et al. (2006b). The assumed SF efficiency is $\nu =0.15 Gyr^{-1}$ and the wind efficiency is $\lambda=5$. Two different histories of SF have been tested here: the one of Dolphin (2002) (continuous line) and that of Rizzi et al. (2003) (long dashed line), but this does not produce important differences in the results. The main difference between the two histories of SF is the number 
of bursts (3 in Dolphin and 4 in Rizzi et al.). Data from Koch et al. (2006). Lower panel: predicted
stellar metallicity distribution for Carina compared with the predicted G-dwarf 
metallicity distribution in the solar neighbourhood (dashed line).}
\end{figure}

The good fit of the stellar metallicity distribution indicates that both the 
assumed history of SF and the IMF are close to reality.

LM04 predicted the stellar metallicity distribution for all the six dSphs and 
while for Carina the agreement is quite good for Sculptor they
cannot reproduce the bimodal stellar distribution recently suggested by  
Tolstoy et al. (2004) and shown in Figure 59, since their model is a 
one-zone model. 
In Figure 60 are shown the predictions of LM04 for the Sculptor galaxy: it is clear from this Figure that to reproduce the two different stellar populations, one has to assume a multizone model possibly with different efficiencies of SF.
Kawata et al. (2006) explained the bimodality of stellar populations in Sculptor as a consequence of dissipative collapse 
which produces higher metallicities at the center of the galaxy.

\begin{figure}
\includegraphics[width=12cm,height=10cm]{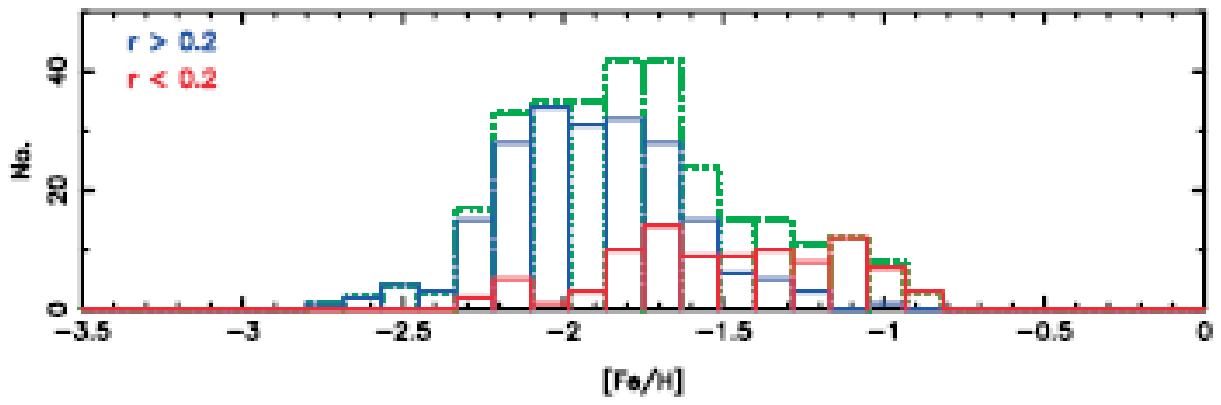}
\hfill
\caption{Observed stellar metallicity distribution for Sculptor. Data and figure are from Tolstoy et al. (2004): all the stars of Sculptor are indicated by the dotted line. The central stars are those indicated by the lower histogram with continuous line, whereas the stars beyond $R> 0.2$ Kpc are indicated by the upper histogram with continuous line.}
\end{figure}

\begin{figure}
\includegraphics[width=12cm,height=10cm]{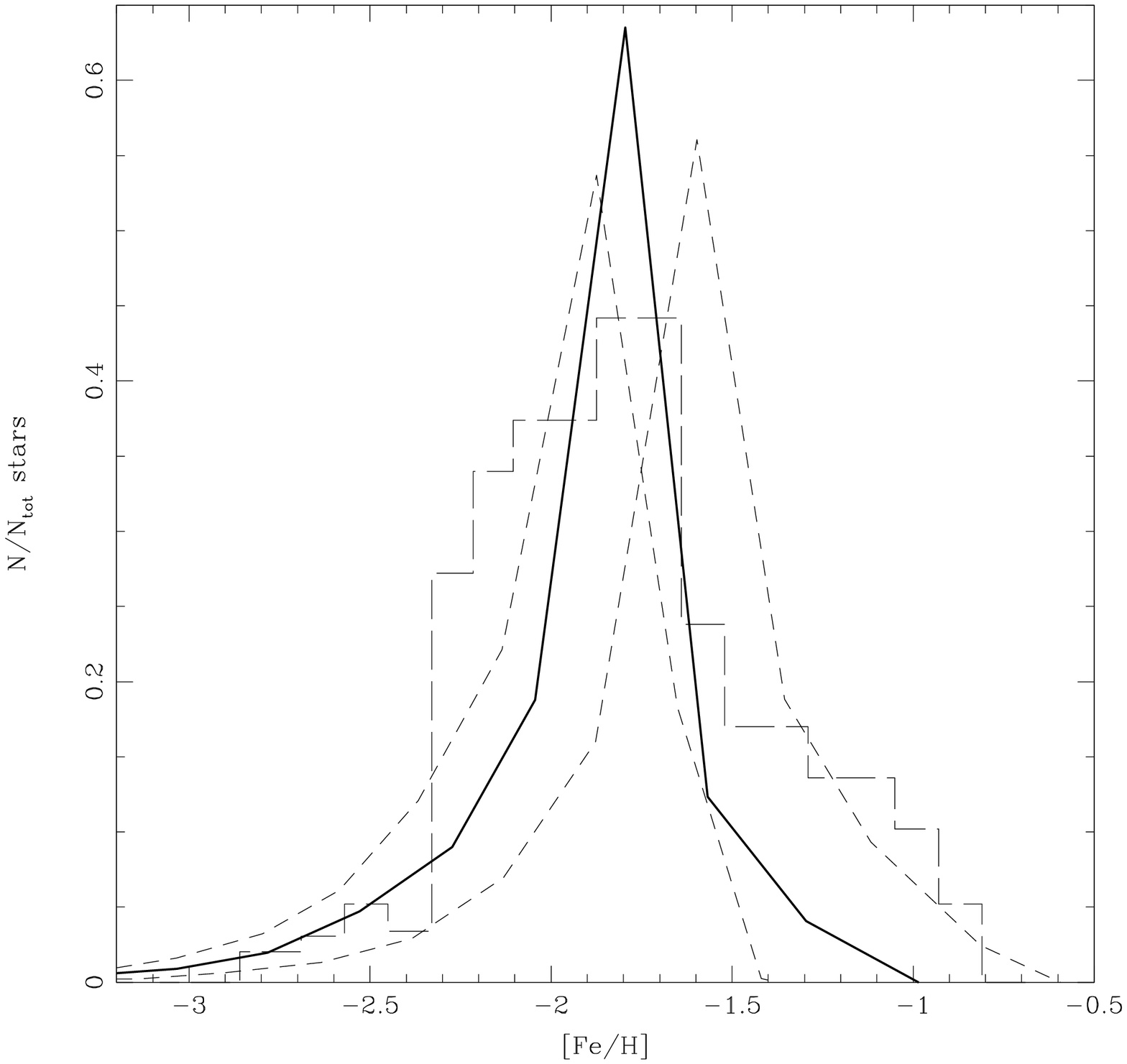}
\hfill
\caption{Observed and predicted stellar metallicity distribution for Sculptor. 
The data (the dotted line of Figure 59) are represented by the histogram (long dashed). The models are from LM04: the solid line represents the best model, whereas the dashed lines represent models with higher (the curve on the right) or lower (the curve on the left) SF efficiency.}
\end{figure}

\section{What have we learned about dSphs?}

From the study of the chemical evolution of dSphs and the Milky Way we can derive the following conclusions:
\begin{itemize}

\item By comparing the [$\alpha$/Fe] ratios in the Milky Way and in dSphs of the Local Group we can conclude that these systems had different histories of 
SF.

\item The [$\alpha$/Fe] ratios in dSphs are always lower than in the Milky Way at the same [Fe/H]. This is a consequence of the time-delay model which predicts this behaviour for systems which suffered a lower star formation than the solar vicinity.

\item  The occurrence of strong galactic winds or gas loss in general is necessary to keep the SF low and it produces the steep decrease of the [$\alpha$/Fe] ratio observed in dSphs (see Lanfranchi \& Matteucci 2007).

\item Good agreement is found both for [s/Fe] and [r/Fe] abundance ratios. 
The [s/Fe] ratios are predicted to be higher than the same ratios in Milky Way stars with the same [Fe/H]. This is again a consequence of the time-delay model.

\item The dSphs of the Local Group contain very old stars but they suffered extended periods of SF, far beyond the reionization epoch. This is suggested both from the color-magnitude diagrams of these galaxies and from the level of the abundances of s-process elements such as Ba, which could not have been observed if the SF had stopped at the reionization epoch (Fenner et al. 2006).

\item All the previous conclusions suggest that it is unlikely that the dSphs 
have been the building blocks of the Milky Way, as predicted by current CDM models (see review by Geisler et al. 2007 for a detailed discussion of this point).
Robertson et al. (2005) studied the formation of the Galactic stellar halo by means of different accretion histories for the dark matter halo of the Milky Way in the framework of the $\lambda$CDM model. They concluded, on the basis of the [$\alpha$/Fe] ratios in Galactic halo stars, in dwarf irregulars and dSphs, that it is more likely that the Galactic dark matter  halo was formed by an early accretion of dwarf irregular galaxies, which formed stars for a short time and then were destroyed.  
Concerning dSphs, they suggest that their chemical abundances should have been affected by galactic winds and that they
 should have been accreted and destroyed over the entire Milky Way lifetime.

\end{itemize}

\section{Other spirals}

In this section we will briefly describe the properties of other spirals of the Local Group for which we have enough chemical information.

In external spirals we can measure:
\begin{itemize}
\item  the SFR mainly from $H_{\alpha}$ emission and this suggests a correlation between the SFR and the  total surface gas density, as discussed previously.

\item Abundance gradients are also found in disks of local spirals (see Garnett et al. 1997): in particular, it is found that abundance gradients, expressed in dex/Kpc, are steeper  in smaller disks but the correlation disappears if they are expressed in dex/$R_d$ (where $R_d$ is the disk scalelength).  It is interesting to note that a universal gradient slope per unit scale length may be explained by viscous disk models (e.g. Lin \& Pringle, 1987; Sommer-Larsen \& Yoshii, 1989). Further information on gradients is that they are flatter in disks with bars: probably the bar induces radial flows which can wash-out the abundance gradients if their velocity is high enough (see Tinsley, 1980). 

\item The gas distribution: one finds differences in the gas distribution along the disk betwen field and cluster galaxies, these latter being subject to ram pressure stripping.

\item Integrated colors of galaxy disks give information on the distribution of stellar populations along the disks. Several authors (Josey \& Arimoto, 1992; Jimenez et al. 1998; Prantzos \& Boissier, 2000) have suggested that color gradients, as well as metallicity gradients, can be reproduced by assuming an inside-out formation for disks, as has been suggested for the disk of the Milky Way.
\item In Figure 61 we show the results of a paper by Boissier et al. (2001).
They conclude that more massive disks are redder, more metal rich and more gas-poor  than smaller ones. On the other hand, their estimated SF efficiency (defined as the SFR per unit mass of gas) seems to be similar among different spirals: this leads them to conclude that more massive disks are older than less massive ones. The various quantities in Figure 61 are plotted as functions of the disk circular velocity which is a measure of the dark matter halo of each galaxy. The various curves are obtained by varying the spin parameter $\lambda$, expressed as a $\lambda/ \lambda_{MW}$, where $\lambda_{MW}$ refers to the Milky Way.
In fact, in the framework of semi-analytical models of galaxy formation the 
evolution of galactic disks can be described by means of scaling laws 
calibrated on the Milky Way with $V_c$ and $\lambda$ as parameters (e.g. Mo et al. 1998).

\begin{figure}
\includegraphics[width=14cm,height=14cm]{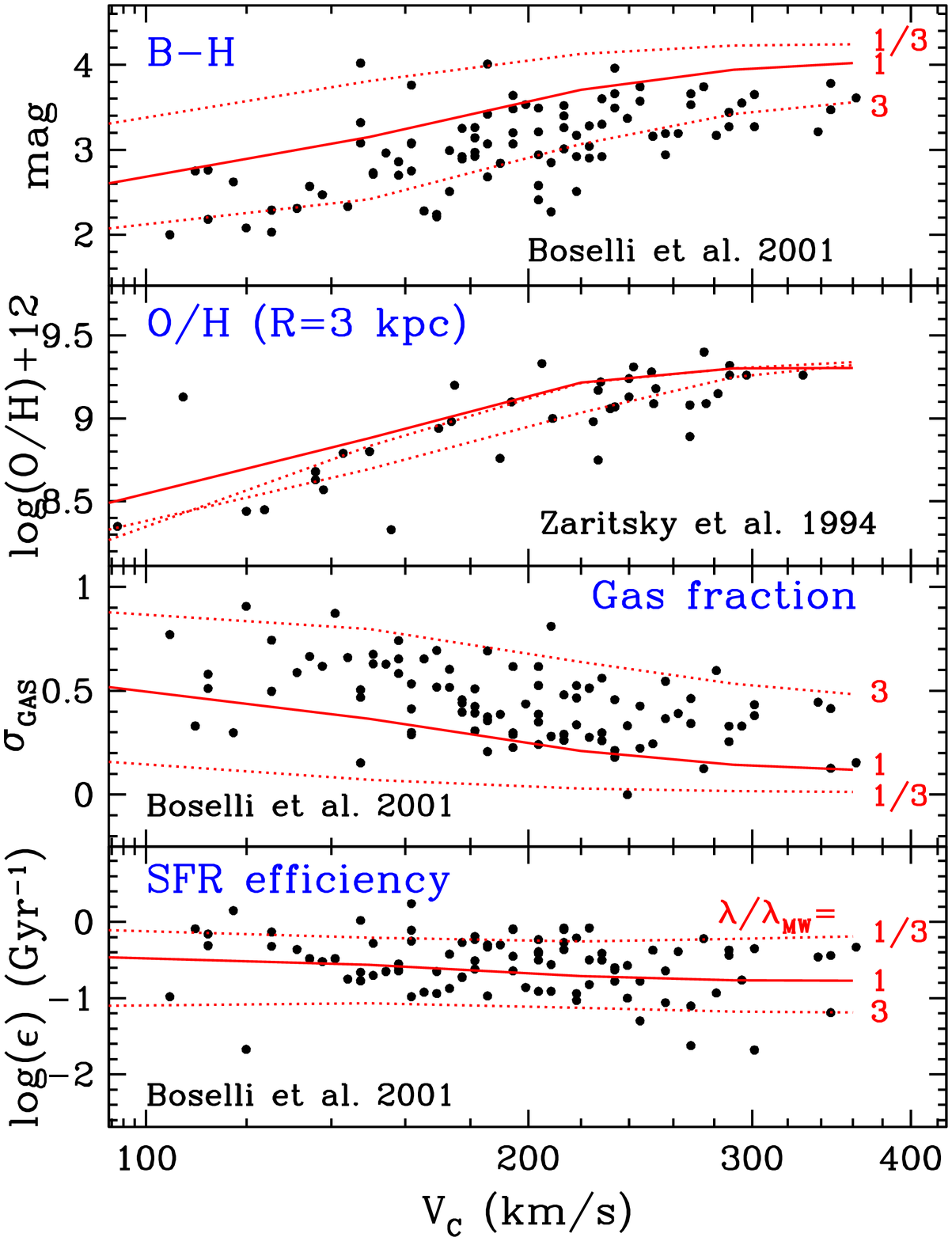}
\hfill
\caption{Distribution of the (B-H) color, the 12+log(O/H), the gas fraction and the SFR efficiency (the SFR per unit mass of gas) versus the circular velocity of disks of local spirals.The different lines refer to different values of the spin parameter $\lambda$. Figure from Boissier et al. (2001) where the references to the data can be found.}
\end{figure}

\end{itemize}

\subsection{Chemical models for external spirals}

Several models of chemical evolution of Local Group spirals have been developed in the past years (e.g. Diaz \& Tosi, 1984; Moll\'a et al. 1996; Chiappini et al. 2003).
Diaz \& Tosi (1984) first modeled the chemical evolution of M31, M33, 
M83 and M101. 
Moll\'a et al. (1996) modeled several spirals of the Local Group (M31, 
NGC300, M33, NGC628, NGC3198, NGC6946).
In Figure 62 we show the predictions, compared to observations, of the Moll\'a et al. model for M31.

\begin{figure}
\includegraphics[width=3.0in,height=4.0in,angle=-90]{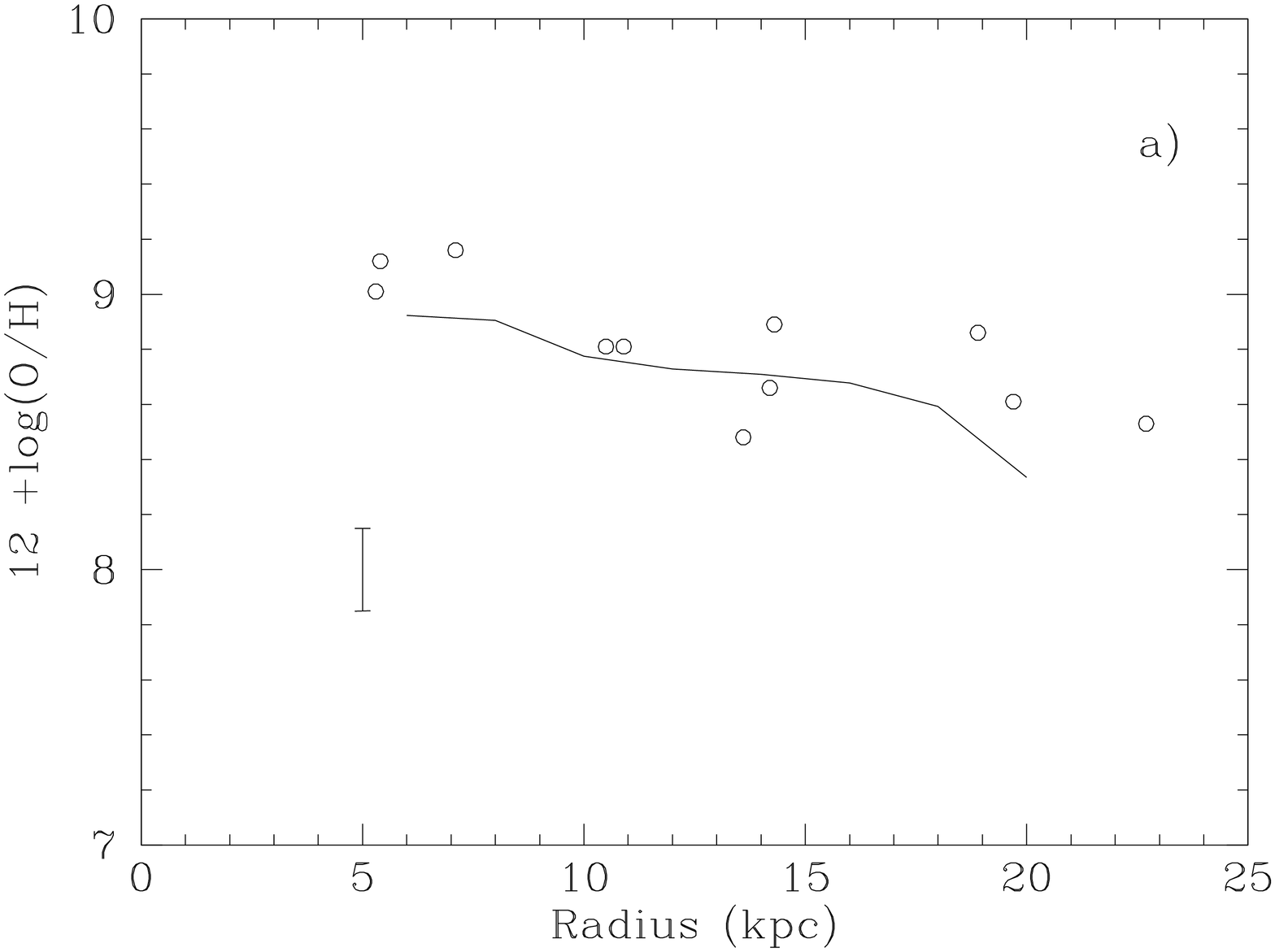}
\includegraphics[width=3.0in,height=4.0in,angle=-90]{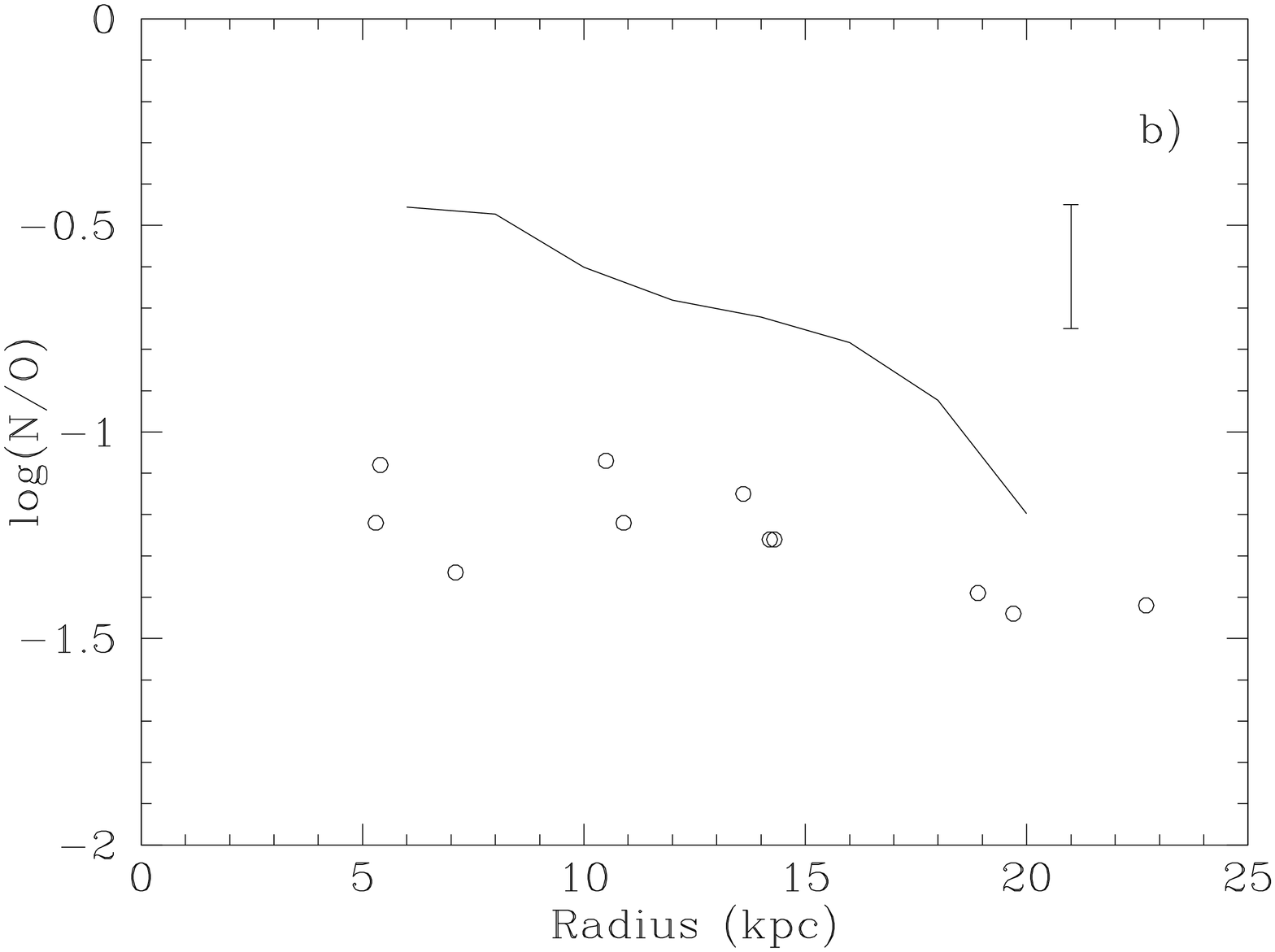}
\caption{ Predicted and observed abundance gradients in M31. Figure and models are from Moll\'a et al. (1996).}
\hfill
\end{figure}

As an another example of abundance gradients and gas distribution in a local spiral galaxy we show in Figure 63 
the observed and predicted gas distribution and abundance gradients for the disk of M101. In this case the gas distribution and the abundance gradients are reproduced with systematically smaller timescales for the disk formation relative to the 
MW (M101 formed faster), and the difference between the timescales of formation of the internal and external regions is smaller ($\tau_{M101}=0.75 r(Kpc) - 0.5$ Gyr, Chiappini et al. 2003) compared 
to eq. (34).

\begin{figure}

\includegraphics[width=4.5in,height=3.0in]{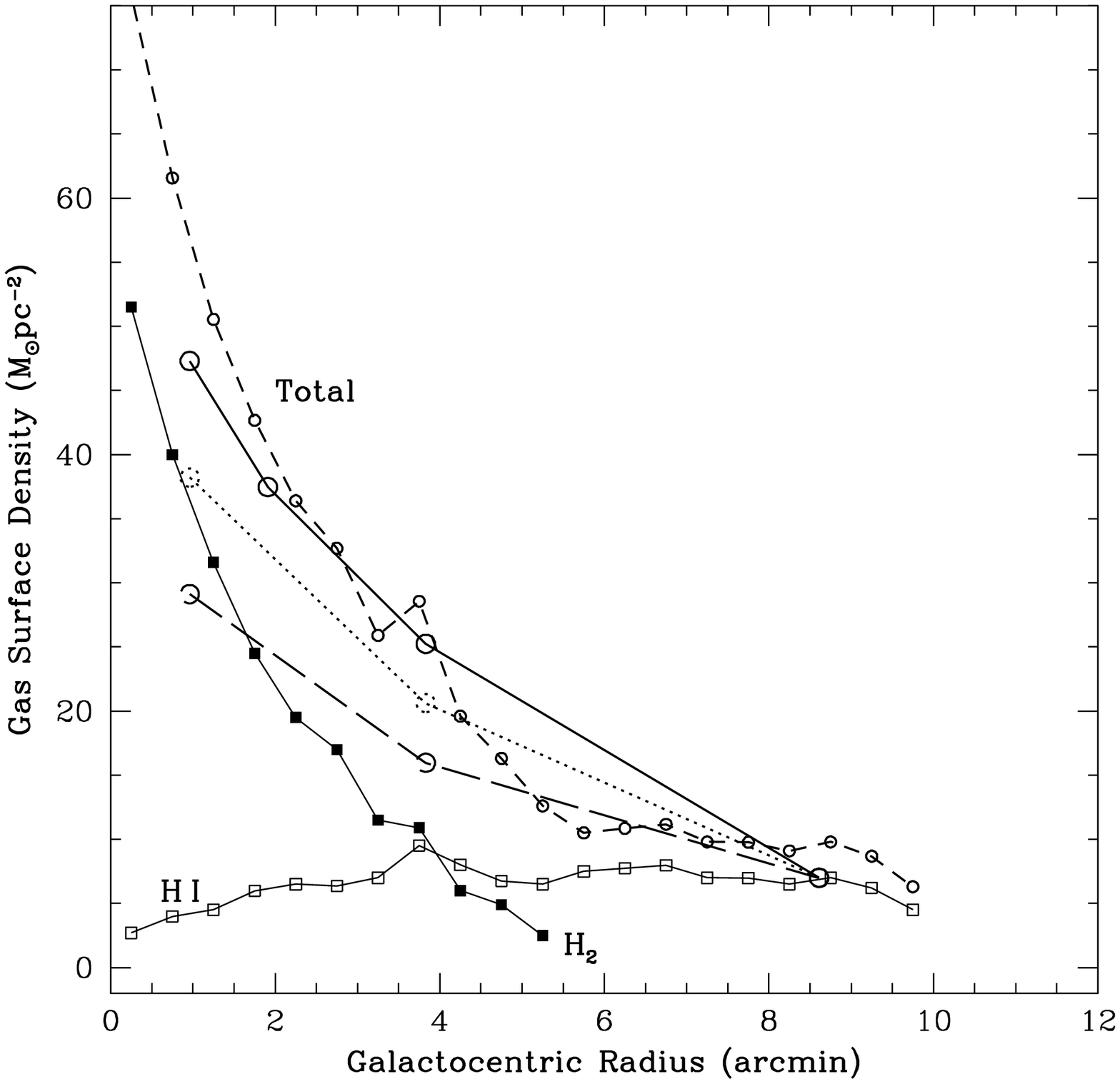}
\includegraphics[width=4.5in,height=3.0in]{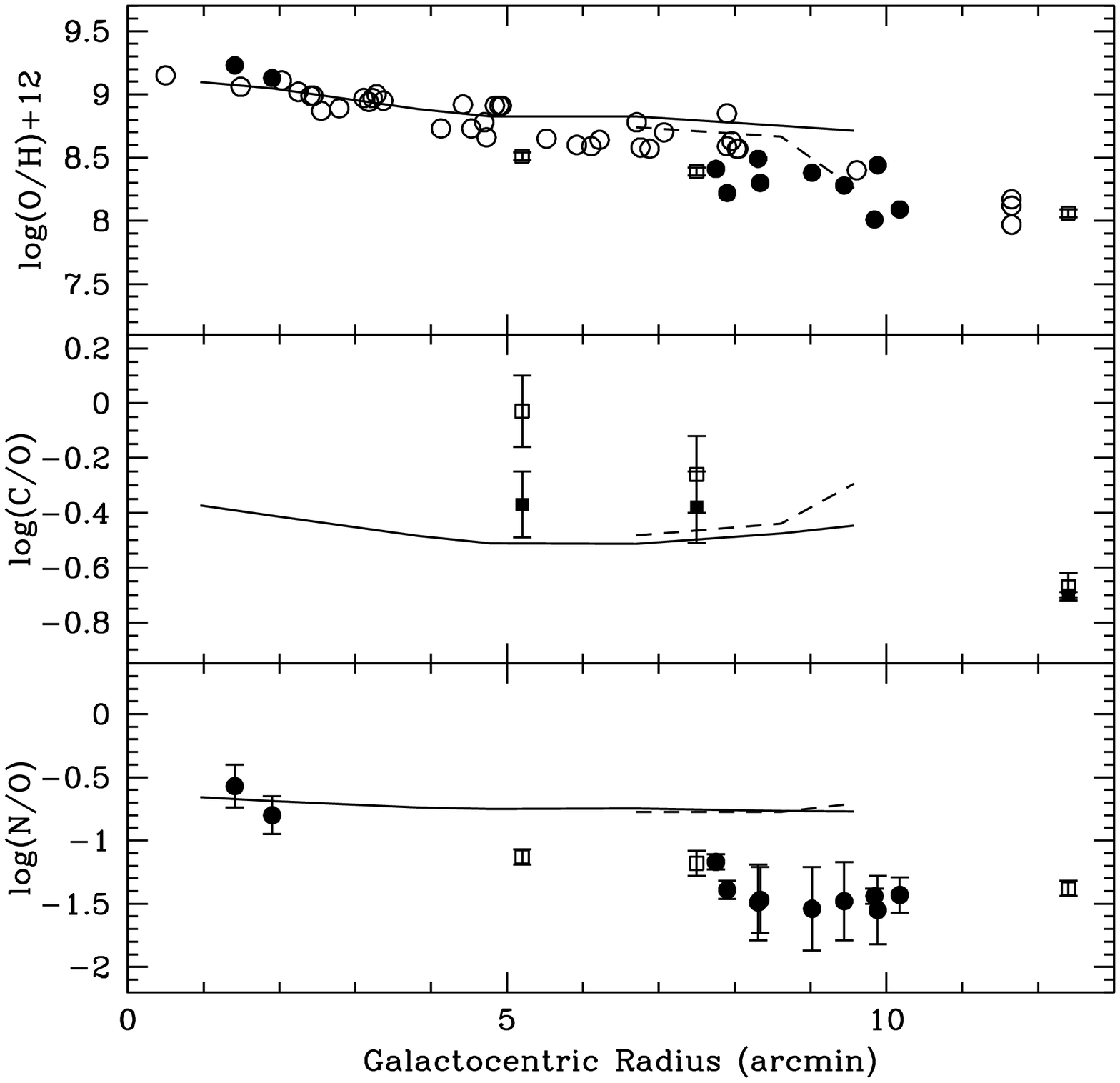}
\hfill
\caption{Upper panel: predicted and observed gas distribution along the disk of M101. The observed HI, $H_2$ and total 
gas are indicated in the Figure.
The large open circles indicate the models: in particular, the open circles connected by a continuous line refer to a model with central surface mass density of 1000$M_{\odot}pc^{-2}$, while the dotted line refers to a model with 800$M_{\odot}pc^{-2}$ and the dashed line to a model with  600$M_{\odot}pc^{-2}$.
Lower panel: predicted and observed abundance gradients of C,N,O elements along the disk of M101. The models are the lines and differ for a different threshold density for SF, being larger in the dashed model. All the models are by Chiappini et al. (2003).}\label{fig} 
\end{figure}

Therefore, in the Chiappini et al. (2003) paper it is suggested that
the fact that more luminous spirals ( e.g. M101) 
tend to have shallower abundance 
gradients than less luminous ones can be interpreted as due to a faster formation (down-sizing) of large spirals, as also hinted at by the results of Boissier et al. (2001).

In summary, from the available studies of spirals in the Local Group we can suggest the following:
\begin{itemize}
\item The disks of spirals have all formed inside-out and the more massive disks have formed faster than the less massive one.
\item This translates into a faster gas accretion rate and consequently into a faster SFR.
\item In other words, the most massive disks are also the oldest, a conclusion which is not in line with classical predictions from CDM models.
\end{itemize}

\section{Cosmic chemical evolution}
With the name cosmic chemical evolution we indicate the chemical evolution taking place in comoving volumes large enough to be representative of the whole universe (Pei \& Fall 1995). The evolution  can be described in terms of comoving densities of gas and stars $\Omega_{gas}$ and $\Omega_{stars}$, both measured in units of the present critical density ($\rho_c = {3H_o^{2} \over 8 \pi G}$) and the mean abundance of heavy elements in the ISM, Z, including dust.
Under IRA we can write, following Pei \& Fall:

\begin{equation}
\dot{\Omega}_{gas} + \dot{\Omega}_{stars} = \dot{\Omega_{f}}
\end{equation}

and

\begin{equation}
\Omega_{gas}\dot{Z}-y_{Z}\dot{\Omega}_{stars}= (Z_f -Z) \dot{\Omega_{f}}
\end{equation}

where the dots represent differentiation with respect to the cosmic time.
The term $\dot{\Omega_f}$ can represent either the infall or the outflow rate according 
to its sign.
For the closed-box model $\dot{\Omega_{f}} =0$ and its solution is:

\begin{equation}
Z = -y_Z ln(\Omega_{gas}/ \Omega_{gas_{\infty}})
\end{equation}

with $\Omega_{gas_{\infty}}$ being the gas comoving density at some suitably 
high redshift when there are still no stars and heavy elements.

By means of these equations Pei \& Fall followed the evolution of DLAs 
(the quasar absorbers). They expressed the quantity $\Omega_{gas}$ in terms 
of the observable properties of DLAs. They assumed IRA.

Since all these cosmic quantities refer to an unitary volume of the universe 
which contains galaxies of all morphological types,

Calura \& Matteucci (2004) proposed another approach to the cosmic chemical evolution, which takes into account galaxies of different morphological type. 
They computed the cosmic chemical enrichment of the universe by means of detailed models of chemical evolution of galaxies of all morphological types, relaxing IRA and assuming for each galaxy type a different history of SF, as discussed in the previous sections.
They defined the comoving cosmic density of stars and gas for galaxies of different morphological type (ellipticals, spirals and irregulars) as:

\begin{equation}
\rho_{*,k}=\rho_{B,k} \cdot (M_{*}/L)_{B,k} \\
\end{equation}
for the stars and

\begin{equation}
\rho_{g,k}=\rho_{B,k} \cdot (M_{g}/L)_{B,k}\\ 
\end{equation}
for the gas. The quantities $(M_{*}/L)_{B,k}$ and $(M_{g}/L)_{B,k}$ are the 
predicted M/L ratios for stars and gas, respectively. 
$L_{B,k}$ is the blue luminosity for each galaxy type ($k$ indicates the morphological type) and $\rho_{B,k}$ is the comoving luminosity for a given galaxian morphological type.

They computed the  mean mass weighted metallicity of galaxies by summing the metallicities predicted for the different morphological  types as:

\begin{equation}
<Z_{galaxies}>=\frac{\sum_{k} \rho_{g,k} \, Z_{g,k} + \sum_{k} 
\rho_{*,k} \, Z_{*,k}}{\sum_{k} (\rho_{g,k} +\rho_{*,k}) }\\
\end{equation}

and obtained:

\begin{equation}
<Z_{galaxies}>=0.0175=0.9 Z_{\odot}
\end{equation}

where $Z_{\odot}=0.02$ and 
with $56\%$, $42\%$ and $2\%$ of metals produced in ellipticals, spirals and 
irregulars, respectively. Therefore, the conclusion is that the average metallicity in galaxies is almost solar and that most of the metals in the universe have been produced by elliptical galaxies.
They also predicted the average [O/Fe] ratios for each galaxy type both in the gas and in stars. 

In particular:

[O/Fe]$_{*,Ellipt} = 0.4$ dex, [O/Fe]$_{gas,Ellip} = -0.33$ dex,
[O/Fe]$_{*,Spiral} = 0.1$ dex and [O/Fe]$_{gas,Spiral} = 0.01$dex.

Then, they computed the metallicity in the intergalactic medium (IGM) by considering all the metals ejected by galaxies (mainly ellipticals) into the IGM:
\begin{equation}
<Z_{IGM}>=\frac{\Omega_{Z,IGM}}{\Omega_{b,IGM}} \\
\end{equation}

with $\Omega_{b,IGM}= \Omega_{b}-\Omega_{b,*}-\Omega_{b,gas} = 0.0753$ being
the baryonic density of the IGM and $\Omega_{b}=0.02h^{-2}$ (from WMAP, Spergel et al. 2003, 2007) being the total baryonic content of the universe.
Therefore, they obtained: 
\begin{equation}
<Z_{IGM}>=6.54 \times 10^{-4} = 0.03 Z_{\odot}.
\end{equation}

Finally, they computed the average metallicity of the universe by accounting for all the metals produced in galaxies over the lifetime of the universe:
\begin{equation}
<Z_{universe}>=\frac{\sum_{k}\Omega_{Z,k}}{\Omega_{b}} = {\Omega_{Z,tot} \over
\Omega_{b}}= 0.0017=0.09 Z_{\odot}
\end{equation}
where $\Omega_{Z,tot}$ represents the sum of all the metals produced in all galaxies and $\Omega_{b}$ represents the the total amount of baryons in the universe.

In summary, the mean metallicity inside galaxies of all morphological types is almost solar, whereas the mean metallicity of the universe is roughly 1/10 solar.

Calura \& Matteucci (2003) computed also cosmic Type Ia SN rates ($SNR_{cosm}$), expressed in SNu (number of $SNe/10^{10} L_{B_{\odot}}$ per century). In particular, they took into account the contribution of all galaxy types in the following way:

\begin{equation}
SNR_{cosm}(z)= {\sum_k{SNR_{k}(z)} \over \sum_k {L_{B_{k}}}}
\end{equation}

where  $SNR_{cosm}$ can represent the Type II, Ib/c, Ia SNe. The sums are over all 
the galactic morphological types, $L_{B_{k}}$ is the total blue luminosity of the $kth$ morphological type. In order to compute the $SNR_{cosm}$ for each galaxy they assumed a SN model progenitor and a cosmic SFR, calculated as:

\begin{equation}
\dot{\rho_{*}}= \sum_k{\rho_{B,k}(z) \cdot (M_{*}/L)_{B,k}(z) \cdot \psi(z)_k},
\end{equation}
where $\rho_{B,k}(z)$ and  $(M_{*}/L)_{B,k}(z)$ have been already defined and $\psi(z)_k$ represents the history of SF of a galaxy of $kth$ morphological type. They assumed the SF histories of Figure 41. 
In Figures 64 and 65 we show some examples of predicted cosmic SN rates by adopting the same cosmic SFR (eq. 55) but different assumptions about the Type Ia progenitor model.

\begin{figure}
\includegraphics[width=9.0in,height=6.0in]{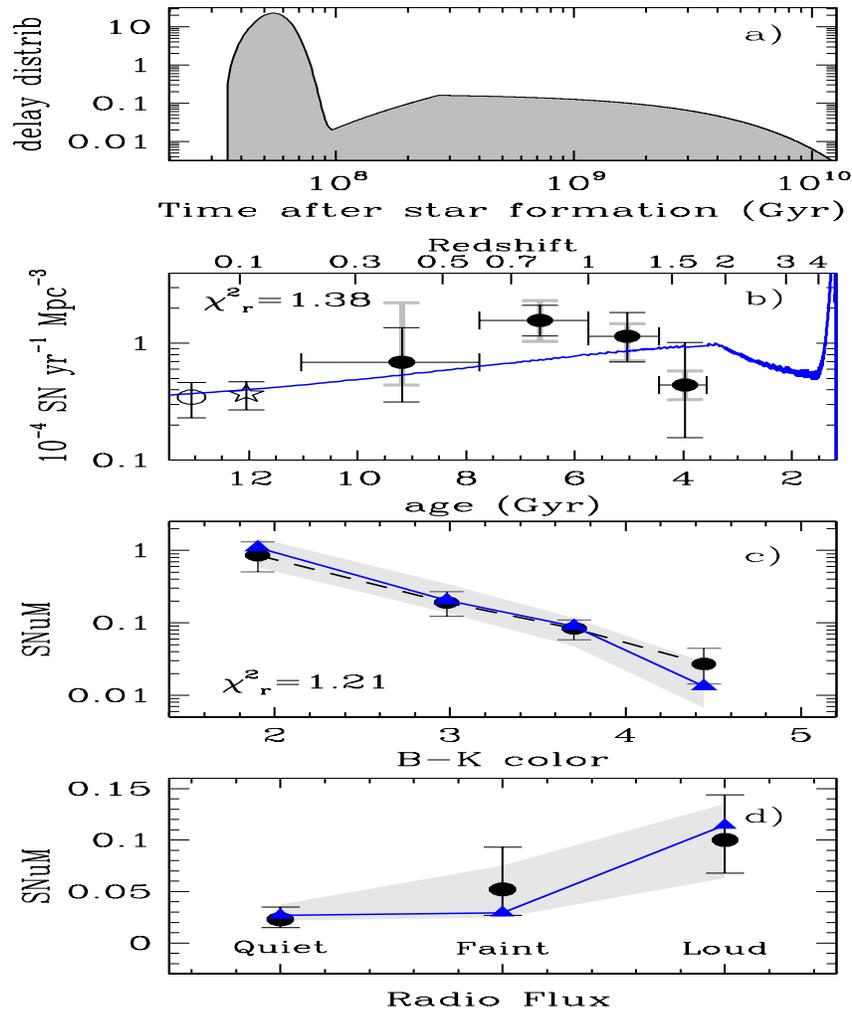}
\hfill
\caption{Theoretical cosmic Type Ia SN rates compared with observational data from Mannucci et al. (2006). The progenitor model adopted assumes the delay-time distribution suggested by Mannucci et al. (2005;2006) (panel a), whereas the cosmic SFR is the one of Calura \& Matteucci (2003) (panel b). In panels c) an d) are shown predictions and data for the Type Ia SN rate per unit galactic mass (SNuM) versus color and radio flux in radio galaxies, respectively. For the references about the data see Mannucci et al. (2006). Figure adapted from Mannucci et al. (2006).}\label{fig} 
\end{figure}

\begin{figure}
\includegraphics[width=9.0in,height=6.0in]{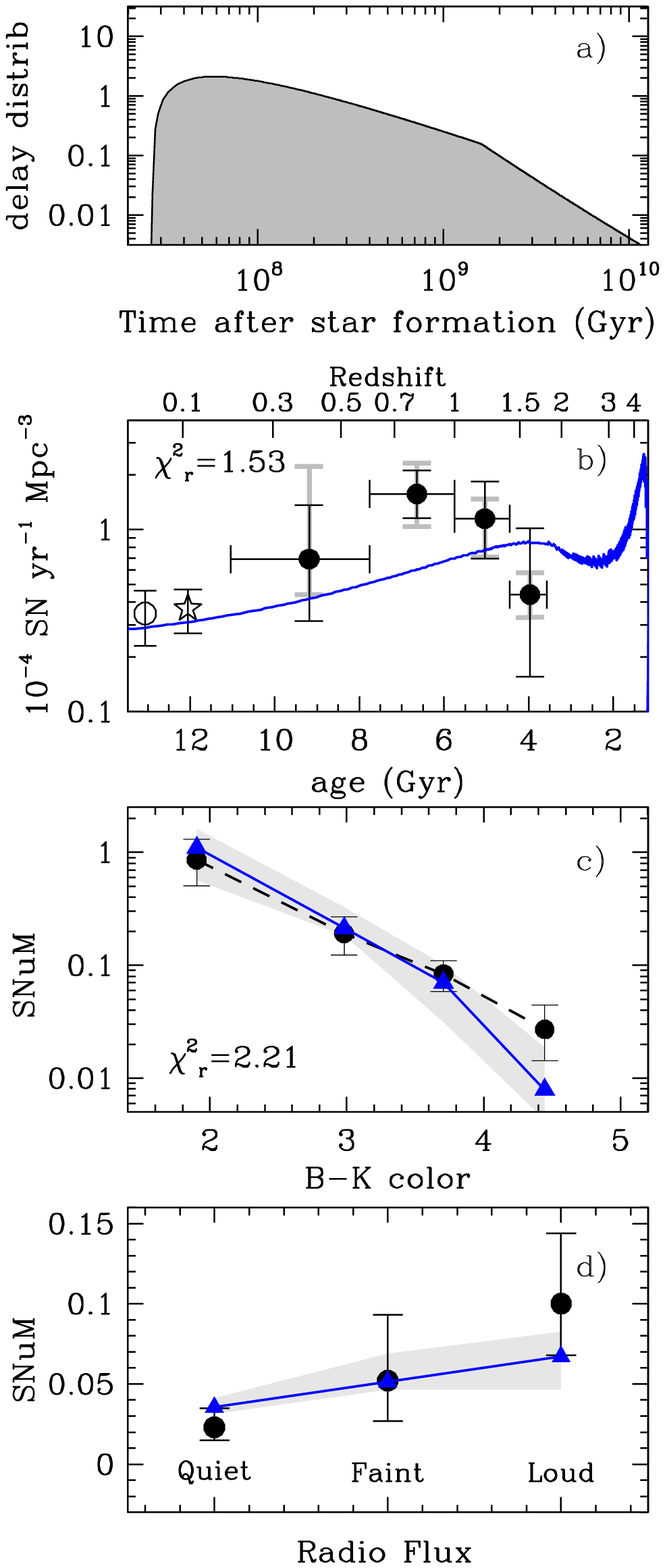}
\hfill
\caption{Theoretical cosmic Type Ia SN rates compared with observational data from Mannucci et al. (2005;2006). The progenitor model adopted assumes the delay-time distribution suggested by Matteucci \& Recchi (2001) (panel a), whereas the cosmic SFR is the one of Calura \& Matteucci (2003)(panel b). In panels c) an d) are shown predictions and data for the Type Ia SN rate per unit galactic mass (SNuM) versus color and radio flux in radio galaxies, respectively. For the references about the data see Mannucci et al. (2006).Figure adapted from Mannucci et al. (2006).
}\label{fig} 
\end{figure}
As one can see from these  Figures, the best DTD appears to be the one of Mannucci et al. (2006), although this conclusion is based only on the fit of the rates in radio galaxies. Concerning the cosmic Type Ia SN rate (panel b) the agreement is good for both  DTDs except for the point at the highest redshift which is highly uncertain. We need more detections of SNe Ia at high redshift before drawing any conclusion on the high redshift cosmic Type Ia SN rate.
In conclusion, the study of the Type Ia SN cosmic rates is very important to impose constraints on the SN progenitor model, the histories of SF in galaxies and, last but not least, the Hubble diagram.

\vfill\eject

\acknowledgement{ I warmly thank Eva Grebel and Ben Moore for inviting me to deliver these lectures in the beautiful village of Murren. I also thank my collaborators, Silvia Kuna Ballero, Francesco Calura, Gabriele Cescutti, Cristina Chiappini, Gustavo Lanfranchi, Antonio Pipino and Simone Recchi, whose precious help has allowed me to put together most of the material presented in these lectures.Finally, I am grateful to I.J. Danziger for his patience in reading the manuscript.}
%

%
%
%
%
%
%
%
%
%
%
%
%
%
 \bibliographystyle{}
\bibliography{}

%


\printindex
\end{document}